\documentclass{aa}
\usepackage[varg]{txfonts}
\usepackage{graphicx}
\usepackage{lscape}
\usepackage{placeins}
\usepackage[breaklinks]{hyperref}
\usepackage{siunitx}
\usepackage[version=4]{mhchem}
\usepackage{multirow}
\usepackage{booktabs}

\renewcommand{\linenumbers}{}
\newcommand{\abs}[1]{\lvert #1\rvert} 
\newcommand{\ur}[1]{\,\mathrm{#1}}
\newcommand{\PD}[2]{\frac{\partial #1}{\partial #2}} 
\newcommand{\PDPD}[3]{\frac{\partial ^{2}#1}{\partial #2\partial #3}} 
\newcommand{\iPD}[2]{\partial #1/\partial #2} 
\newcommand\micron{\mbox{$\mu$m}}

\usepackage{color}
\definecolor{cobalt}{rgb}{0.06, 0.2, 0.65}
\hypersetup{
  colorlinks,
  citecolor=cobalt,
  linkcolor=cobalt,
  urlcolor=cobalt,
}

\defcitealias{Taniguchi2021}{T21}

\begin{document}

\title{MAGIS (Measuring Abundances of red super Giants with Infrared Spectroscopy) project}

\subtitle{I. Establishment of an abundance analysis procedure for red supergiants and its evaluation with nearby stars}

\author{
	Daisuke Taniguchi\inst{\ref{inst:NAOJ},\ref{inst:Hongo}}\and 
	Noriyuki Matsunaga\inst{\ref{inst:Hongo},\ref{inst:LiH}}\and 
	Naoto Kobayashi\inst{\ref{inst:IoA},\ref{inst:Kiso}}\and 
	Mingjie Jian\inst{\ref{inst:Hongo},\ref{inst:Stockholm}}\and 
	Brian Thorsbro\inst{\ref{inst:Hongo},\ref{inst:CotedAzur}}\and 
	Kei Fukue\inst{\ref{inst:LiH},\ref{inst:Shiga-med}}\and 
	Satoshi Hamano\inst{\ref{inst:NAOJ},\ref{inst:LiH}}\and 
	Yuji Ikeda\inst{\ref{inst:LiH},\ref{inst:Photocoding}}\and 
	Hideyo Kawakita\inst{\ref{inst:LiH},\ref{inst:KSU}}\and 
	Sohei Kondo\inst{\ref{inst:LiH},\ref{inst:Kiso}}\and 
	Shogo Otsubo\inst{\ref{inst:LiH}}\and 
	Hiroaki Sameshima\inst{\ref{inst:IoA}}\and 
	Takuji Tsujimoto\inst{\ref{inst:NAOJ}}\and 
	Chikako Yasui\inst{\ref{inst:NAOJ}}
}

\institute{
	National Astronomical Observatory of Japan, 2-21-1 Osawa, Mitaka, Tokyo 181-8588, Japan \\
	\email{d.taniguchi.astro@gmail.com}\label{inst:NAOJ}
	\and 
	Department of Astronomy, Graduate School of Science, The University of Tokyo, 7-3-1 Hongo, Bunkyo-ku, Tokyo 113-0033, Japan\label{inst:Hongo}
	\and 
	Laboratory of Infrared High-resolution spectroscopy (LiH), Koyama Astronomical Observatory, Kyoto Sangyo University, Motoyama, Kamigamo, Kita-ku, Kyoto 603-8555, Japan\label{inst:LiH}
	\and 
	Institute of Astronomy, Graduate School of Science, The University of Tokyo, 2-21-1 Osawa, Mitaka, Tokyo 181-0015, Japan\label{inst:IoA}
	\and 
	Kiso Observatory, Institute of Astronomy, Graduate School of Science, The University of Tokyo, 10762-30 Mitake, Kiso-machi, Kiso-gun, Nagano 397-0101, Japan\label{inst:Kiso}
	\and 
	Department of Astronomy, Stockholm University, AlbaNova University centre, Roslagstullsbacken 21, 114 21 Stockholm, Sweden\label{inst:Stockholm}
	\and 
	Observatoire de la C\^ote d'Azur, CNRS UMR 7293, BP4229, Laboratoire Lagrange, F-06304 Nice Cedex 4, France\label{inst:CotedAzur}
	\and 
	Education Center for Medicine and Nursing, Shiga University of Medical Science, Seta Tsukinowa-cho, Otsu, Shiga, 520-2192, Japan\label{inst:Shiga-med}
	\and 
	Photocoding, 460-102 Iwakura-Nakamachi, Sakyo-ku, Kyoto 606-0025, Japan\label{inst:Photocoding}
	\and 
	Department of Astrophysics and Atmospheric Sciences, Faculty of Science, Kyoto Sangyo University, Motoyama, Kamigamo, \\ Kita-ku, Kyoto 603-8555, Japan\label{inst:KSU}
}

\date{Received Day Month 202X / accepted Day Month 202X}

\abstract
{
Given their high luminosities~($L\gtrsim 10^{4}L_{\sun }$), red supergiants~(RSGs) are good tracers of the chemical abundances of the young stellar population in the Milky Way and nearby galaxies. 
However, previous abundance analyses tailored to RSGs suffer some systematic uncertainties originating in, most notably, the synthesized molecular spectral lines for RSGs. 
}
{
We establish a new abundance analysis procedure for RSGs that circumvents difficulties faced in previous works, and test the procedure with ten nearby RSGs observed with the near-infrared high-resolution spectrograph WINERED~($0.97\text{--}1.32\,\micron $, $R=28\,000$). 
The wavelength range covered here is advantageous in that the molecular lines contaminating atomic lines of interest are mostly weak. 
}
{
We first determined the effective temperatures ($T_{\mathrm{eff}}$) of the targets with the line-depth ratio (LDR) method, and calculated the surface gravities ($\log g$) according to the Stefan-Boltzmann law. 
We then determined the microturbulent velocities ($v_{\mathrm{micro}}$) and metallicities ([Fe/H]) simultaneously through the fitting of individual \ion{Fe}{i} lines. 
Finally, we also determined the abundance ratios ([X/Fe] for element X) through the fitting of individual lines. 
}
{
We determined the [X/Fe] of ten elements (\ion{Na}{i}, \ion{Mg}{i}, \ion{Al}{i}, \ion{Si}{i}, \ion{K}{i}, \ion{Ca}{i}, \ion{Ti}{i}, \ion{Cr}{i}, \ion{Ni}{i}, and \ion{Y}{ii}). 
We estimated the relative precision in the derived abundances to be $0.04\text{--}0.12\ur{dex}$ for elements with more than two lines analyzed (e.g., \ion{Fe}{i} and \ion{Mg}{i}) and up to $0.18\ur{dex}$ for the other elements (e.g., \ion{Y}{ii}). 
We compared the resultant abundances of RSGs with the well-established abundances of another type of young star, namely the Cepheids, in order to evaluate the potential systematic bias in our abundance measurements, assuming that the young stars (i.e., both RSGs and Cepheids) in the solar neighborhood have common chemical abundances. 
We find that the determined RSG abundances are highly consistent with those of Cepheids within ${\lesssim }0.1\ur{dex}$ for some elements (notably [Fe/H] and [Mg/Fe]), which means the bias in the abundance determination for these elements is likely to be small. 
In contrast, the consistency is worse for some other elements (e.g., [Si/Fe] and [Y/Fe]). 
Nevertheless, the dispersion of the chemical abundances among our target RSGs is comparable with the individual statistical errors on the abundances. 
Hence, the procedure is likely to be useful to evaluate the relative difference in chemical abundances among RSGs. 
}
{}

\keywords{
	stars: abundances -- 
	stars: massive -- 
	stars: late-type -- 
	infrared: stars --
	Galaxy: abundances -- 
	methods: data analysis
}

\titlerunning{Chemical abundances of red supergiants. I}

\maketitle


\section{Introduction}\label{sec:IntroRSGFeH}

The Milky Way is the ``closest'' galaxy in the Universe, and provides us with unique opportunities to investigate the properties of a galaxy in great detail. 
Indeed, we are able to obtain full ``7D'' information on Galactic stars (with distances within several kiloparsecs): the position, velocity, and chemical abundances. 
In particular, chemical abundances provide clear information on stellar age and star-formation history, and thereby play an essential role in decoding the formation and merger history of the Galaxy~\citep{Helmi2020}. 

In the present paper, we focus on the young stellar population (younger than a few hundred million years) in the Galaxy, whose chemical abundances have been used as a tracer of the present-day gas~\citep[e.g.,][]{Grisoni2018,Esteban2022}. 
The chemical abundances of the young population are usually traced with \ion{H}{ii} regions, young open clusters, classical Cepheid variables, OB-type stars, and red supergiants (RSGs)~\citep[][and references therein]{Esteban2022,Magrini2023,Trentin2024,Braganca2019,Luck2014}. 
Among them, an increasing number of RSGs~\citep[ages ${\lesssim }50\ur{Myr}$;][]{Ekstrom2012} have recently been found in many parts of the Galaxy~\citep[e.g.,][]{Sellgren1987,Figer2006,Messineo2019} and in nearby galaxies~\citep[e.g.,][]{Massey2021,Ren2021b}. 

The metallicities indicated by the iron abundance [Fe/H] (and abundance ratios [X/Fe] for an element X) of RSGs in the Galaxy have been determined with high-resolution spectroscopy, that is, in the solar neighborhood~\citep{Luck1989,Luck2014,Carr2000,Ramirez2000,AlonsoSantiago2017,AlonsoSantiago2018,AlonsoSantiago2019,AlonsoSantiago2020,Negueruela2021,Fanelli2022}, the \object{Galactic center}~\citep{Carr2000,Ramirez2000,Cunha2007,Davies2009a}, and at the tip of the Galactic bar~\citep{Davies2009b,Origlia2013,Origlia2016,Origlia2019}. 
It has also been demonstrated that near-infrared (NIR) \textit{J}-band low-resolution spectroscopy of RSGs is useful for investigating the metallicities of young stars in galaxies, with notable applications to the solar neighborhood~\citep{Davies2010,Gazak2014}, the inner Galactic disk~\citep{Asad2020}, \object{the Magellanic Clouds}~\citep{Davies2015,Patrick2016}, \object{NGC 300}~\citep{Gazak2015}, \object{NGC 6822}~\citep{Patrick2015}, \object{NGC 55}~\citep{Patrick2017}, and \object{IC 1613}~\citep{Chun2022}. However, there remain some problems in the conventional abundance analysis procedures for RSGs adopted in these works, as highlighted below. 

\citet{Luck1989}, \citet{Luck2014}, and collaborators determined the stellar parameters and [Fe/H] of Galactic RSGs with the classical equivalent-width (EW) method~\citep[e.g.,][]{Jofre2019}, using \ion{Fe}{i} and \ion{Fe}{ii} lines in optical high-resolution spectra. 
\citet{Carr2000} also determined the effective temperatures $T_{\mathrm{eff}}$ and microturbulent velocities $v_{\mathrm{micro}}$ using the EW method, but with lines of the \ce{CO} molecule in the NIR \textit{HK} band. 
Some other works~\citep{Lambert1984,Davies2009a,Davies2009b,Origlia2013,Origlia2016,AlonsoSantiago2017,AlonsoSantiago2018,AlonsoSantiago2019,AlonsoSantiago2020} also used the EW method to measure chemical abundances after determining stellar parameters in some other ways. 
Whereas the EW method is often useful for late-type stars, EWs of RSGs are easily overestimated because broad absorption lines in RSGs tend to be severely contaminated with other lines, especially molecular lines. 
Thus, the stellar parameters and abundances derived with the EW method may also be biased~\citep{Cunha2007}. 

A method to overcome the contamination problem in the EW method is to fit individual lines, whereby observed and synthesized spectra are matched around the lines. 
It is important to use synthesized spectra that well reproduce the observed spectra. 
By fitting individual \ion{fe}{i} lines, \citet{Ramirez2000} and \citet{Fanelli2022} determined $v_{\mathrm{micro}}$ and [Fe/H] of (a part of) their target RSGs. 
\citet{Cunha2007}, \citet{Origlia2019}, \citet{Fanelli2022}, and \citet{Guerco2022} also fitted lines of various elements and determined chemical abundances. 

Still, it is difficult to resolve the degeneracy between stellar parameters when only using the fitting or the EW measurement of individual iron lines, especially in the case of RSGs. 
For this reason, many previous works determined some of the stellar parameters in an independent way to mitigate the difficulty before analyzing iron lines. 
For example, $T_{\mathrm{eff}}$ has often been determined on the basis of the relations between $T_{\mathrm{eff}}$ and the strengths of \ce{TiO} molecular lines in the optical~\citep{Levesque2005} or \ce{CO} and/or \ce{H2O} lines in the \textit{HK} bands~\citep{Ramirez2000,Blum2003,Cunha2007,Davies2008}. 
These relations were often calibrated with the RSGs whose $T_{\mathrm{eff}}$ are measured with interferometry or with the so-called \ce{TiO} method~\citep[e.g.,][]{Blum2003}. 
Alternatively, $T_{\mathrm{eff}}$ has also been determined with the \ce{C}-thermometer method proposed by \citet{Fanelli2021,Fanelli2022}, in which the balance between the carbon abundance derived with \ion{C}{i} and \ce{CO} lines is imposed. 
However, the results of any of these methods are, to a greater or lesser extent, affected by the CNO abundances, the discrepancy between the molecular spectra of real stars and synthesized spectra based on a simplified model, and/or potential systematic errors in the adopted $T_{\mathrm{eff}}$ values~\citep[see, e.g.,][]{Taniguchi2021}. 
Regarding other stellar parameters, $\log g$ is usually determined using the Stefan-Boltzmann law~\citep{Lambert1984,Carr2000,Ramirez2000,Cunha2007,Fanelli2022} because the small number of lines of ionized species in the spectra of RSGs makes it challenging to employ the so-called ionization equilibrium method~\citep[e.g.,][]{Jofre2019}. 
Another parameter, $v_{\mathrm{micro}}$ , has in some cases been determined with a relation of $v_{\mathrm{micro}}$ to $T_{\mathrm{eff}}$ and/or $\log g$ calibrated with observations or a 3D simulation~\citep{Ramirez2000,AlonsoSantiago2017,AlonsoSantiago2018,AlonsoSantiago2019,AlonsoSantiago2020,Negueruela2021}. 
The relations for RSGs have often been estimated by extrapolating those for giants and/or dwarfs. 

Another strategy for abundance analysis is to use global spectral synthesis. 
With optical spectra, \citet{AlonsoSantiago2017,AlonsoSantiago2018,AlonsoSantiago2019,AlonsoSantiago2020} and \citet{Negueruela2021} determined $T_{\mathrm{eff}}$, $\log g$, and [Fe/H] simultaneously by fitting narrow ranges of the spectra around many iron lines using the \textsc{SteParSyn} code~\citep{Tabernero2022}. 
With \textit{K}-band spectra, \citet{Cunha2007} fitted several \ion{Fe}{i} lines and determined $v_{\mathrm{micro}}$ and [Fe/H]. 
\citet{Davies2010,Davies2015}, \citet{Gazak2014,Gazak2015}, \citet{Patrick2015,Patrick2016,Patrick2017}, and \citet{Asad2020} fitted several lines of \ion{Fe}{i}, \ion{Mg}{i}, \ion{Si}{i}, and \ion{Ti}{i} and determined $T_{\mathrm{eff}}$, $\log g$, $v_{\mathrm{micro}}$, and [Fe/H] simultaneously, assuming $\text{[X/Fe]}=0.0\ur{dex}$. 
Similarly, \citet{Davies2009a,Davies2009b} and \citet{Origlia2013,Origlia2016,Origlia2019} determined $T_{\mathrm{eff}}$, $\log g$, and $v_{\mathrm{micro}}$ by matching the observed strengths and shapes of absorption bands of three molecules (\ce{CO}, \ce{OH}, and \ce{CN} in the NIR) with synthesized ones. 
These methods are useful when synthesized spectra that well reproduce observed ones are available, which is usually not the case for RSGs. 

In summary, conventional abundance analysis procedures of RSGs are subject to uncertainties related to at least one of the following points: (1)~molecular lines (or $T_{\mathrm{eff}}$ values of RSGs in the literature), (2)~EW measurement, (3)~an extrapolated $\log g$--$v_{\mathrm{micro}}$ relation, and (4)~the assumption on the chemical-abundance ratios for some elements. 
Any one of these four points may result in a systematic bias on the derived stellar parameters. 
Moreover, most of the conventional procedures have not been well tested with RSGs with the known reliable abundances or at least with the abundances that can be predicted. 
Such a test is crucial when analyzing spectra of types for which the analysis procedure has not been well established, such as NIR spectra of late-type stars and the spectra of M-type stars~\citep[e.g.,][]{Smith2013,Ishikawa2020,Nandakumar2023}. 

Here, circumventing all the above problems, we establish a procedure to derive the chemical abundances of RSGs from observed spectra based on fitting individual atomic lines, and test this procedure with real stars. 
Specifically, we use high-resolution spectra of ten nearby RSGs in the NIR \textit{YJ} bands ($0.97\text{--}1.32\,\micron $; Sect.~\ref{sec:ObsReduc}). 
The wavelength range used in this procedure is   advantageous in that it is the least affected by molecular lines in the optical and NIR wavelength ranges~\citep{Coelho2005,Davies2010}. 
With these spectra, we present our procedure for the abundance analysis of RSGs~(Sect.~\ref{sec:method}), and extensively evaluate the procedure~(Sect.~\ref{sec:results}).


\section{Observations and data reduction}\label{sec:ObsReduc}

\begin{figure*}
\centering 
\includegraphics[width=18cm]{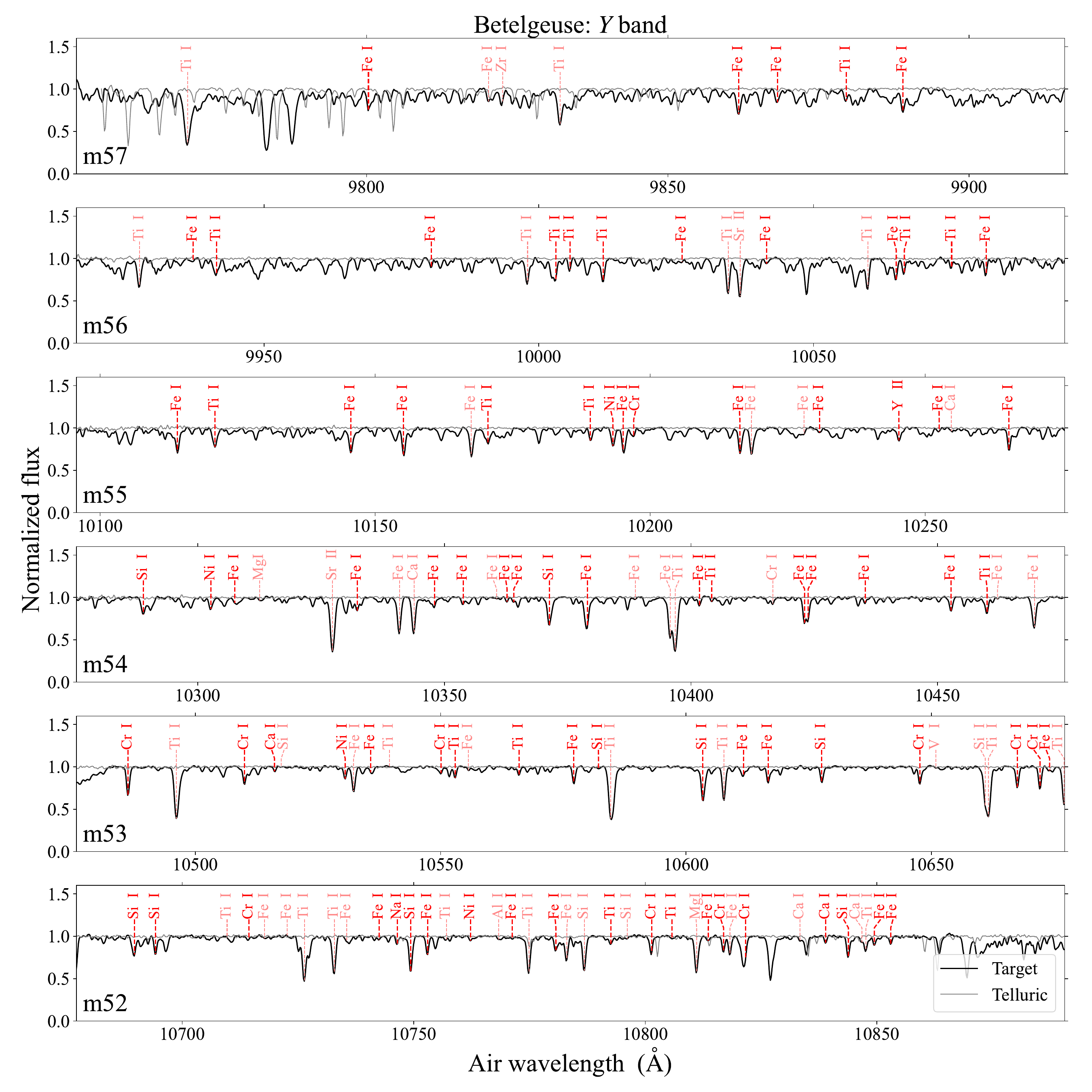}
\caption{Example of a RSG spectrum observed with WINERED: namely that of \object{Betelgeuse} in the \textit{Y} band (echelle orders 57-52). Black thick lines show the reduced spectrum of \object{Betelgeuse}, after telluric lines were removed. Gray thin lines show the spectrum of the corresponding telluric standard A0V star, \object{HIP 27830}, after the stellar lines were removed. Red thick vertical dashed lines near the top edge of each panel indicate the wavelengths of the lines from the VALD3 and/or MB99 line list used for measuring [X/H]. Light-red thin vertical dashed lines indicate the wavelengths of the candidate lines preselected in Sects.~\ref{ssec:LineSelection} and \ref{sssec:AnalysisXH} from VALD3 and/or MB99 but eventually rejected in Sects.~\ref{sssec:DetVFeH} and \ref{sssec:AnalysisXH} for both line lists. }
\label{fig:BetelgeuseSpec1}
\end{figure*}

\begin{figure*}
\centering 
\includegraphics[width=18cm]{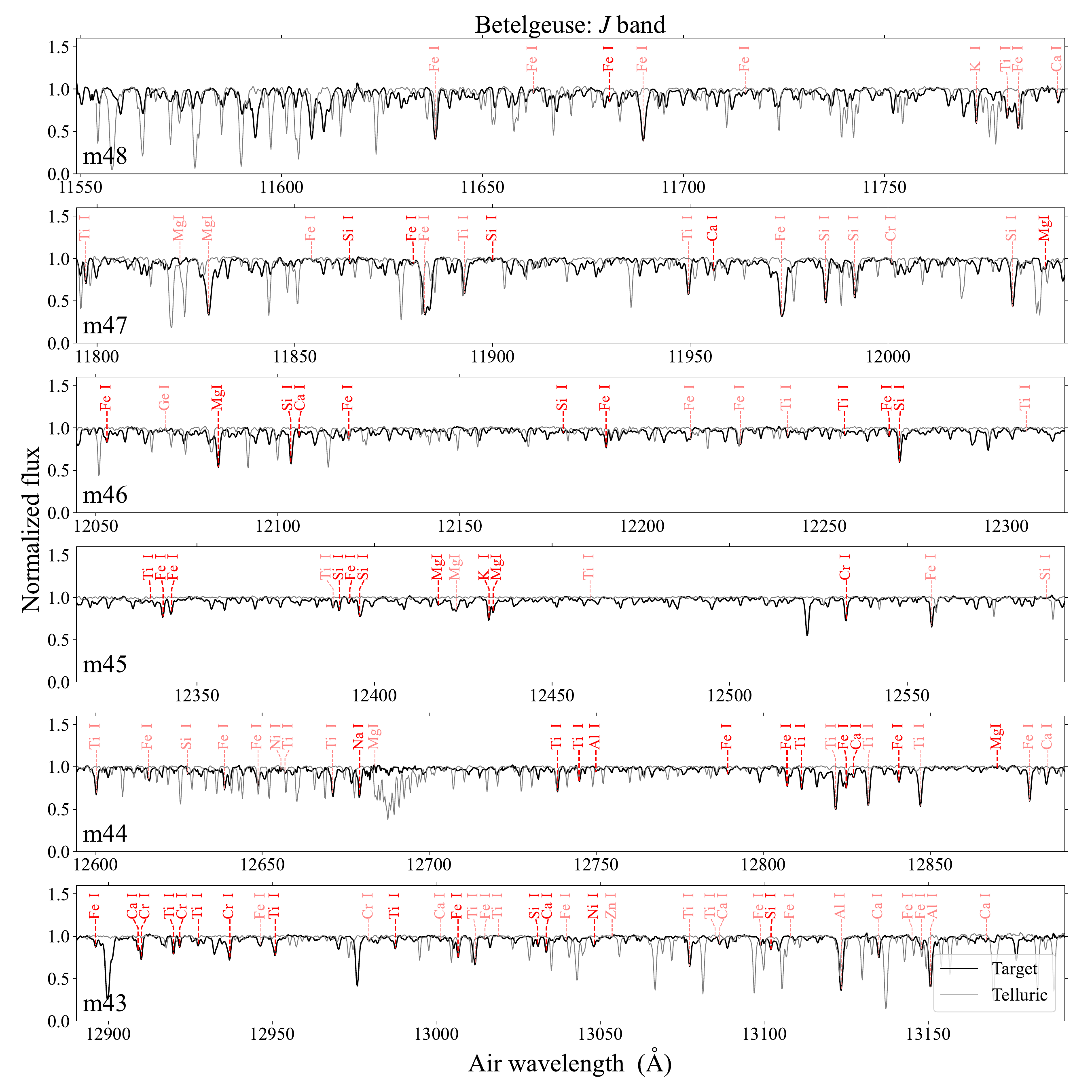}
\caption{Same as Fig.~\ref{fig:BetelgeuseSpec1} but for the \textit{J} band (echelle orders 48--43). }
\label{fig:BetelgeuseSpec2}
\end{figure*}

In this paper, we use the NIR high-resolution spectra of ten nearby RSGs observed by \citet[hereafter \citetalias{Taniguchi2021}]{Taniguchi2021}. 
The RSGs are located within ${\sim }2\ur{kpc}$ of the Sun, and their locations are translated into galactocentric distances ($R_{\mathrm{GC}}$) of $8\lesssim R_{\mathrm{GC}}\lesssim 10\ur{kpc}$. 
Their $T_{\mathrm{eff}}$ and bolometric luminosity $L$ were determined by \citetalias{Taniguchi2021}. 

All the objects were observed using the NIR high-resolution spectrograph WINERED installed on the Nasmyth platform of the $1.3\ur{m}$ Araki Telescope at Koyama Astronomical Observatory of Kyoto Sangyo University in Japan~\citep{Ikeda2022}. 
Spectra covering a wavelength range from $0.90$ to $1.36\,\micron $~(\textit{z$^{\prime }$}, \textit{Y} and \textit{J} bands) with a spectral resolution of $R=28\,000$ were collected using the WINERED WIDE mode with the nodding pattern of A--B--B--A or O--S--O. 
All the targets are bright~($-3.0\leq J\leq 3.0\ur{mag}$), and the total integration time for each target within the slit ranged between $3\text{--}180\ur{sec}$, with which a S/N per pixel of $100$ or higher (${>}200$ for most echelle orders of most stars) was achieved. 
Telluric standard stars~\citep[slow-rotating A0V stars in most cases; see][]{Sameshima2018} were also observed, and their spectra were used to subtract the telluric absorption. 
Table~\ref{table:ObsLog} summarizes the observation log. 

As in \citetalias{Taniguchi2021}, we analyzed the echelle orders 57--52~(\textit{Y} band; $0.97\text{--}1.09\,\micron $) and 48--43~(\textit{J} band; $1.15\text{--}1.32\,\micron $) only among the available orders 61--42 because stellar atomic lines in the unselected orders are severely contaminated with lines of the telluric and/or stellar \ce{CN} molecule. 

The initial steps of the spectral reduction were performed with WINERED Automatic Reduction Pipeline~\citep[WARP;][]{Hamano2024}\footnote{\url{https://github.com/SatoshiHamano/WARP}}. 
Then, the telluric absorption lines were removed, using the observed spectra of the A0V stars after their intrinsic lines had been removed with the method described in \citet{Sameshima2018}. 
We did not remove the telluric lines for the 55th--53rd orders~($1.01$ to $1.07\,\micron $) of the objects taken in winter, in which almost no significant telluric lines were present. 
Finally, the radial velocities were measured by comparing the observed and synthesized spectra, the wavelength scale was adjusted to the one in the standard air at rest using the formula given by \citet{Ciddor1996}, and the continuum was renormalized. 
An example of the reduced spectrum is presented in Figs.~\ref{fig:BetelgeuseSpec1} and \ref{fig:BetelgeuseSpec2}. 

\begin{table}
\centering 
\caption{Observation log of our sample RSGs observed by \citetalias{Taniguchi2021}. }
\label{table:ObsLog}
\begin{tabular}{lrlc}\hline \hline 
Name & HD & Sp. type\tablefootmark{a} & Obs. date \\ \hline 
\object{$\zeta $  Cep} & 210745 & K1.5Ib & 2015-08-08 \\
\object{41 Gem} & 52005 & K3--Ib & 2015-10-28 \\
\object{$\xi $ Cyg} & 200905 & K4.5Ib--II & 2016-05-14 \\
\object{V809 Cas} & 219978 & K4.5Ib & 2015-10-31 \\
\object{V424 Lac} & 216946 & K5Ib & 2015-07-30 \\
\object{$\psi ^{1}$ Aur} & 44537 & K5--M1Iab--Ib & 2013-02-22 \\
\object{TV Gem} & 42475 & M0--M1.5Iab & 2016-01-19 \\
\object{BU Gem} & 42543 & M1--M2Ia--Iab & 2016-01-19 \\
\object{Betelgeuse} & 39801 & M1--M2Ia--Iab & 2013-02-22 \\
\object{NO Aur} & 37536 & M2Iab & 2015-10-28 \\
\hline 
\end{tabular}
\tablefoot{
\tablefoottext{a}{Taken from SIMBAD~\citep{Wenger2000} on 2020 April 26. }
}
\end{table}


\section{Chemical abundance analysis: Method}\label{sec:method}

In our procedure, we first determine $T_{\mathrm{eff}}$, using the line-depth ratio (LDR) method~(Sect.~\ref{ssec:Teff}), which neither relies on molecular lines nor is not calibrated against literature $T_{\mathrm{eff}}$ of RSGs. 
Then, we estimate $\log g$, using the Stefan-Boltzmann law, as has been done in many other works~(Sect.~\ref{ssec:logg}). 
Next, we determine $v_{\mathrm{micro}}$ and [Fe/H] simultaneously, fitting small wavelength ranges of spectra around individual \ion{Fe}{i} lines under the assumption that the derived iron abundances from individual lines are independent of the line strength~(Sect.~\ref{ssec:vmicFeH}). 
Finally, we determine [X/Fe] of elements other than iron by the fitting for individual lines~(Sect.~\ref{sssec:AnalysisXH}). 

For the spectral analysis in this paper, we developed the \textsc{Python3} code named \textsc{Octoman} (Optimization Code To Obtain Metallicity using Absorption liNes), which is a wrapper of the spectral synthesis code MOOG~\citep{Sneden1973,Sneden2012}. 
The code mainly comprises of two functions: spectral synthesis and fitting of individual lines, as detailed in Appendices~\ref{app:MOOG} and \ref{app:Octoman}, respectively. 
The code has already been used in some studies for the abundance analysis of late-type stars~\citep{Matsunaga2023,Elgueta2024}. 
In this work, we used the MARCS spherical model atmospheres with $M=5M_{\odot }$~\citep{Gustafsson2008}. 
We used the VALD3 and MB99 line lists and compared the results to identify potential differences if any. 
We adopted the solar abundance pattern and isotope ratios presented by \citet{Asplund2009} throughout the paper unless otherwise specified.

\subsection{Effective temperature ($T_{\mathrm{eff}}$)}\label{ssec:Teff}

We adopted $T_{\mathrm{eff}}$ of the sample RSGs determined in \citetalias{Taniguchi2021} using the LDR method~\citep{Gray1991}. 
In \citetalias{Taniguchi2021}, they used $11$ LDR--$T_{\mathrm{eff}}$ relations calibrated against nine solar-metallicity red giants to determine $T_{\mathrm{eff}}$ of the RSGs. 
\citetalias{Taniguchi2021} estimated the resultant precision of $T_{\mathrm{eff}}$ to be ${\sim }40\ur{K}$ when analyzing a high-S/N spectrum, although they might be less precise, depending on several parameters including S/N, $T_{\mathrm{eff}}$, and macroturbulent velocity $v_{\mathrm{macro}}$. 
\citetalias{Taniguchi2021} also estimated the systematic bias in the derived $T_{\mathrm{eff}}$ due to effects of $\log g$, $v_{\mathrm{micro}}$, line broadening, and non-local thermodynamic equilibrium (non-LTE) to be ${\sim }100\ur{K}$. 
In the present work, we adopted the recalculated $T_{\mathrm{eff}}$, using re-reduced spectra of the RSGs. 
The updates in $T_{\mathrm{eff}}$ values are mostly within $\lesssim 30\ur{K}$, which is much smaller than the systematic bias of ${\sim }100\ur{K}$.

\subsection{Surface gravity ($\log g$)}\label{ssec:logg}
The \textit{YJ}-band spectra of RSGs contain no useful \ion{Fe}{ii} lines and only a small number of lines of ionized atoms other than iron. 
Hence, it is difficult to determine $\log g$ of RSGs with the ionization balance method. 
Also, no asteroseismic measurement was available for $\log g$ of the target RSGs. 
We thus estimated evolutionary $\log g$, using the Stefan-Boltzmann law instead, as described below. 

First, we determined the bolometric luminosity $L$ in the way described in Sect.~4.3 of \citetalias{Taniguchi2021}; i.e., we calculated $L$ of each target RSG with 
\begin{equation}
\log (L/L_{\sun })=\frac{K_{\mathrm{s}}-A(K_{\mathrm{s}})+\mathrm{BC}_{K_{\mathrm{s}}}+5\log \varpi -10-M_{\mathrm{bol},\sun }}{-2.5}\text{,}
\end{equation}
where $K_{\mathrm{s}}$ is the \textit{K}$_{\mathrm{s}}$-band magnitude taken from the 2MASS point source catalog~\citep{Cutri2003,Skrutskie2006}, $A(K_{\mathrm{s}})$ is the extinction in the \textit{K}$_{\mathrm{s}}$ band converted from $A(V)$ listed in \citet{Levesque2005} according to the reddening law $A(K)/A(V)=0.1137$ given by \citet{Cardelli1989} where we assumed $R_{V}=3.1$, $\mathrm{BC}_{K_{\mathrm{s}}}$ is the bolometric correction estimated by means of interpolation of the relation between $T_{\mathrm{eff}}$ and $\mathrm{BC}_{K}$ presented by \citet{Levesque2005}, $\varpi $ is the parallax in mas taken from the \textsc{Hipparcos} catalog~\citep{vanLeeuwen2007} for \object{Betelgeuse} and from the \textit{Gaia} DR3~\citep{Gaia2016,GaiaVallenari2023} for the others, where we corrected for the systematic bias according to the recipe presented by \citet{Lindegren2021b}, and $M_{\mathrm{bol},\sun }=4.74\ur{mag}$~\citep[IAU 2015 recommendation;][]{Prsa2016} is the bolometric magnitude of the Sun. 

As discussed in \citetalias{Taniguchi2021}, $T_{\mathrm{eff}}$ and $\log (L/L_{\sun })$ that we determined were in good agreement with the Geneva's stellar evolution model with rotation presented by \citet{Ekstrom2012} on the HR diagram; i.e., the pair of our estimated values $(T_{\mathrm{eff}},L)$ fell in the region where RSGs are expected to stay for a long period. 
Then, we estimated the current masses, $M$, of the RSGs, by means of the visual inspection of the HR diagram. 
With these masses, we calculated evolutionary surface gravity $\log g$ for the gravity $g$ in the cgs unit system, using the Stefan-Boltzmann law, as 
\begin{equation}
\log g=\log (M/M_{\sun })+4\log T_{\mathrm{eff}}-\log (L/L_{\sun })-C\text{,}
\end{equation}
where $C$ represents $\log L_{\sun }/(4\pi \sigma GM_{\sun })=10.607$ with the Stefan-Boltzmann constant $\sigma $. 

Table~\ref{table:loggRSG} summarizes the results of the calculations. 
In the calculations, the errors and the median values were computed with the Monte Carlo method~\citep{Anderson1976}, with excluding samples with $\varpi <0$ and/or $A(K_{\mathrm{s}})<0$. 
We ignored the systematic errors in the input parameters mentioned in literature, which could, if properly taken into account, increase the errors in $\log g$ that we determined. 
Nevertheless, the systematic effect would not affect the final results of the abundance analysis for most elements because varying $\log g$ by, e.g., $0.5$, has little effect~($<0.1\ur{dex}$) on the resultant [Fe/H]~\citep{Origlia2019,Kondo2019}. 
Also, we ignored the turbulent pressure~\citep{Chiavassa2011b}, which would decrease $\log g$ by up to $0.3$~\citep{Davies2015}. 

\begin{table*}
\centering 
\caption{Derived $\log g$ and related values. }
\label{table:loggRSG}

\begin{tabular}{l crclcccr}\hline \hline
Name & $\varpi $  (mas) & \multicolumn{1}{c}{$K_{\mathrm{s}}$  (mag)} & $A(V)$  (mag) & \multicolumn{1}{c}{$T_{\mathrm{eff}}$  (K)} & $\mathrm{BC}_{K_{\mathrm{s}}}$  (mag) & $\log (L/L_{\sun })$ & $M/M_{\sun }$ & \multicolumn{1}{c}{$\log g$} \\ \hline
$\zeta $ Cep & $3.319\pm 0.146$ & $0.343\pm 0.170$ & $0.00\pm 0.15$ & $4073\pm 31$ & $2.49\pm 0.02$ & $3.73^{+0.08}_{-0.08}$ & $8\text{--}9$ & $1.03^{+0.08}_{-0.08}$ \\
41 Gem & $0.754\pm 0.091$ & $2.107\pm 0.336$ & $0.00\pm 0.15$ & $3962\pm 27$ & $2.56\pm 0.02$ & $4.28^{+0.17}_{-0.17}$ & $11\text{--}14$ & $0.60^{+0.18}_{-0.18}$ \\
$\xi $ Cyg & $2.859\pm 0.127$ & $-0.038\pm 0.202$ & $0.00\pm 0.15$ & $3893\pm 26$ & $2.61\pm 0.02$ & $3.96^{+0.09}_{-0.09}$ & $8\text{--}10$ & $0.75^{+0.10}_{-0.10}$ \\
V809 Cas & $1.030\pm 0.039$ & $0.788\pm 0.176$ & $2.17\pm 0.15$ & $3799\pm 36$ & $2.68\pm 0.03$ & $4.58^{+0.08}_{-0.08}$ & $13\text{--}15$ & $0.28^{+0.08}_{-0.08}$ \\
V424 Lac & $1.429\pm 0.113$ & $0.724\pm 0.178$ & $0.31\pm 0.15$ & $3767\pm 48$ & $2.71\pm 0.04$ & $4.23^{+0.10}_{-0.10}$ & $9\text{--}12$ & $0.49^{+0.12}_{-0.12}$ \\
$\psi ^{1}$ Aur & $0.478\pm 0.110$ & $0.577\pm 0.186$ & $0.62\pm 0.15$ & $3777\pm 60$ & $2.70\pm 0.05$ & $5.26^{+0.24}_{-0.20}$ & $9\text{--}25$ & $-0.35^{+0.28}_{-0.36}$ \\
TV Gem & $0.507\pm 0.135$ & $0.947\pm 0.188$ & $2.17\pm 0.15$ & $3739\pm 101$ & $2.73\pm 0.08$ & $5.12^{+0.28}_{-0.22}$ & $9\text{--}21$ & $-0.29^{+0.29}_{-0.36}$ \\
BU Gem & $0.607\pm 0.125$ & $0.806\pm 0.230$ & $2.01\pm 0.15$ & $3896\pm 70$ & $2.61\pm 0.05$ & $5.06^{+0.22}_{-0.19}$ & $9\text{--}21$ & $-0.15^{+0.25}_{-0.31}$ \\
Betelgeuse & $6.55\pm 0.83$ & $-4.378\pm 0.186$ & $0.62\pm 0.15$ & $3633\pm 37$ & $2.81\pm 0.03$ & $4.92^{+0.14}_{-0.13}$ & $15\text{--}19$ & $-0.06^{+0.14}_{-0.15}$ \\
NO Aur & $0.961\pm 0.093$ & $0.971\pm 0.196$ & $1.39\pm 0.15$ & $3663\pm 30$ & $2.79\pm 0.02$ & $4.49^{+0.12}_{-0.11}$ & $10\text{--}13$ & $0.21^{+0.13}_{-0.13}$ \\
\hline 
References & \multicolumn{1}{c}{1,2} & \multicolumn{1}{c}{3} & \multicolumn{1}{c}{4} & \multicolumn{1}{c}{TW} & \multicolumn{1}{c}{TW} & \multicolumn{1}{c}{TW} & \multicolumn{1}{c}{TW} & \multicolumn{1}{c}{TW} \\
\hline
\end{tabular}
\tablefoot{
See main text for the definitions of the listed quantities. Three quantities, $\varpi $, $K_{\mathrm{s}}$, and $A(V)$, and their respective errors were taken from the literature, $M/M_{\odot }$ was estimated by visual inspection of the HR diagram, and the remaining quantities were computed using the Monte Carlo method. 
}
\tablebib{
(1)~\citet{GaiaVallenari2023}; (2)~\citet{vanLeeuwen2007}; (3)~\citet{Cutri2003}; (4)~\citet{Levesque2005}; (TW)~This work. 
}
\end{table*}

\subsection{Line selection for the abundance analysis}\label{ssec:LineSelection}

For the abundance measurements, we chose the atomic lines that are comparatively free from contamination from surrounding lines among all the neutral and first-ionized atomic lines in the VALD3 and MB99 line lists. 
We considered lines in wavelength ranges of $9,760\text{--}10,860\,\text{\AA }$ for the \textit{Y} band and $11,620\text{--}13,170\,\text{\AA }$ for the \textit{J} band~(Sect.~\ref{sec:ObsReduc}). 
Since the MB99 list contains only the lines with the wavelengths longer than $10,000\,\text{\AA }$, the spectra within $9,760\text{--}10,000\,\text{\AA }$ were analyzed only with VALD3. 
We excluded the lines of carbon, nitrogen, and oxygen\footnote{There are three \ion{C}{i} lines in VALD3 ($\lambda 10685.34$, $10707.32$, and $10729.53\,\text{\AA }$) and six in MB99 ($\lambda 10683.09$, $10685.36$, $10691.26$, $10707.34$, $10729.54$, $11895.78\,\text{\AA }$) that satisfy the line-selection conditions. Regarding the other lines from neutral or ionized carbon, nitrogen, and oxygen, four \ion{N}{i} lines in VALD3 ($\lambda 10397.738$, $10398.155$, $10407.169$, and $10407.587\,\text{\AA }$) were deeper than $0.01$ in the synthesized spectra of a theoretical RSG having the stellar parameters of RSG3 but with a nitrogen-rich abundance pattern as seen sometimes for RSGs~\citep[e.g.,][]{Lambert1984,Carr2000}. Whichever, these four lines are severely contaminated with other lines and were not detected in our observed spectra of the target RSGs. }, along with hydrogen and helium, because the CNO abundances had been adjusted in such a way that the synthesized \ce{CN} spectra well reproduced the observed ones as we see later (Sect.~\ref{ssec:CNfitting}). 
We note that during line selection, we assumed $\ce{^{12}C}/\ce{^{13}C}=10$ as the typical isotope ratio of carbon for RSGs~\citep{Hinkle1976,Milam2009,Fanelli2022} and used solar isotope ratios from \citet{Asplund2009} for the other elements. 

In order to evaluate the amount of contamination for each atomic line, we considered synthesized spectra of a theoretical RSG, RSG3 defined in Table~3 of \citetalias{Taniguchi2021}, having the solar metallicity and $T_{\mathrm{eff}}=3850\ur{K}$. 
Specifically, we synthesized three types of spectra for RSG3 for the wavelength range around each line with different groups of lines: (1)~\texttt{All} --- all the atomic and molecular lines, (2)~\texttt{OneOut} --- all the lines except for the line of interest \citep[see Fig.~3 in][for three examples of \texttt{OneOut} spectra]{Kondo2019}, and (3)~\texttt{OnlyOne} --- only the line of interest. 
With these synthesized spectra, we first measured the depth $d_{\text{\texttt{OnlyOne}}}$ from unity in the wavelength $\lambda _{0}$ of the line in \texttt{OnlyOne}, excluding the lines shallower than $0.03$ for \ion{Fe}{i} lines and $0.01$ for the other species. 
Then, following \citet{Kondo2019}, we computed two EWs $W_{1}^{\mathtt{\alpha }}$ and $W_{2}^{\mathtt{\alpha }}$, where $\mathtt{\alpha }$ indicates \texttt{All} or \texttt{OneOut}, around the line, as defined by 
\begin{equation}
W_{i}^{\mathtt{\alpha }}\equiv \int _{\lambda _{0}(1-\Delta _{i}/2c)}^{\lambda _{0}(1+\Delta _{i}/2c)}(1-f_{\mathtt{\alpha }})\,d\lambda \text{,}
\end{equation}
where $f_{\texttt{All}}(\lambda )$ and $f_{\texttt{OneOut}}(\lambda )$ indicate the synthesized spectra for \texttt{All} and \texttt{OneOut}, respectively, and $c$ indicates the speed of light. 
In the computation, we considered two wavelength ranges ($\Delta _{1}$ and $\Delta _{2}$ corresponding to $40$ and $80\ur{\si{km.s^{-1}}}$, respectively). 
Both $\Delta _{1}$ and $\Delta _{2}$ were larger than those for red giants used by \citet{Kondo2019} considering that $v_{\mathrm{macro}}$ of RSGs are larger than those of red giants. 
With these EWs, we defined two indices $\beta _{1}$ and $\beta _{2}$ as 
\begin{equation}
\beta _{1}\equiv W_{1}^{\text{\texttt{OneOut}}}/W_{1}^{\text{\texttt{All}}},\quad \beta _{2}\equiv (W_{2}^{\texttt{OneOut}}-W_{1}^{\texttt{OneOut}})/W_{1}^{\text{\texttt{All}}}\text{.}\label{eq:beta1beta2}
\end{equation}
The two indices measure the degrees of contamination by other lines in the core part of the line ($\beta _{1}$) and in the continuum region ($\beta _{2}$). 
We chose the lines with $\beta _{1}<0.5$ and $\beta _{2}<1.0$ to exclude the highly contaminated lines. 
Furthermore, we removed the lines around which either the hydrogen Paschen series nor Helium $10830\,\text{\AA }$ lines is present within $\pm 60\ur{\si{km.s^{-1}}}$. 
We note that when two or more lines from an element were located within $(\Delta _{1}+\Delta _{2})/2=60\ur{\si{km.s^{-1}}}$, only the line with the largest $d_{\text{\texttt{OnlyOne}}}$ was used. 

\begin{figure*}
\centering 
\includegraphics[width=18cm]{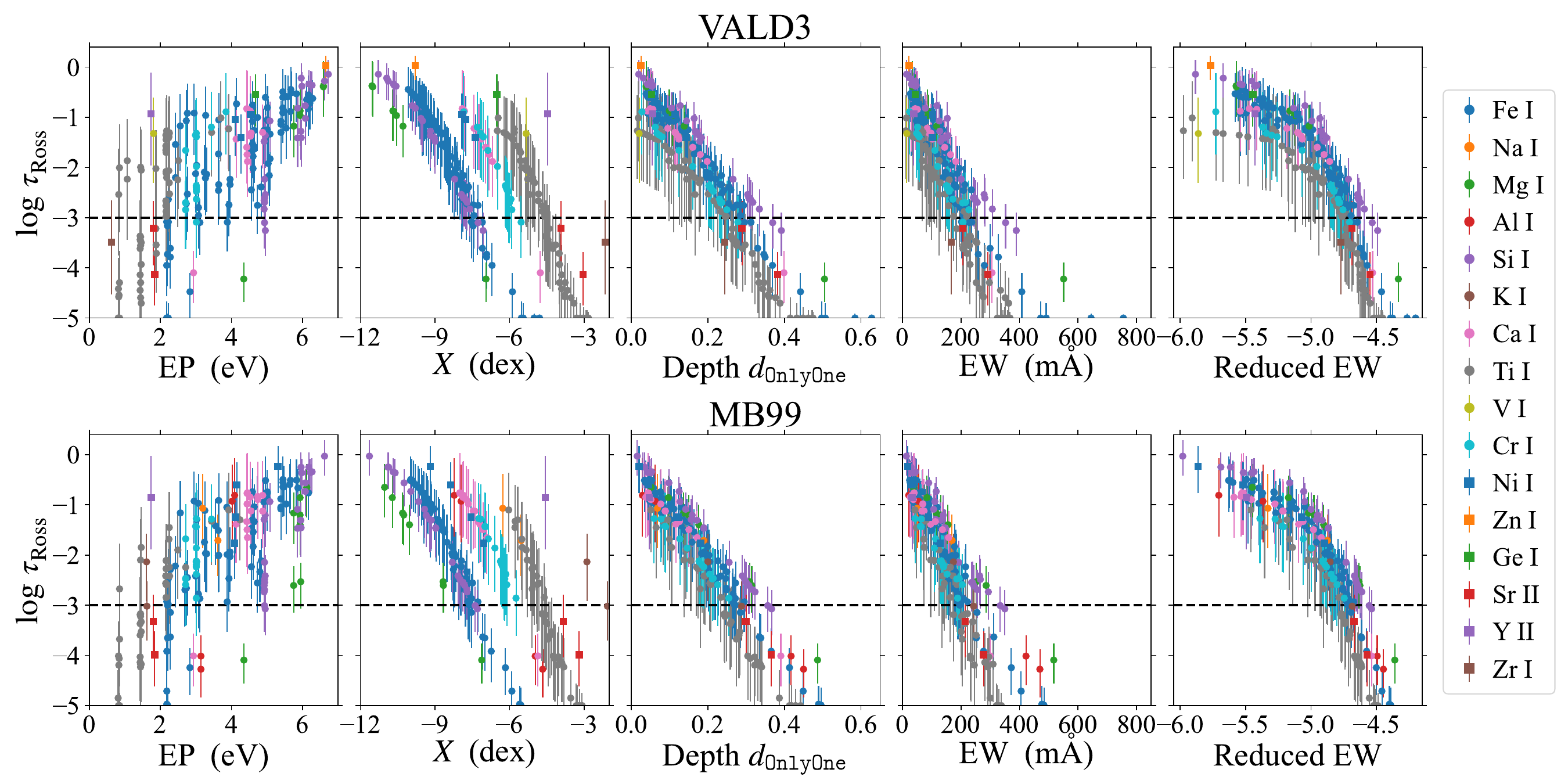}
\caption{$\log \tau _{\mathrm{Ross}}^{}$ of the line-forming layers of the lines preselected in Sect.~\ref{ssec:LineSelection} as functions of the EP, the $X$ index at $3850\ur{K}$, the model depth $d_{\text{\texttt{OnlyOne}}}$, the model EW, and the reduced model EW. Top and bottom panels show the results with employed line lists of VALD3 and MB99, respectively. The vertical error bar represents the range of Rosseland-mean optical depth where the contribution function for the line is larger than the half of the maximum value at $\log \tau _{\mathrm{Ross}}^{}$. Horizontal black dashed line at $\log \tau _{\mathrm{Ross}}^{}=-3.0$ indicates the final criteria of our line selection. }
\label{fig:CFRSG3}
\end{figure*}

Applying these criteria to our sample atomic lines left lines of \ion{Mg}{i}, \ion{Si}{i}, \ion{Ca}{i}, \ion{Ti}{i}, \ion{V}{i}, \ion{Cr}{i}, \ion{Fe}{i}, \ion{Ni}{i}, \ion{Zn}{i}, \ion{Ge}{i}, \ion{Sr}{ii}, \ion{Y}{ii}, and \ion{Zr}{i} for VALD3 and \ion{Na}{i}, \ion{Mg}{i}, \ion{Al}{i}, \ion{Si}{i}, \ion{K}{i}, \ion{Ca}{i}, \ion{Ti}{i}, \ion{Cr}{i}, \ion{Fe}{i}, \ion{Ni}{i}, \ion{Sr}{ii}, and \ion{Y}{ii} for MB99. 
Especially, the criteria left $51$ and $32$ \ion{Fe}{i} lines in the \textit{Y} and \textit{J} bands, respectively, for VALD3, and $42$ and $30$ lines for MB99. 
Fig.~\ref{fig:CFRSG3} shows $\log \tau _{\mathrm{Ross}}$ calculated with the RSG3 model as functions of some line parameters: excitation potential (EP), $X$ index described in Sect.~\ref{sssec:DetVFeH}, $d_{\text{\texttt{OnlyOne}}}$, EW, and reduced EW. 
We note that one of our line-selection conditions, $\log \tau _{\mathrm{Ross}}>-3$, corresponds to $\log (\mathrm{EW}/\lambda )\lesssim -4.8\text{--}-4.6$ (or $\mathrm{EW}\lesssim 150\text{--}350\ur{m\text{\AA }}$), and the exact threshold depends mainly on the species (and wavelength) of interest.

\subsection{Adjustment of the strengths of \ce{CN} lines}\label{ssec:CNfitting}

Since \textit{YJ}-band spectra of RSGs contain many \ce{CN} lines, which contaminate the atomic lines of our interest for the abundance analysis, the difference in the strengths of \ce{CN} lines between the observed and synthesized spectra, if exists, affects the chemical abundance measurements using atomic lines. 
The strengths of \ce{CN} lines depend on two abundance ratios, [C/O] and [N/H], together with stellar parameters and metallicity~(Appendix~\ref{app:CN}). 
Since the two ratios affect the strengths of weak and strong \ce{CN} lines differently, we optimized the two ratios, together with $\ce{^{12}C}/\ce{^{13}C}$, for each star to well reproduce the observed \ce{CN} strengths with synthesized ones for each of the VALD3 and MB99 line lists in the following procedure. 
We note that this process is simply for the purpose of adjusting the \ce{CN} line strengths and is not intended for determining CNO abundances. 

First, we chose the \ce{CN} lines that are relatively free from contamination by lines of other species. 
For this, we synthesized two types of spectra, \texttt{All} and \texttt{OnlyOne} defined in Sect.~\ref{ssec:LineSelection}, around all the \ce{CN} lines listed by \citet{Sneden2014}. 
We also synthesized another type of spectra named \texttt{SameElIonOut} (abbreviating same-element-ion-out) that is synthesized with the list of all the lines except for those of the species of interest, which is \ce{^{12}C^{14}N}, \ce{^{13}C^{14}N}, or \ce{^{12}C^{15}N} in this case. 
With these spectra, we defined two indices, $\beta _{3}$ and $\beta _{4}$, as 
\begin{align}
&\beta _{3}\equiv W_{1}^{\text{\texttt{SameElIonOut}}}/W_{1}^{\text{\texttt{All}}} \\
&\beta _{4}\equiv (W_{2}^{\texttt{SameElIonOut}}-W_{1}^{\texttt{SameElIonOut}})/W_{1}^{\text{\texttt{All}}}\text{.}
\end{align}
The two indices mimic $\beta _{1}$ and $\beta _{2}$ defined in Eq.~(\ref{eq:beta1beta2}), but they use the spectrum \texttt{SameElIonOut} instead of \texttt{OneOut} to measure the degree of contamination to a \ce{CN} line by surrounding lines of species other than \ce{CN}. 
Then, we imposed three criteria to filter out weak and/or heavily-contaminated \ce{^{12}C^{14}N} lines: $d_
{\text{\texttt{OnlyOne}}}>0.03$, $\beta _{3}<0.3$, and $\beta _{4}<0.3$. 
We used the same conditions on the line depth $d_{\text{\texttt{OnlyOne}}}$ for \ce{^{13}C^{14}N} and \ce{^{12}C^{15}N} lines, but we relaxed the condition on the contamination fractions: $\beta _{3}<0.5$, and $\beta _{4}<1.0$. 
When two or more lines are located within $80\ur{\si{km.s^{-1}}}$ (which is different from $60\ur{\si{km.s^{-1}}}$ used for \ion{Fe}{i} lines in Sect.~\ref{ssec:LineSelection}), only the line with the largest $d_{\text{\texttt{OnlyOne}}}$ was used. 
Application of these criteria left $43$ and $110$ \ce{^{12}C^{14}N} lines in the \textit{Y} and \textit{J} bands, respectively, for VALD3, and $49$ and $133$ lines for MB99. 
There are $5$ and $6$ \ce{^{13}C^{14}N} lines left in the \textit{J} band for VALD3 and MB99, respectively, to determine $\ce{^{12}C}/\ce{^{13}C}$, and no \ce{^{12}C^{15}N} lines left for the both lists. 

Fitting the selected \ce{CN} lines simultaneously, we determined [C/O], [N/H], and $\ce{^{12}C}/\ce{^{13}C}$ together with the full-width at half maximum (FWHM) of the line broadening, $v_{\mathrm{broad}}$, as functions of $v_{\mathrm{micro}}$ spanning $0.6\text{--}4.4\ur{\si{km.s^{-1}}}$ with a step of $0.2\ur{\si{km.s^{-1}}}$ for each star, as follows. 
We used small wavelength ranges ($\pm \Delta _{2}/2=\pm 40\ur{\si{km.s^{-1}}}$) of the spectrum around the selected \ce{CN} lines to fit with a synthesized one. 
Some ranges having unrealistic flux values caused by, e.g., poor continuum normalization, were excluded. 
For a given set of [C/O], [N/H], $\ce{^{12}C}/\ce{^{13}C}$, and $v_{\mathrm{broad}}$, we determined the constant continuum level of each range that minimized the residual between the observed and synthesized spectra. 
Then, we calculated the residual of pixels that appeared in any ranges. 
We determined [C/O], [N/H], $\ce{^{12}C}/\ce{^{13}C}$, and $v_{\mathrm{broad}}$ that minimized the residual using \texttt{SciPy} package~\citep{SciPy}. 
We note that we assumed in the procedure the chemical abundances of all the elements other than carbon and nitrogen to be solar. 
Finally, we interpolated [C/O] [N/H], and $\ce{^{12}C}/\ce{^{13}C}$ on the $v_{\mathrm{micro}}$ set with a polynomial function and used the interpolated [C/O], [N/H], and $\ce{^{12}C}/\ce{^{13}C}$ as functions of $v_{\mathrm{micro}}$ together with the fixed $\text{[O/H]}=0.0\ur{dex}$ in the subsequent analyses for each star.

\subsection{Microturbulence ($v_{\mathrm{micro}}$) and metallicity ([Fe/H])}\label{ssec:vmicFeH}
In this section, we describe our procedure to determine $v_{\mathrm{micro}}$ and [Fe/H] simultaneously with the fitting of individual \ion{Fe}{i} lines listed in Table~\ref{table:LineListVALDMelendez}. 
We mainly follow the procedure given by \citet{Kondo2019} and \citet{Fukue2021}, who analyzed the spectra of two K-type red giants, Arcturus and $\mu $~Leo, observed with the WINERED spectrograph, though we modified the procedure in many points in order to fit it to analyze RSGs.

\subsubsection{Metallicity measurement with individual lines}\label{sssec:MeasureFeH}

For each \ion{Fe}{i} line selected in Sect.~\ref{ssec:LineSelection} of a star, we estimated the metallicity, [Fe/H], with which the synthesized spectrum reproduced a small part of an observed spectrum around the line. 
The wavelength range with a width of $\Delta _{2}=80\ur{\si{km.s^{-1}}}$ (i.e., $\pm 40\ur{\si{km.s^{-1}}}$ from the wavelength $\lambda _{0}$ of the line) was used to fit each \ion{Fe}{i} line. 
During the fitting, we fixed $T_{\mathrm{eff}}$ and $\log g$ at the respective values determined in Sects.~\ref{ssec:Teff} and \ref{ssec:logg} and [C/H], [N/H], and [O/H] at those determined in Sect.~\ref{ssec:CNfitting}. 
In contrast, we varied $v_{\mathrm{micro}}$ from $1.0\ur{\si{km.s^{-1}}}$ to $4.0\ur{\si{km.s^{-1}}}$ with a step of $0.1\ur{\si{km.s^{-1}}}$, and we then examined the dependence of the derived [Fe/H] on $v_{\mathrm{micro}}$ to determine the appropriate $v_{\mathrm{micro}}$ value. 

The basic algorithm of our fitting procedure implemented in \textsc{Octoman} followed that of \citet{Takeda1995a}; its detailed process is described in Appendix~\ref{app:Octoman}. 
Briefly, we fitted the observed spectrum with a synthesized one until the end condition was satisfied, allowing four parameters to vary in an iterative way: the metallicity [Fe/H], FWHM $v_{\mathrm{broad}}$ of the line broadening under the assumption of a Gaussian profile, velocity offset $\Delta \mathrm{RV}$, and continuum normalization factor $C$. 
The end condition is met, if the variation of the fitting parameters is below a certain threshold. 
If the condition was not satisfied within $40$ iterations, we considered the fitting for the line of the star a failure. 

For each fitting run to converge well, the choice of the initial values for the free parameters (for a line) in the fitting matters. 
To determine the initial parameter set, we first examined a specific case of $v_{\mathrm{micro}}=2.5\ur{\si{km.s^{-1}}}$ among the above-mentioned set of $v_{\mathrm{micro}}$. 
With the $v_{\mathrm{micro}}$ value, we tried to run the fitting procedure with nine sets of the initial parameters; three values for [Fe/H], $-0.3$, $0.0$, and $+0.3\ur{dex}$, three values for $v_{\mathrm{broad}}$, $14$, $17$, and $20\ur{\si{km.s^{-1}}}$, and $\Delta \mathrm{RV}=0\ur{\si{km.s^{-1}}}$. 
Then, we selected the parameter set that gave the smallest residual as the initial parameter set for $v_{\mathrm{micro}}=2.5\ur{\si{km.s^{-1}}}$. 
We subsequently gave as the initial parameter set for each of the $v_{\mathrm{micro}}$ grid points the best-fitting parameter set at the closest $v_{\mathrm{micro}}$ with which fitting had been successfully performed. 

Applying the algorithm to the spectra of all of our observed RSGs, we obtained [Fe/H] as a function of $v_{\mathrm{micro}}$ spanning $1.0\text{--}4.0\ur{\si{km.s^{-1}}}$ for each line of each star.

\subsubsection{Simultaneous determination of $v_{\mathrm{micro}}$ and [Fe/H]}\label{sssec:DetVFeH}

\begin{figure*}
\centering 
\includegraphics[width=18cm]{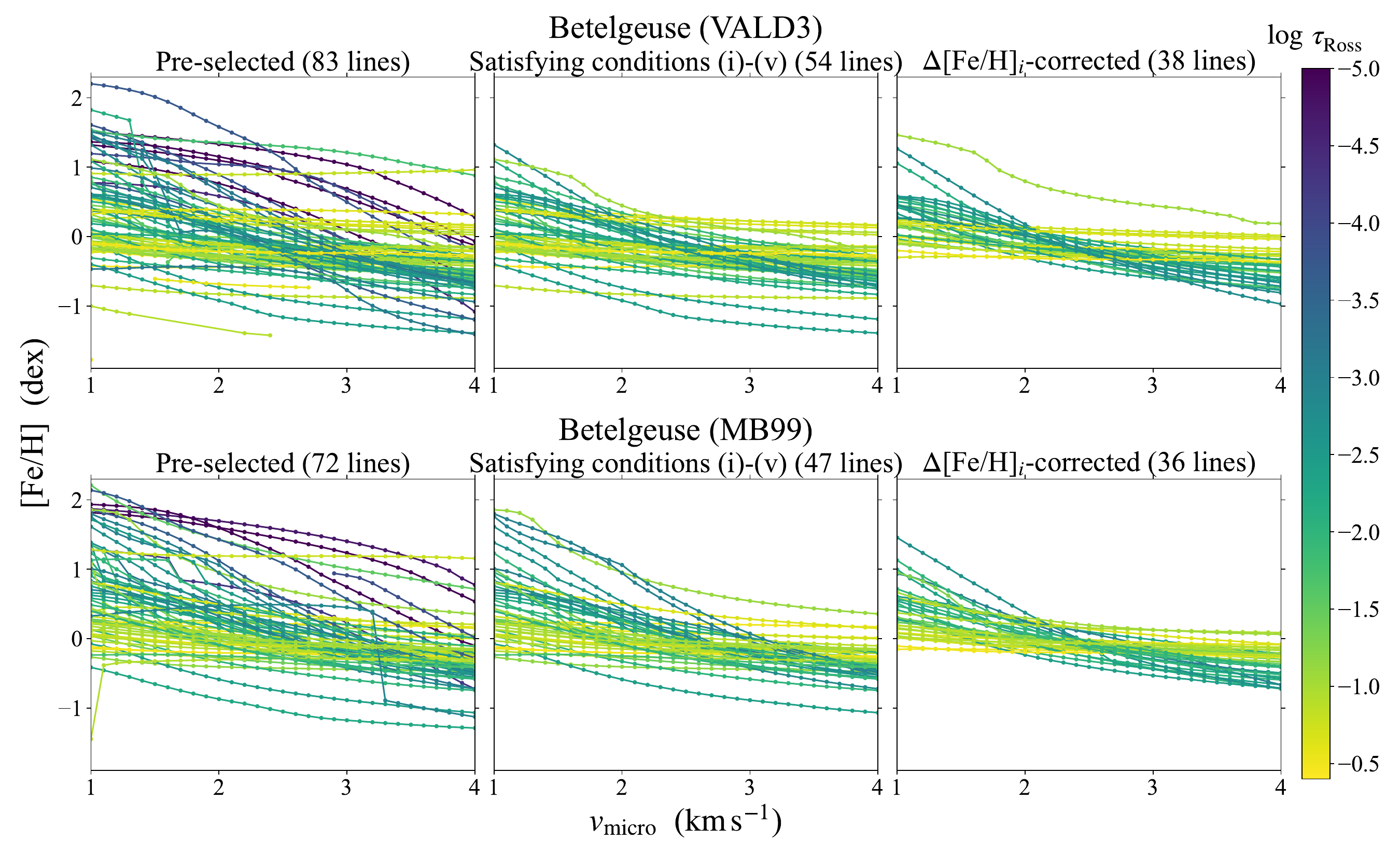}
\caption{Examples of how [Fe/H] is determined from each absorption line as a function of $v_{\mathrm{micro}}$: here for the case for \object{Betelgeuse}. Top and bottom panels show the results with VALD3 and MB99, respectively. Left panels show the measurements for all the \ion{Fe}{i} lines preselected in Sect.~\ref{ssec:LineSelection}, and middle panels do for the lines satisfying conditions (i)--(v) in the main text~(Sect.~\ref{sssec:DetVFeH}) and eventually used. Right panels show the measurements after the correction term $\Delta \text{[Fe/H]}_{i}$ defined in Eq.~(\ref{eq:CorrectionTerm}) subtracted. Each curve in the figures corresponds to each absorption line color-coded according to $\log \tau _{\mathrm{Ross}}^{}$ of the line-forming layer of the line; a lighter color corresponds to a larger $\log \tau _{\mathrm{Ross}}^{}$ and thus a smaller EW. The dot on each curve indicates the [Fe/H] value determined with the corresponding $v_{\mathrm{micro}}$. No dots are plotted where the fitting for a line with a value of $v_{\mathrm{micro}}$ failed. }
\label{fig:mpfitFeH}
\end{figure*}

Using a series of $\text{[Fe/H]}$ as functions of $v_{\mathrm{micro}}$ for all the analyzed lines of a star, we searched for the combination of $v_{\mathrm{micro}}$ and [Fe/H] that gives no correlation between line strength and [Fe/H] determined through the fitting of individual lines. 
Our search took four step: (1)~excluding poorly-fitted lines, (2)~setting the initial guess of $v_{\mathrm{micro}}$ and [Fe/H], (3)~estimating and subtracting the correction term to [Fe/H] of each line, and (4)~determining the final $v_{\mathrm{micro}}$ and [Fe/H]. 

\begin{figure*}
\centering 
\includegraphics[width=17.3cm]{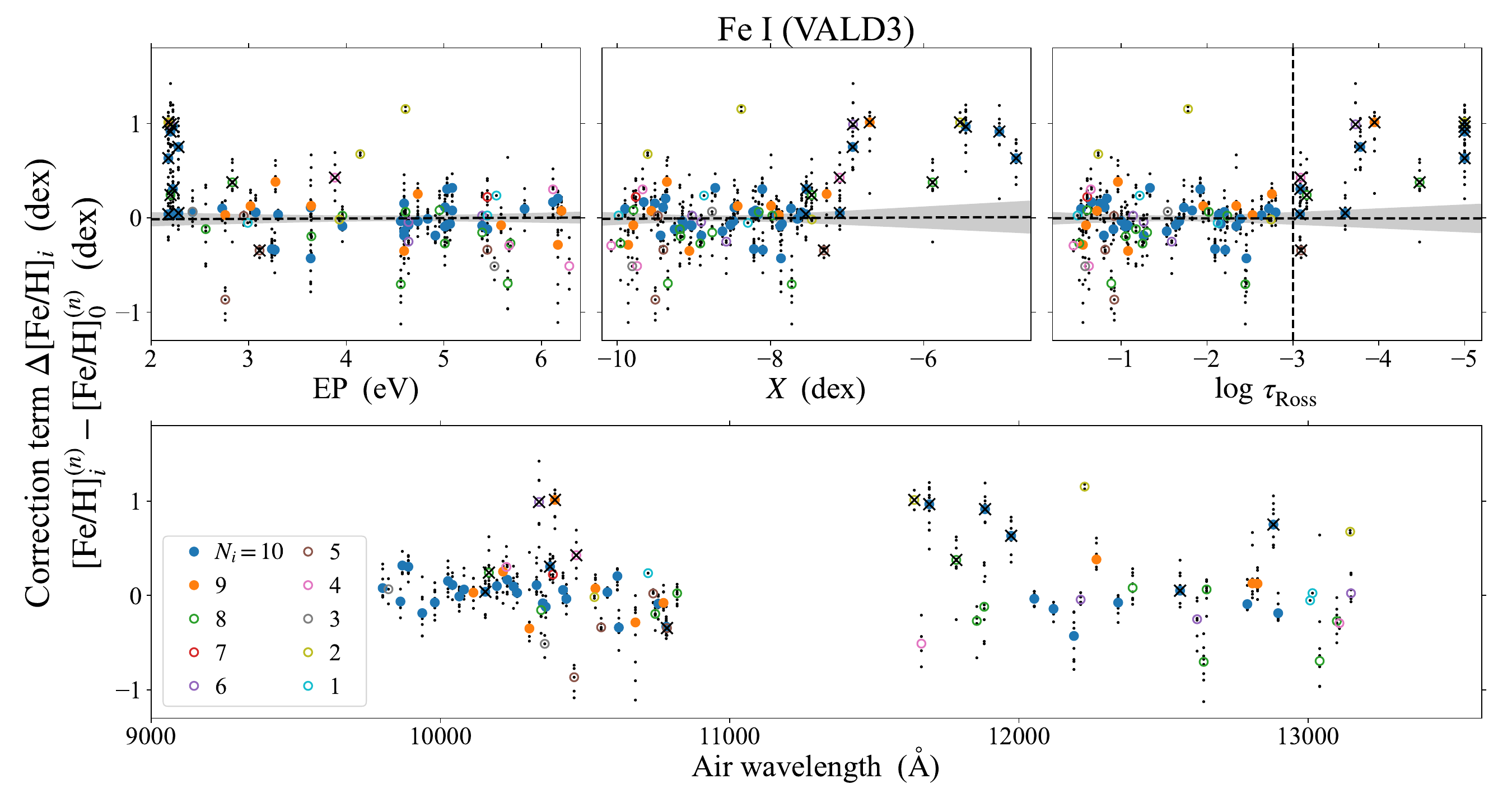}
\caption{Correction term to the determined [Fe/H] values for each line for VALD3 as functions of EP (top-left panel), $X$ index at $3850\ur{K}$ (top-middle), $\log \tau _{\mathrm{Ross}}^{}$ (top-right), and wavelength in the standard air (bottom). Black dots show $\text{[Fe/H]}_{i}^{(n)}-\text{[Fe/H]}_{0}^{(n)}$ (see text). Circles show their mean values $\Delta \text{[Fe/H]}_{i}$ among the sample stars, where the color of each circle indicates the number $N_{i}$ of stars for which the corresponding line was successfully fitted. The lines indicated by open circles (i.e., $N_{i}<9$; those in colors other than blue or orange) were excluded in the analysis. Circles overplotted by black crosses indicate the lines that do not satisfy the condition on the line strength ($\log \tau _{\mathrm{Ross}}>-3$ indicated by the vertical black dashed line in the top right panel) and thus are not used in the analysis. Horizontal black dashed lines in the top panels indicate the linear regression between the values of the $x$ and $y$ axes for the lines marked with filled circles, together with the gray-shaded areas indicating the $1\sigma $-confidence intervals. }
\label{fig:loggfCorrectionVALD}
\end{figure*}

\begin{figure*}
\centering 
\includegraphics[width=17.3cm]{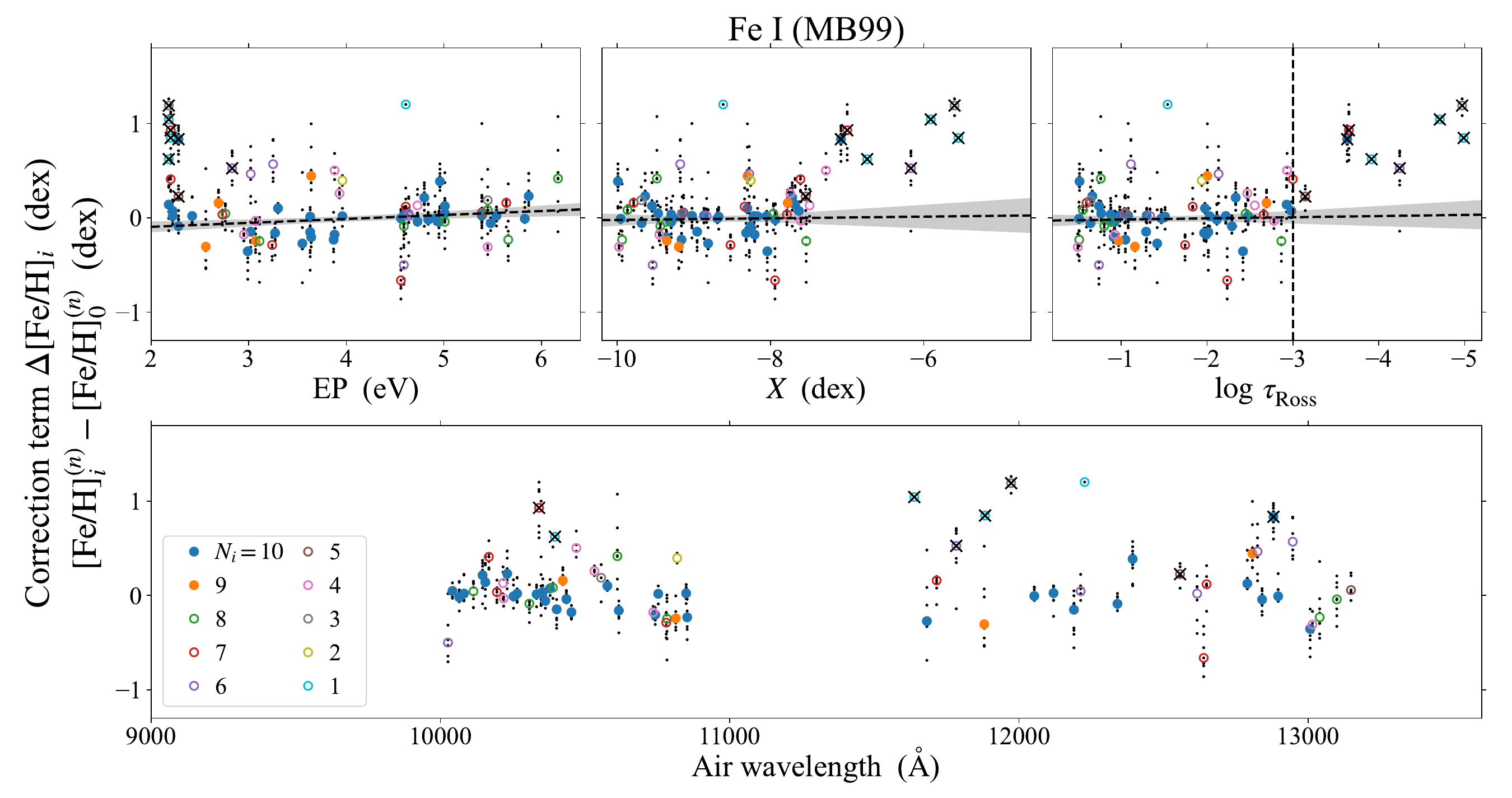}
\caption{Same as Fig.~\ref{fig:loggfCorrectionVALD} but for MB99. }
\label{fig:loggfCorrectionMelendez}
\end{figure*}

In the first step, we examined how [Fe/H] was distributed against the set of $v_{\mathrm{micro}}$ for all the $83$ (VALD3) and $72$ (MB99) \ion{Fe}{i} lines selected in Sect.~\ref{ssec:LineSelection} for each star, with the aim of excluding some lines that are unsuitable to the abundance analysis for the star. 
Left panels in Fig.~\ref{fig:mpfitFeH} show the results for \object{Betelgeuse} as an example. 
Though [Fe/H] of each absorption line was expected to be a smooth function of $v_{\mathrm{micro}}$, some of them showed anomalous variations. 
Furthermore, we failed to determine [Fe/H] with a considerable number of the $v_{\mathrm{micro}}$ values for some lines. 
These problems mainly occurred when the line was severely contaminated with other lines. 
In order to filter out undesirable lines due to these or some other reasons, we set five conditions for a line to be accepted. 
The first and second ones are that (i)~[Fe/H] were successfully determined with more than $26$ among $31$ $v_{\mathrm{micro}}$ values, and (ii)~all the slopes between the adjacent $v_{\mathrm{micro}}$ values are within a range from $-2.0\ur{\si{dex.(km.s^{-1})^{-1}}}$ to $+0.1\ur{\si{dex.(km.s^{-1})^{-1}}}$. 
These two conditions were imposed to filter out the lines that show anomalous variations. 
Remaining data gaps, if exist, were filled by means of linear interpolations (or extrapolations). 
The third one is that (iii)~the Rosseland-mean optical depth $\log \tau _{\mathrm{Ross}}^{}$ that gives the largest contribution function for the line~\citep{Gurtovenko2015} satisfies $\log \tau _{\mathrm{Ross}}^{}>-3.0$. 
This condition was required because strong lines were often highly affected by non-LTE effects and imperfect modeling of damping wings~\citep[see Sect.~5.2 in][]{Kondo2019}. 
Indeed, as we show below, the [Fe/H] of strong lines that did not meed Condition (iii) tend to be larger than those of the weak lines by ${\sim }1\ur{dex}$~(see top-right panels of Figs.~\ref{fig:loggfCorrectionVALD} and \ref{fig:loggfCorrectionMelendez}). 
The forth one is that (iv)~the median value of [Fe/H] among all the $v_{\mathrm{micro}}$ values is between $-1.5$ and $+1.0\ur{dex}$. 
This condition was required because there was a disagreement between [Fe/H] and [M/H] in our spectral synthesis when $\text{[Fe/H]}<-1.55$ or $\text{[Fe/H]}>+0.95\ur{dex}$, as mentioned in Appendix~\ref{app:Octoman}. 
Furthermore, the lines that fail to satisfy it, i.e., those with a very high or low [Fe/H], would be totally unexpected, which would be attributed to a poor match between the observed and synthesized spectra. 
The fifth, and final one is that (v)~the median value of [Fe/H] among all the $v_{\mathrm{micro}}$ values is within $3\sigma $ of the median values of all the remaining lines for the star. 
Applications of the five conditions filtered our ${\sim 30}$ and ${\sim }20$ lines for the VALD3 and MB99 lists, respectively, though the exact number of lines varied slightly, depending on the star. 
Middle panels of Fig.~\ref{fig:mpfitFeH} show example results of the line fitting after the five conditions were imposed. 
As expected, the figure shows only lines with smooth relations between [Fe/H] and $v_{\mathrm{micro}}$, without very high or low [Fe/H] values, and with weak or moderate line strengths. 

From the remaining lines, we further excluded those that were accepted for fewer than nine out of ten stars; that is, we used the lines that were unusable for only zero or one stars. 
The excluded lines were not used in the subsequent analysis for all the stars together with the lines that were rejected for each star. 
The resultant number $N_{\mathrm{line}}$ of the lines to be used was $38$ for VALD3 and $36$ for MB99. 

In the second step in the search for the $v_{\mathrm{micro}}$ and [Fe/H] combination, we determined the tentative values of $v_{\mathrm{micro}}$ and [Fe/H] of each star with the method described in \citet{Kondo2019}. 
Briefly, we searched for the $v_{\mathrm{micro}}$ value that gives no correlation between the $X$ index~\citep{Magain1984} indicating the line strength and [Fe/H]. 
In general, evaluating the line strength from observed spectra of RSGs is not straightforward due to the severe line contamination. 
In the long-established abundance analysis of optical spectra of late-type giants and dwarfs without severe line contamination, the strength of each line is evaluated with either the observed or expected EW. 
The systematic errors in the resultant abundances depending on the choice of the two types of EWs have been extensively examined~\citep[e.g.,][]{Magain1984,Mucciarelli2011,Hill2011}. 
In the case of either spectra of RSGs or NIR spectra of late-type stars (including RSGs), however, most of the lines to be used for abundance analysis are more or less contaminated by other lines, and thus it is usually difficult to accurately measure EWs observationally. 
We thus used, instead of observed EWs, the so-called $X$ index in our analysis. 
The $X$ index is often adopted for the abscissa of the curve-of-growth~\citep[e.g.,][]{Gray2008}. 
\citet{Kondo2019} successfully applied it in the analysis of NIR spectra of red giants. 
The $X$ of each line in a spectrum is defined as 
\begin{equation}\label{eq:Xindex}
X\equiv \log gf-\mathrm{EP}\times \Theta _{\mathrm{exc}}\text{,}
\end{equation}
where $\Theta _{\mathrm{exc}}$ is the inverse temperature of the atmosphere layer from which the line originates. 
We adopted an approximation formula of $\Theta _{\mathrm{exc}}=5040\ur{K}/(0.86T_{\mathrm{eff}})$ following the work by \citet{Gratton2006}. 
We note that the target type of stars analyzed by \citet{Gratton2006} was metal-rich red clump stars and thus different from ours, solar-metallicity RSGs. 
We also note that the value of $\Theta _{\mathrm{exc}}T_{\mathrm{eff}}$ is not a constant as we assume and depends on the line and/or spectral type of the star~\citep{Gray2008}. 
Nevertheless, we consider that the same formula should be applicable because the skewed $X$ scale as a result of the variation of $\Theta _{\mathrm{exc}}T_{\mathrm{eff}}$ would not change $v_{\mathrm{micro}}$ that gives no correlation between $X$ and [Fe/H] of individual lines. 

Here we describe the detailed procedure for the second step. 
For each star, we first prepared $10^{6}$ bootstrap samples of the chosen lines, that is, we resampled the lines randomly from the original list of lines, allowing each line to be selected more than once. 
Then, for each bootstrap sample, the relation between the $X$ index indicating the line strength and [Fe/H] for each $v_{\mathrm{micro}}$ was fitted with a linear regression function, $\text{[Fe/H]}=a(v_{\mathrm{micro}})X+b(v_{\mathrm{micro}})$. 
The slopes of the regression lines, $a(v_{\mathrm{micro}})$, were linearly interpolated to determine the microturbulent velocity $v_{\mathrm{micro},0}$ that gives $a(v_{\mathrm{micro},0})=0\ur{\si{dex.dex^{-1}}}$. 
The corresponding [Fe/H] value, $b(v_{\mathrm{micro},0})=\text{[Fe/H]}_{0}$, was also obtained with the linear interpolation of $b(v_{\mathrm{micro}})$. 
Then, we considered the medians of $\text{[Fe/H]}_{0}$ and $v_{\mathrm{micro},0}$ among the entire bootstrap samples as the best estimates of [Fe/H] and $v_{\mathrm{micro}}$, respectively, of the star at this step. 
Finally, we calculated the $15.9\%$ and $84.1\%$ percentiles, i.e., $1\sigma $ intervals for a Gaussian distribution, of the bootstrap samples as the standard errors of the best estimates. 

The determined [Fe/H] for most of the stars had a scatter of $0.3\text{--}0.4\ur{dex}$; the middle panels of Fig.~\ref{fig:mpfitFeH} show an example case. 
A significant amount of the large scatter was likely to be attributed to errors in $\log gf$~\citep{Andreasen2016,Kondo2019}, and the systematic error in the line fitting originating in the line contamination. 
The two types of probable sources of errors are expected to add systematic errors to the [Fe/H] measurements for all the stars for each line. 
Thus, in the third step in the search for the $v_{\mathrm{micro}}$ and [Fe/H] combination, we took an approach similar to the differential analysis~\citep{Ramirez2014,Nissen2018}, as we describe in the following two paragraphs, to remove this type of systematic errors. 

In the usual differential analysis, [Fe/H] of individual lines of a target star are compared with those of a standard star, such as the Sun. 
Then, the offsets in [Fe/H] between the two stars are used to determine the differential metallicity (and some of the stellar parameters) of the target. 
In our case, however, none of the target stars had [Fe/H] measurements for all the lines of interest. 
Furthermore, none of the targets had a well-known [Fe/H] to be used as a standard star. 
Thus, the above-mentioned standard-star method was unsuitable in our derivation of the abundances. 
Instead, we calculated a ``correction term'' (or ``line-by-line systematics'') to [Fe/H] of each line, using all the available [Fe/H] measurements for the targets, and used it. 
Specifically, we calculated the correction term $\Delta \text{[Fe/H]}_{i}$ for line $i$, using [Fe/H] of the line $i$ of each star $n$ with $v_{\mathrm{micro},0}$, which is denoted as $\text{[Fe/H]}_{i}^{(n)}$, as 
\begin{equation}
\Delta \text{[Fe/H]}_{i}=\frac{1}{N_{i}}\sum _{n=1}^{N_{i}}\left(\text{[Fe/H]}_{i}^{(n)}-\text{[Fe/H]}_{0}^{(n)}\right)\label{eq:CorrectionTerm}\text{,}
\end{equation}
where $N_{i}$ indicates the number of the stars having the [Fe/H] measurement for line $i$. 
We then ``corrected'' or removed the line-by-line systematic from $\text{[Fe/H]}_{i}^{(n)}$ as $\text{[Fe/H]}_{i}^{(n)}\mapsto \text{[Fe/H]}_{i}^{(n)}-\Delta \text{[Fe/H]}_{i}$. 
Middle and right panels of Fig.~\ref{fig:mpfitFeH} show an example case before and after the correction, respectively. 

In the fourth and final step in the search for the $v_{\mathrm{micro}}$ and [Fe/H] combination, we recalculated $v_{\mathrm{micro},0}$ and $\text{[Fe/H]}_{0}$ along with their standard errors, using the same method employed for obtaining the tentative values but with the corrected [Fe/H] values for individual lines. 
The scatter in $\text{[Fe/H]}_{i}^{(n)}$ for each star was confirmed to be smaller than the scatter in $\abs{\Delta \text{[Fe/H]}_{i}}$~(Figs.~\ref{fig:loggfCorrectionVALD} and \ref{fig:loggfCorrectionMelendez}). 
Accordingly, we conclude that the correction terms improved the precision in the determined [Fe/H] as expected.

\subsection{Abundances of elements other than iron}\label{sssec:AnalysisXH}

Having determined $T_{\mathrm{eff}}$, $\log g$, CNO abundances, $v_{\mathrm{micro}}$, and [Fe/H] in previous sections, we determined the chemical abundances of elements other than iron. 
We used basically the same procedure as we had determined [Fe/H], but with some modifications. 

For each line selected in Sect.~\ref{ssec:LineSelection} of an element X, we derived the abundance [X/H] for a parameter set of $v_{\mathrm{micro}}$ ranging from $1.0$ to $4.0\ur{\si{km.s^{-1}}}$ with a step of $0.1\ur{\si{km.s^{-1}}}$, using the \textsc{Octoman} code in the same way as for the iron abundance~(see Sect.~\ref{sssec:MeasureFeH}). 
During the fitting for the element X, we fixed the global metallicity [M/H] of the model atmosphere and the abundances of elements other than C, N, O, and X to the value of [Fe/H] that we had determined, and allowed [X/H], $v_{\mathrm{broad}}$, $\Delta \mathrm{RV}$, and continuum normalization $C$ to vary. 
After the fitting for the entire set of $v_{\mathrm{micro}}$, we excluded the lines failing to satisfy any of the Conditions (i)--(iv) that had been applied in determining the iron abundance in Sect.~\ref{sssec:DetVFeH}. 
Then, we interpolated the set of $v_{\mathrm{micro}}$ and derived the abundance of an element X for a given line with $v_{\mathrm{micro},0}$. 

We then calculated the correction term $\Delta \text{[X/H]}_{i}$ for each line in the same way as for the iron abundance, subtracted the correction term from $\text{[X/H]}_{i}^{(n)}$, and calculated the mean of [X/H] of all the remaining lines. 
Consequently, we have determined the abundances of \ion{Mg}{i}, \ion{Si}{i}, \ion{Ca}{i}, \ion{Ti}{i}, \ion{Cr}{i}, \ion{Ni}{i}, and \ion{Y}{ii} for VALD3 and \ion{Na}{i}, \ion{Mg}{i}, \ion{Al}{i}, \ion{Si}{i}, \ion{K}{i}, \ion{Ca}{i}, \ion{Ti}{i}, \ion{Cr}{i}, \ion{Ni}{i}, and \ion{Y}{ii} for MB99.

\subsection{Error budget for abundance measurements}\label{ssec:errorbudget}

\subsubsection{Error budget for [Fe/H]}\label{sssec:errorbudgetFeH}

We consider two sources of errors in the derived [Fe/H] of each star: (1)~$\Delta _{\mathrm{b}}$ --- the confidence interval in the determination of [Fe/H] and $v_{\mathrm{micro}}$ in the bootstrap and (2)~$\Delta _{T_{\mathrm{eff}}}$ and $\Delta _{\log g}$ --- the errors propagated from the errors in $T_{\mathrm{eff}}$ and $\log g$, respectively. 
The total error $\Delta _{\mathrm{total}}$ in the final [Fe/H] measurement were calculated as 
\begin{equation}
\Delta _{\mathrm{total}}\equiv \sqrt{{\Delta _{\mathrm{b}}}^{2}+{\Delta _{T_{\mathrm{eff}}}}^{2}+{\Delta _{\log g}}^{2}}\text{.}
\end{equation}
In more detail, $\Delta _{\mathrm{b}}$ was estimated using the bootstrap method as described in Sect.~\ref{sssec:DetVFeH}. 
The error includes both the standard error due to the scatter in [Fe/H] determined for individual lines and the error propagated from the error in $v_{\mathrm{micro}}$ because we determined [Fe/H] and $v_{\mathrm{micro}}$ simultaneously with the bootstrap method. 
To determine $\Delta _{T_{\mathrm{eff}}}$ with numerical error propagation for each star, we fitted again all the lines used for the final [Fe/H] determination, totaling $38$ and $36$ lines for VALD3 and MB99, respectively, with the determined $v_{\mathrm{micro}}$ and with three different effective temperatures assumed: the best estimate $T_{\mathrm{eff}}$, and the best estimate plus or minus its error, $T_{\mathrm{eff}}\pm \Delta T_{\mathrm{eff}}$. 
Then, we estimated the error $\Delta _{T_{\mathrm{eff}}}$ by calculating the bootstrapped median of the differences between $\text{[Fe/H]}_{i}^{(n)}(T_{\mathrm{eff}}\pm \Delta T_{\mathrm{eff}})$ and $\text{[Fe/H]}_{i}^{(n)}(T_{\mathrm{eff}}$). 
We estimated the error $\Delta _{\log g}$ in the same way.

\subsubsection{Error budget for [X/H] other than [Fe/H]}\label{sssec:errorbudgetXH}

We consider two sources of errors in [X/H] of each star: (1)~${\Delta '}_{\mathrm{sca}}$\footnote{We use a prime symbol to denote the error in [X/H]; when an error variable symbol is not accompanied by a prime, it indicates the error in [Fe/H]. } --- the standard error of the line-by-line scatter and (2)~${\Delta '}_{v_{\mathrm{micro}}}$, ${\Delta '}_{T_{\mathrm{eff}}}$, ${\Delta '}_{\log g}$, and ${\Delta '}_{\text{[Fe/H]}}$ --- the errors propagated from the errors in $v_{\mathrm{micro}}$, $T_{\mathrm{eff}}$, $\log g$, and [Fe/H], respectively. 
Ignoring covariance terms, the total errors ${\Delta '}_{\mathrm{total}}^{\text{[X/H]}}$ in the final [X/H] was calculated as 
\begin{equation}
{\Delta '}_{\mathrm{total}}^{\text{[X/H]}}\equiv \sqrt{{{\Delta '}_{\mathrm{sca}}}^{2}+{{\Delta '}_{v_{\mathrm{micro}}}}^{2}+{{\Delta '}_{T_{\mathrm{eff}}}}^{2}+{{\Delta '}_{\log g}}^{2}+{{\Delta '}_{\text{[Fe/H]}}}^{2}}
.\end{equation}
In more detail, ${\Delta '}_{\mathrm{sca}}$ was simply calculated as the standard error of [X/H] from individual lines for the elements where the number of the lines $N_{\mathrm{line}}^{(n)}$ for the element X for the star $n$ is $5$ or larger. 
In cases where $N_{\mathrm{line}}^{(n)}$ is smaller than $5$, however, the standard error of the [X/H] values is inaccurate, and thus, we multiplied the standard deviation of the measured [Fe/H] values by $1/\sqrt{N_{\mathrm{line}}^{(n)}}$ to estimate ${\Delta '}_{\mathrm{sca}}$, assuming that the errors in [X/H] and [Fe/H] measurements from individual lines are approximately equal. 
The other error terms were estimated with numerical error propagation in the same way as the estimation of the errors $\Delta _{T_{\mathrm{eff}}}$ and $\Delta _{\log g}$ of [Fe/H]. 
We also determined the total error in [X/Fe] in a similar way.


\section{Chemical abundance analysis: Results}\label{sec:results}

\begin{figure}
\centering 
\includegraphics[width=9cm]{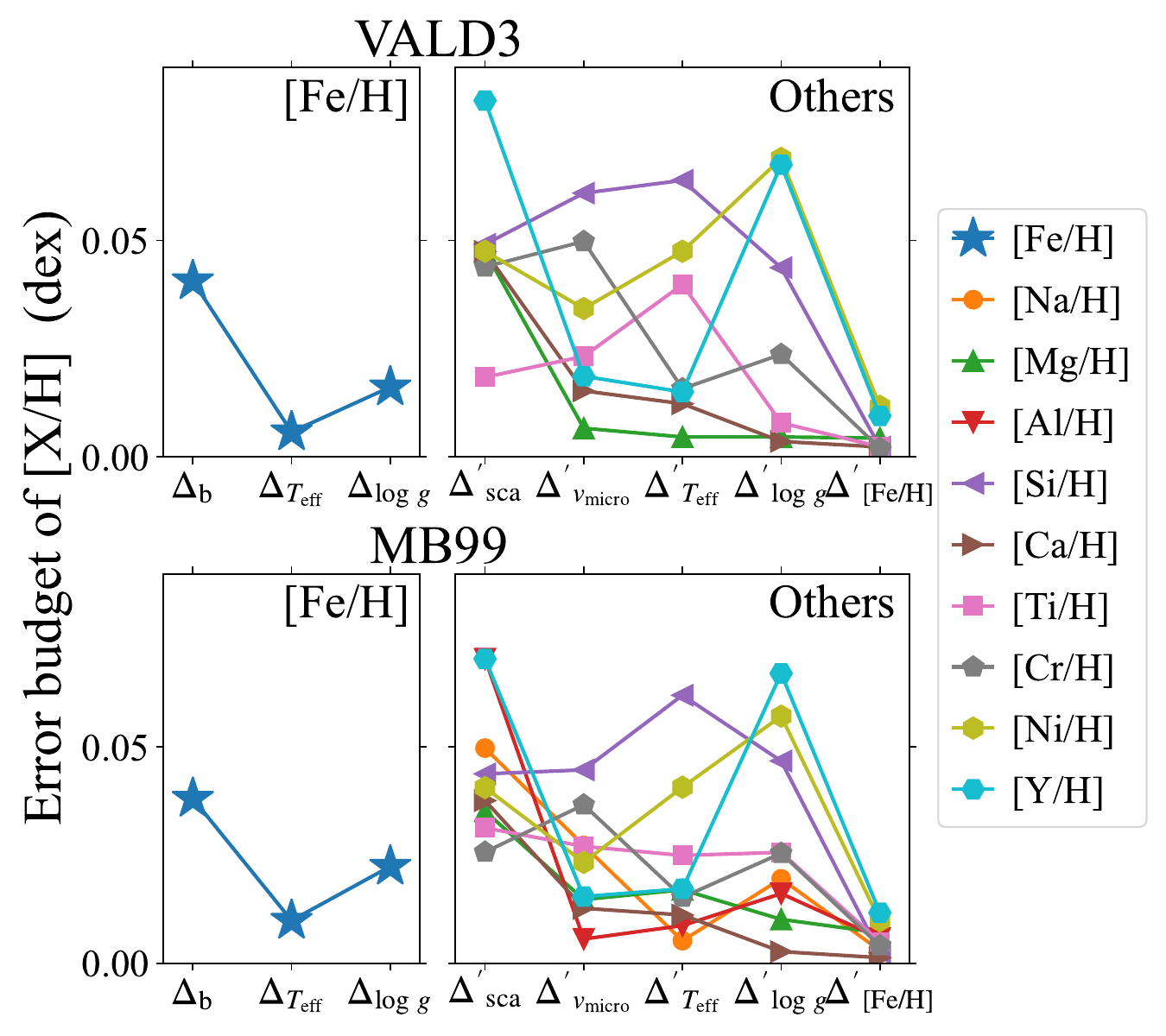}
\caption{Error budget of [X/H] measurements. Left panels show the medians of the absolute values of the three sources of errors ($\Delta _{\mathrm{b}}$, $\Delta _{T_{\mathrm{eff}}}$, and $\Delta _{\log g}$) in the [Fe/H] determination (see Sect.~\ref{sssec:errorbudgetFeH} for the definitions) among our ten target RSGs. Right panels show the medians of the absolute values of the five sources of errors (${\Delta '}_{\mathrm{sca}}$, ${\Delta '}_{v_{\mathrm{micro}}}$, ${\Delta '}_{T_{\mathrm{eff}}}$, ${\Delta '}_{\log g}$, and ${\Delta '}_{\text{[Fe/H]}}$) in the [X/H] determination other than [Fe/H] (see Sect.~\ref{sssec:errorbudgetXH} for the definitions). }
\label{fig:errorbudgetall}
\end{figure}

We summarize the resultant stellar parameters and [Fe/H] of the target RSGs in Table~\ref{table:AtmosRes} and the chemical abundances in Tables~\ref{table:AbnRes} and \ref{table:AbnFeRes}. 
The typical precision $\Delta _{\mathrm{total}}$ in the determined [Fe/H] is ${\sim }0.05\ur{dex}$, which is dominated by $\Delta _{\mathrm{b}}$ in most cases~(left panels of Fig.~\ref{fig:errorbudgetall}). 
This level of precision is comparable with, or better than, the previous works of RSGs mentioned in Sect.~\ref{sec:IntroRSGFeH}. 
The errors ${\Delta '}_{\mathrm{total}}^{\text{[X/H]}}$ in the determined [X/H] other than [Fe/H] are dominated by $\Delta _{\mathrm{sca}}$ for most of the elements, especially the elements with a small number of measured lines~(right panels of Fig.~\ref{fig:errorbudgetall}). 
Considering the high sensitivity of [X/H] on $T_{\mathrm{eff}}$ and $\log g$ for some of the elements, especially, [Si/H], [Ni/H], and [Y/H], the high precision of $T_{\mathrm{eff}}$ and $\log g$ in this work (${\sim }30\text{--}100\ur{K}$ for $T_{\mathrm{eff}}$ and ${\sim }0.1\text{--}0.3$ for $\log g$) is essential for the high precision in the [X/H] measurements. 

In this section, we evaluate the results with VALD3 and MB99. 
As we demonstrate in this section, both the results turn out to be similar in terms of the precision and systematic bias, and thus we conclude that the two results are equally reliable. 

\begin{table*}
\centering 
\caption{Derived stellar parameters and [Fe/H]. }
\label{table:AtmosRes}
\begin{tabular}{l lr cr cr} \hline \hline 
 & & & \multicolumn{2}{c}{VALD3} & \multicolumn{2}{c}{MB99} \\ \cmidrule(rl){4-5} \cmidrule(rl){6-7}
Name & \multicolumn{1}{c}{$T_{\mathrm{eff}}$} & \multicolumn{1}{c}{$\log g$} & $v_{\mathrm{micro}}$ & \multicolumn{1}{c}{[Fe/H]} & $v_{\mathrm{micro}}$ & \multicolumn{1}{c}{[Fe/H]} \\
 & \multicolumn{1}{c}{(K)} & & \multicolumn{1}{c}{(\si{km.s^{-1}})} & \multicolumn{1}{c}{(dex)} & \multicolumn{1}{c}{(\si{km.s^{-1}})} & \multicolumn{1}{c}{(dex)} \\
\hline 
$\zeta $ Cep & $4073\pm 31$ & $1.03^{+0.08}_{-0.08}$ & $2.51^{+0.20}_{-0.19}$ & $-0.099^{+0.041}_{-0.038}$ & $2.32^{+0.11}_{-0.11}$ & $0.087^{+0.042}_{-0.038}$ \\
41 Gem & $3962\pm 27$ & $0.60^{+0.18}_{-0.18}$ & $1.91^{+0.09}_{-0.09}$ & $-0.076^{+0.042}_{-0.037}$ & $1.91^{+0.11}_{-0.10}$ & $0.065^{+0.050}_{-0.045}$ \\
$\xi $ Cyg & $3893\pm 26$ & $0.75^{+0.10}_{-0.10}$ & $1.82^{+0.07}_{-0.06}$ & $-0.096^{+0.030}_{-0.027}$ & $1.63^{+0.07}_{-0.08}$ & $0.109^{+0.040}_{-0.035}$ \\
V809 Cas & $3799\pm 36$ & $0.28^{+0.08}_{-0.08}$ & $2.11^{+0.11}_{-0.12}$ & $-0.065^{+0.028}_{-0.024}$ & $2.26^{+0.10}_{-0.10}$ & $0.037^{+0.028}_{-0.024}$ \\
V424 Lac & $3767\pm 48$ & $0.49^{+0.12}_{-0.12}$ & $1.98^{+0.13}_{-0.11}$ & $-0.039^{+0.039}_{-0.035}$ & $1.94^{+0.11}_{-0.09}$ & $0.078^{+0.045}_{-0.039}$ \\
$\psi ^{1}$ Aur & $3777\pm 60$ & $-0.35^{+0.28}_{-0.36}$ & $2.40^{+0.19}_{-0.14}$ & $-0.259^{+0.047}_{-0.054}$ & $2.21^{+0.13}_{-0.14}$ & $-0.081^{+0.067}_{-0.052}$ \\
TV Gem & $3739\pm 101$ & $-0.29^{+0.29}_{-0.36}$ & $2.31^{+0.38}_{-0.28}$ & $-0.148^{+0.095}_{-0.107}$ & $2.31^{+0.18}_{-0.18}$ & $-0.025^{+0.089}_{-0.065}$ \\
BU Gem & $3896\pm 70$ & $-0.15^{+0.25}_{-0.31}$ & $2.24^{+0.33}_{-0.23}$ & $-0.289^{+0.075}_{-0.091}$ & $2.07^{+0.20}_{-0.18}$ & $-0.129^{+0.046}_{-0.045}$ \\
Betelgeuse & $3633\pm 37$ & $-0.06^{+0.14}_{-0.15}$ & $2.19^{+0.16}_{-0.17}$ & $-0.111^{+0.076}_{-0.061}$ & $2.37^{+0.14}_{-0.16}$ & $-0.064^{+0.050}_{-0.042}$ \\
NO Aur & $3663\pm 30$ & $0.21^{+0.13}_{-0.13}$ & $2.07^{+0.14}_{-0.15}$ & $-0.078^{+0.050}_{-0.046}$ & $2.33^{+0.11}_{-0.12}$ & $-0.056^{+0.055}_{-0.050}$ \\
\hline 
\end{tabular}
\end{table*}

\subsection{Direct comparison with previous results}\label{ssec:CompPrevious}

\begin{figure}
\centering 
\includegraphics[width=9cm]{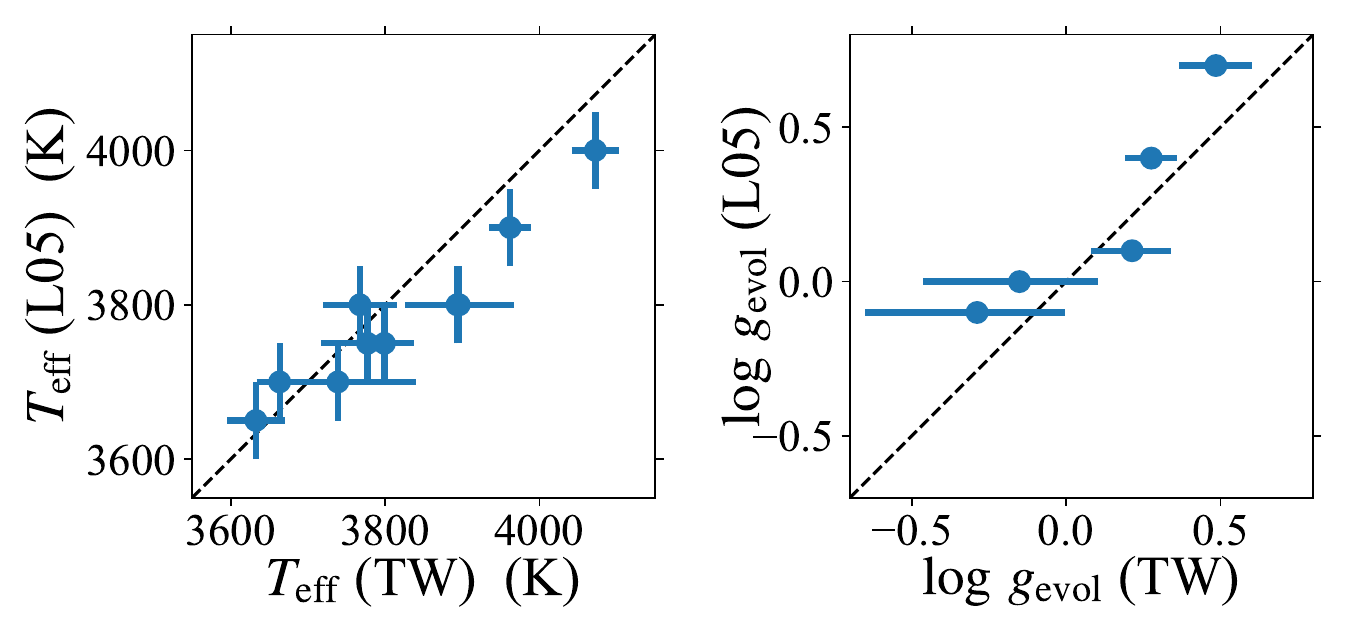}
\caption{Comparison of our results and those of \citet{Levesque2005} for the RSGs included in both samples: $T_{\mathrm{eff}}$ and $\log g$. }
\label{fig:CompLevesqueTefflogg}
\end{figure}

\begin{figure}
\centering 
\includegraphics[width=9cm]{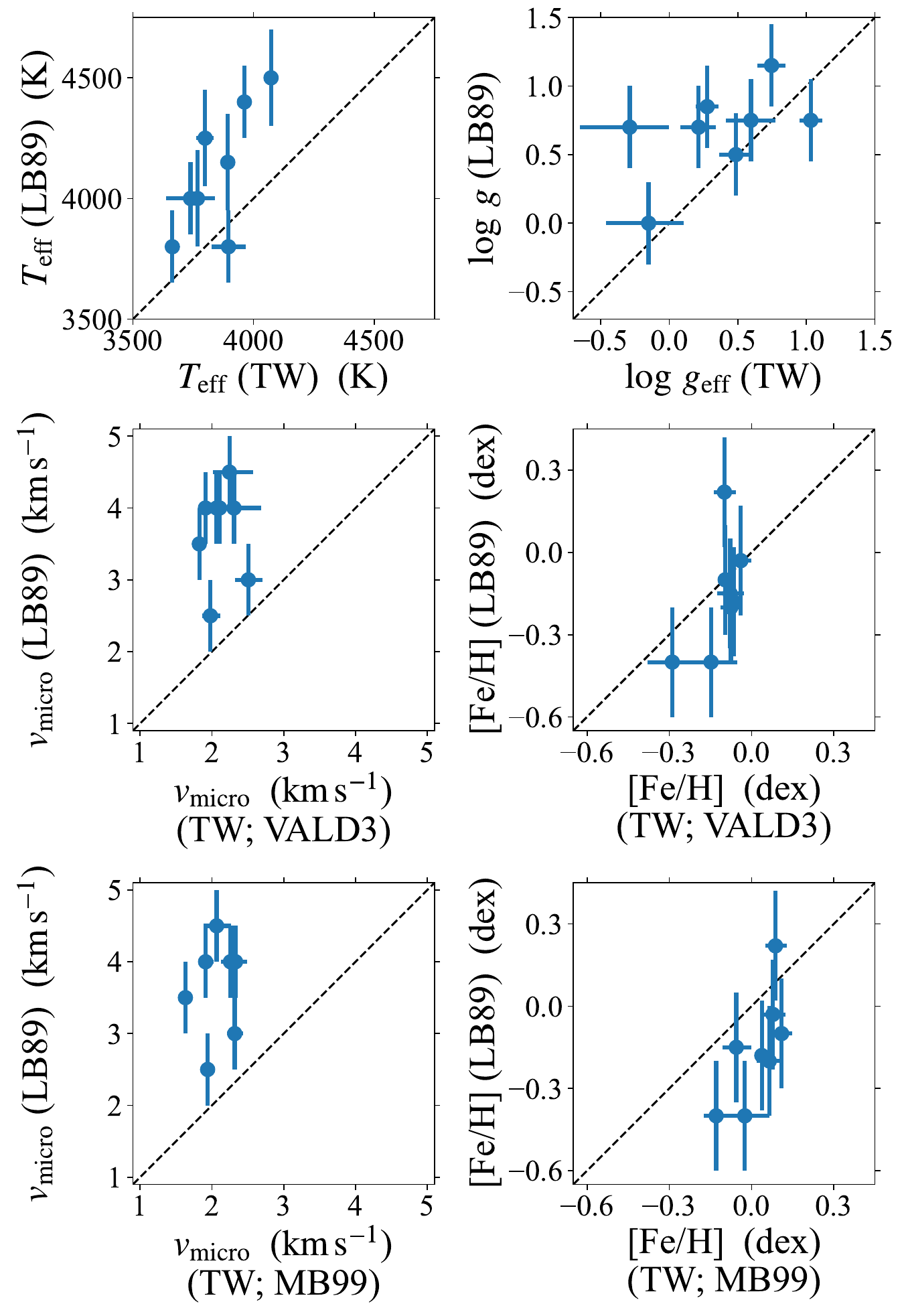}
\caption{Comparison of our results and those of \citet{Luck1989} for the RSGs included in both samples: stellar parameters and [Fe/H]. Top panels show $T_{\mathrm{eff}}$ and $\log g$, which are used in common with VALD3 and MB99. Middle and bottom panels show $v_{\mathrm{micro}}$ and [Fe/H] determined with VALD3 and MB99, respectively. }
\label{fig:CompLuck}
\end{figure}

Some previous works determined the stellar parameters and/or chemical abundances of the ten RSGs that we analyzed in this paper. 
In this section, we compare our results with previous measurements. 

Figure~\ref{fig:CompLevesqueTefflogg} compares $T_{\mathrm{eff}}$ and $\log g$ in this work with those determined by \citet{Levesque2005}. 
\citet{Levesque2005} determined $T_{\mathrm{eff}}$ of all our ten target RSGs, but they only determined $\log g$ of five RSGs among them (\object{V809 Cas}, \object{V424 Lac}, \object{TV Gem}, \object{BU Gem}, and \object{NO Aur}). 
We find that the difference in our results of $T_{\mathrm{eff}}$ and theirs are smaller than $100\ur{K}$, which is almost within the error bars~(see the detailed discussion in Sect.~4.2 of \citetalias{Taniguchi2021}). 
We also find a good agreement between our results of $\log g$ and theirs, which is expected, given that \citet{Levesque2005} and we used similar methods in determining $\log g$. 
These consistencies support the reliability of our $T_{\mathrm{eff}}$ and $\log g$ measurements. 

Figure~\ref{fig:CompLuck} compares stellar parameters and [Fe/H] in this work and those determined by \citet{Luck1980} and \citet{Luck1982b,Luck1982a} and summarized by \citet{Luck1989}. 
Our and their samples include eight common RSGs (\object{$\zeta $  Cep}, \object{41 Gem}, \object{$\xi $ Cyg}, \object{V809 Cas}, \object{V424 Lac}, \object{TV Gem}, \object{BU Gem}, and \object{NO Aur}). 
The comparison reveals large differences in the derived stellar parameters, especially in $T_{\mathrm{eff}}$ and $v_{\mathrm{micro}}$, which might be attributed to differences in the derivation methods and the model atmospheres employed; the procedure employed by \citet{Luck1989} was based on EW measurements of individual lines in the optical. 
There is no simple way to determine which one of the two results is more accurate. 
Nevertheless, at least, our determined $T_{\mathrm{eff}}$ and $\log L$ are in good agreement with a stellar evolution model by \citet{Ekstrom2012}~(see Sect.~4.3 of \citetalias{Taniguchi2021}). 
Moreover, the dependence of the correction term $\Delta \text{[Fe/H]}_{i}$ on EP is consistent with zero for both VALD3 and MB99 line lists: $+0.004\pm 0.028$ and $+0.042\pm 0.025\ur{dex/eV}$, respectively~(top-left panels of Figs.~\ref{fig:loggfCorrectionVALD} and \ref{fig:loggfCorrectionMelendez}). 
In other words, our $T_{\mathrm{eff}}$ values determined using the LDR method, and thus independent of the abundance measurement through the line fitting, satisfy the condition known as the excitation equilibrium~\citep[e.g.,][]{Jofre2019}. 
These facts reassure for the accuracy of our result. 
In the next section, we further use some well-established relations to discuss the reliability of our abundance analysis results.

\subsection{Validation of the abundance analysis results}\label{ssec:Validation}

\begin{figure}
\centering 
\includegraphics[width=9cm]{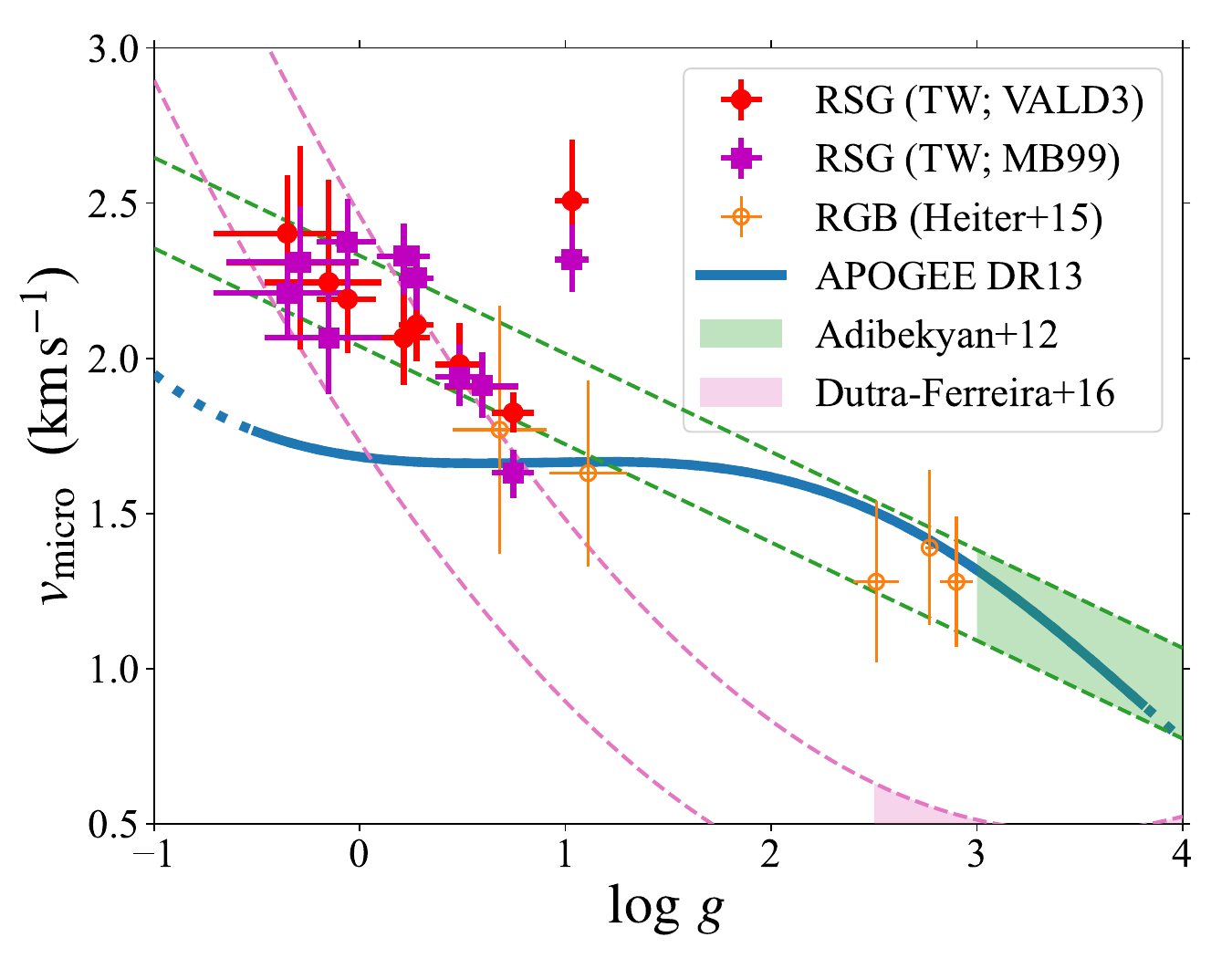}
\caption{Relation between $\log g$ and $v_{\mathrm{micro}}$. Red closed circles and magenta closed squares indicate the values that we determined for the target RSGs with VALD3 and MB99, respectively. Orange open circles indicate the values for the five solar-metallicity red giants among the \textit{Gaia} FGK benchmark stars~\citep{Heiter2015} used in \citetalias{Taniguchi2021}. Blue solid lines show the relation used in the ASPCAP code for APOGEE DR13~\citep{Holtzman2018} for the $\log g$ ranges of their calibrating sample, with the extrapolated relation indicated by blue dotted lines. Green and pink dashed lines indicate the relations calibrated by \citet{Adibekyan2012} and \citet{DutraFerreira2016}, respectively, for $T_{\mathrm{eff}}=3500$ and $4000\ur{K}$, with the $\log g$ ranges of their calibrating samples indicated by shades in the respective colors. }
\label{fig:loggvmicroRSG}
\end{figure}

In this section, we validate our results on two points: (i)~the relation between $\log g$ and $v_{\mathrm{micro}}$~(Sect.~\ref{sssec:loggvmicro}), and (ii)~comparison with the Galactic radial metallicity/abundance gradients traced with Cepheids~(Sects.~\ref{sssec:MetalGradCep} and \ref{sssec:AbnGradCep}, respectively).

\subsubsection{Relation between $\log g$ and $v_{\mathrm{micro}}$}\label{sssec:loggvmicro}

We show the relation between $\log g$ and $v_{\mathrm{micro}}$ in Fig.~\ref{fig:loggvmicroRSG} to examine the reliability of our determined $v_{\mathrm{micro}}$ values, which affects the resultant abundances. 
It is known that $v_{\mathrm{micro}}$ can be, in general, approximated by a function of $\log g$ and some additional parameters~\citep[e.g.,][]{Holtzman2018}. 
Indeed, Fig.~\ref{fig:loggvmicroRSG} shows an overall (negative) correlation between $\log g$ and $v_{\mathrm{micro}}$ that we derived for our target RSGs in both the results with the VALD3 and MB99 line lists. 
Moreover, $\log g$ and $v_{\mathrm{micro}}$ of the RSGs obtained in this work and red giants obtained in a previous work~\citep{Heiter2015} seem to form a continuous relation over a large $\log g$ range even though there is no guarantee that RSGs and red giants follow a single $\log g$--$v_{\mathrm{micro}}$ relation. 
With these results, we conclude that there is no evidence of an apparent systematic bias in our $v_{\mathrm{micro}}$ determination. 

We then compare the relation between $\log g$ and $v_{\mathrm{micro}}$ with those in literature. 
Figure~\ref{fig:loggvmicroRSG} overlays three relations from literature: one calibrated and used by \citet{Holtzman2018} for APOGEE DR13, one calibrated using observational samples of $v_{\mathrm{micro}}$ measurements by \citet{Adibekyan2012}, and one calibrated using the CIFIST grid of 3D hydrodynamic models~\citep{Ludwig2009} by \citet{DutraFerreira2016}. 
We note that the second among the three relations was used by \citet{AlonsoSantiago2017} and the third one by \citet{AlonsoSantiago2018,AlonsoSantiago2019} and \citet{Negueruela2021} to estimate $v_{\mathrm{micro}}$ of RSGs. 
We find an overall agreement between our results and the previously-reported three relations around $-0.5\lesssim \log g\lesssim 0.5$. 
Nevertheless, some systematic differences are present between the relations. 
The differences might be attributed to the fact that the previously-reported relations are not optimized for the $\log g$ range of RSGs. 
Indeed, the covered ranges for the stellar parameters of the calibrating samples are $3<\log g<5$ and $4500<T_{\mathrm{eff}}<6500\ur{K}$ for the work by \citet{Adibekyan2012} and $2.5\leq \log g\leq 4.5$ and $4400<T_{\mathrm{eff}}<6500\ur{K}$ for that by \citet{DutraFerreira2016}. 
The sample of APOGEE DR13 covers a wider range, $-0.5<\log g<3.8$, which include the $\log g$ range of RSGs; nevertheless their sample is mostly concentrated in a relatively narrow range $1.5\lesssim \log g\lesssim 3.5$~\citep[Fig.~6 of][]{Holtzman2018}. 
Thus, the relation for APOGEE DR13 for lower $\log g$ stars may have considerable systematic uncertainty. 
In fact, the stars in the APOGEE DR14~\citep{Holtzman2018} with $T_{\mathrm{eff}}$ and $\log g$ comparable with those of our target RSGs ($T_{\mathrm{eff}}\lesssim 4000\ur{K}$ and $\log \lesssim 0.5\ur{dex}$) have $v_{\mathrm{micro}}>2.0\ur{\si{km.s^{-1}}}$. 
These $v_{\mathrm{micro}}$ values are inconsistent with the $\log g$--$v_{\mathrm{micro}}$ relation that was adopted for APOGEE DR13 but are consistent with the $v_{\mathrm{micro}}$ values of RSGs determined here. 
A grid of 3D hydrodynamic models for RSGs is required to examine further the reliability of the estimated $v_{\mathrm{micro}}$, which is beyond the scope of this work.

\subsubsection{Radial metallicity gradient compared with Cepheids}\label{sssec:MetalGradCep}

\begin{figure*}
\centering 
\includegraphics[width=16.2cm]{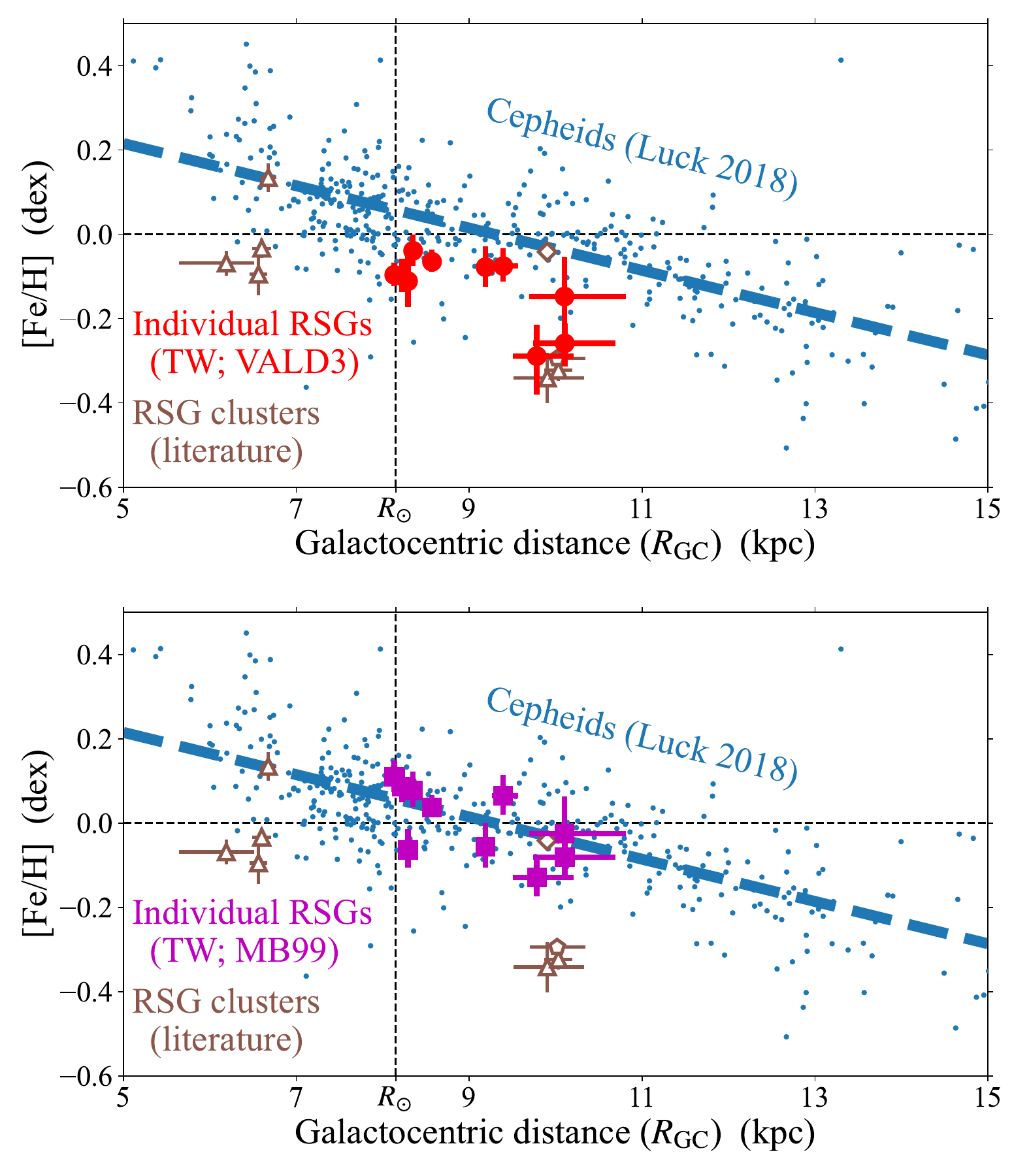}
\caption{Metallicities of RSGs compared with the radial metallicity gradient of Cepheids for the Galactocentric distance ($R_{\mathrm{GC}}$). Filled red circles (top panel) and filled magenta squares (bottom panel) show the derived [Fe/H] of our target RSGs for VALD3 and MB99, respectively. Blue dots show [Fe/H] of Cepheids~\citep{Luck2018} within $5<R<14\ur{kpc}$ as an indicator of the radial metallicity gradient of young stars. Blue dashed lines show the linear fit to the Cepheids's metallicities. Open brown symbols show the weighted-mean metallicities and their corresponding standard errors of RSGs in star clusters or star-forming complexes within $6\lesssim R_{\mathrm{GC}}\lesssim 10\ur{kpc}$ measured by some previous works: \citet{Gazak2014} depicted with a diamond, \citet{AlonsoSantiago2017,AlonsoSantiago2018,AlonsoSantiago2019,AlonsoSantiago2020} and \citet{Negueruela2021} with triangles, and \citet{Fanelli2022} with a pentagon. }
\label{fig:MetalGrad}
\end{figure*}

In this section and the next section, we compare the chemical abundances of RSGs with those of another type of young stars, Cepheids. 
Ideally, we should compare the abundances of RSGs and Cepheids in a single cluster to ensure that both objects have a common abundances. 
However, the number of clusters encompassing RSGs and Cepheids~\citep[e.g.,][]{Negueruela2020,AlonsoSantiago2020} is rather limited. 
Thus, instead, we have compared the derived chemical abundances of RSGs with the radial abundance gradients traced with Cepheids using the abundance measurements presented by \citet{Luck2018}. 
Considering the young ages of RSGs (${\lesssim }50\ur{Myr}$) and Cepheids (${\lesssim }300\ur{Myr}$), the abundances of both RSGs and Cepheids are expected to follow the common gradients, assuming that there is no mechanism favoring the formation of low- or high-metallicity RSGs and/or Cepheids. 
In fact, \citet{Esteban2022} demonstrated that some of the young objects in the solar-neighborhood (\ion{H}{ii} regions, B-type stars, classical Cepheids, and young open clusters) have the metallicity consistent with each other within $0.1\ur{dex}$. 

Figure~\ref{fig:MetalGrad} plots the metallicities of our target RSGs obtained in this work, along with the metallicities of Cepheids reported by \citet{Luck2018} as a function of the Galactocentric distance $R_{\mathrm{GC}}$. 
Also shown are some of the metallicity measurements of RSGs in star clusters or star-forming complexes from previous works~\citep{AlonsoSantiago2017,AlonsoSantiago2018,AlonsoSantiago2019,AlonsoSantiago2020, Negueruela2021,Gazak2014,Fanelli2022}, as we focus on RSG clusters in forthcoming papers. 
We calculated the $R_{\mathrm{GC}}$ values of all the plotted objects assuming the distance to the Galactic Center of $R_{\odot }=8.15\ur{kpc}$~\citep{Reid2019}, which is different from $7.9\ur{kpc}$ adopted by \citet{Luck2018} for gradient calculations. 
Accordingly, we recalculated the radial metallicity gradient of Cepheids, after five iterations of three-sigma clipping, using the [Fe/H] values reported by \citet{Luck2018} and the Bayesian distance estimates using the \textit{Gaia} DR2 parallax data by \citet{Bailer-Jones2018} with excluding some stars: those with negative \textit{Gaia} DR2 parallaxes following \citet{Luck2018}, five stars (\object{HK Cas}, \object{BC Aql}, \object{QQ Per}, \object{EK Del}, and \object{EQ Lac}) as recommended by \citet{Luck2018}, and \object{SU Cas} as recommended by \citet{Kovtyukh2022b} and \citet{Matsunaga2023}. 
We also note that we rescaled the [Fe/H] values presented in the previous works to the solar abundances reported by \citet{Asplund2009}, when the differential analysis against the solar spectrum might not have been performed. 

Consequently, we find a good agreement between [Fe/H] of the RSGs that we obtained using MB99 and those of Cepheids; the difference in the gradients between the two is $0.004\ur{dex}$. 
In contrast, [Fe/H] of the RSGs obtained using VALD3 is slightly, by $0.125\ur{dex}$, smaller than those of Cepheids. 
This level of discrepancy is as expected given the difference in the $\log gf$ values in the two line lists~\citep[see Fig.~7 in][]{Kondo2019}. 
In fact, analyzing NIR \textit{YJ}-band spectra of two red giants, Arcturus and $\mu $~Leo, \citet{Kondo2019} found that [Fe/H] of the two stars determined with the MB99 list were well consistent with literature values, but [Fe/H] using VALD3 were smaller than those using MB99 by $0.20$ and $0.11\ur{dex}$ for Arcturus and $\mu $~Leo, respectively. 
These consistencies support the reliability of our [Fe/H] measurements, especially when using the MB99 list, indicating that our [Fe/H] measurements should be accurate within ${\sim }0.1\ur{dex}$. 

In contrast, [Fe/H] of RSGs determined by some previous works among those plotted in Fig.~\ref{fig:MetalGrad}~\citep{AlonsoSantiago2018,AlonsoSantiago2019,AlonsoSantiago2020,Negueruela2021,Fanelli2022} are found to be systematically lower than those of Cepheids by $0.2\text{--}0.3\ur{dex}$. 
Such low [Fe/H] values have been often found in cool giants with low $\log g$~\citep[e.g.,][]{Casali2020,Magrini2023,GaiaRecioBlanco2023}. 
A part of the systematic differences, especially of \citet{Fanelli2022}, could possibly be explained with the $v_{\mathrm{micro}}$ values that they adopted, as discussed below, considering the strong degeneracy between [Fe/H] and $v_{\mathrm{micro}}$. 

\begin{figure}
\centering 
\includegraphics[width=9cm]{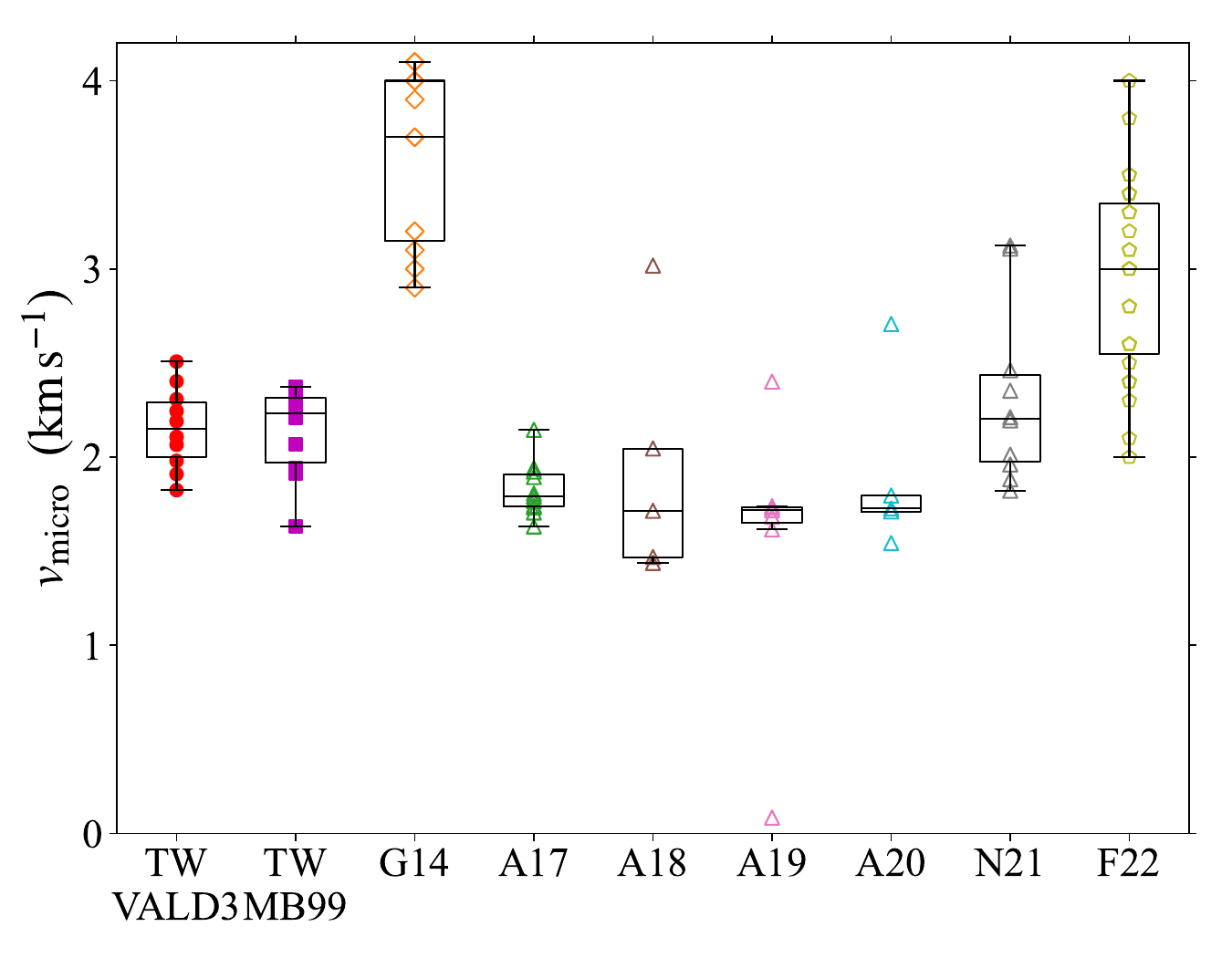}
\caption{Box plot of $v_{\mathrm{micro}}$ determined in this work (marked as TW) and previous works for RSGs plotted in Fig.~\ref{fig:MetalGrad}: \citet{Gazak2014} as G14, \citet{AlonsoSantiago2017,AlonsoSantiago2018,AlonsoSantiago2019,AlonsoSantiago2020} as A17--A20, respectively, \citet{Negueruela2021} as N21, and \citet{Fanelli2022} marked as F22. }
\label{fig:compvmic}
\end{figure}

We show in Fig.~\ref{fig:compvmic} $v_{\mathrm{micro}}$ adopted by this work and the previous works cited in Fig.~\ref{fig:MetalGrad} to highlight their differences to help understand the discrepancies in [Fe/H] among the works in conjunction with $v_{\mathrm{micro}}$. 
Our $v_{\mathrm{micro}}$ (TW in the figure) are found to be concentrated at around ${\sim }2\ur{\si{km.s^{-1}}}$ and are similar to those reported by \citet{AlonsoSantiago2017,AlonsoSantiago2018,AlonsoSantiago2019,AlonsoSantiago2020} and \citet{Negueruela2021} (designated as A17, A18, A19, A20, and N21, respectively, in the figure). 
In contrast, those reported by \citet{Gazak2014} and \citet{Fanelli2022} (G14 and F22, respectively) are significantly higher than our values. 

In the work by \citet{Fanelli2022} among those cited in Fig.~\ref{fig:compvmic}, they analyzed optical and NIR spectra of RSGs in the Perseus Complex. 
We find that their resultant [Fe/H] are systematically ${\sim }0.3\ur{dex}$ lower than the metallicity gradient of Cepheids~(squares in Fig.~\ref{fig:MetalGrad}), and we discuss here its possible connection to the $v_{\mathrm{micro}}$ values that they adopted. 
They adopted $v_{\mathrm{micro}}$ of ${\sim }1\ur{\si{km.s^{-1}}}$ higher than ours, maybe because they included strong \ion{Fe}{i} lines in their analysis; they used \ion{Fe}{i} lines having $-4\lesssim \log \tau _{\mathrm{Ross}}^{}\lesssim -1$, as opposed to our line-selection criterion of $\log \tau _{\mathrm{Ross}}^{}>-3$. 
Their larger $v_{\mathrm{micro}}$ could result in a ${\sim }0.2\text{--}0.4\ur{dex}$ smaller [Fe/H] than ours. 
In fact, recalculation of $v_{\mathrm{micro}}$ of our target RSGs with the criteria of $\log \tau _{\mathrm{Ross}}^{}>-4$ instead of $-3$ yields an increase in $v_{\mathrm{micro}}$ by ${\sim }0.8$ and $0.3\ur{\si{km.s^{-1}}}$ for VALD3 and MB99, respectively, which results in [Fe/H] smaller by ${\sim }0.18$ and $0.06\ur{dex}$, respectively. 
This positive systematic bias in $v_{\mathrm{micro}}$ is caused by the large positive systematic errors in the measured [Fe/H] of strong lines ($\log \tau _{\mathrm{Ross}}^{}<-3$) as shown in the top right panels of Figs.~\ref{fig:loggfCorrectionVALD} and \ref{fig:loggfCorrectionMelendez}. 
The difference in the $T_{\mathrm{eff}}$ values could also in part contribute to the difference in the resultant [Fe/H], but it would be smaller because the sensitivity of [Fe/H] to $T_{\mathrm{eff}}$ is low: $\Delta _{T_{\mathrm{eff}}}/\Delta T_{\mathrm{eff}}\sim 0.02\ur{dex}/100\ur{K}$~(See $\Delta T_{\mathrm{eff}}$ in the left panels of Fig.~\ref{fig:errorbudgetall}). 

In the work by \citet{AlonsoSantiago2017,AlonsoSantiago2018,AlonsoSantiago2019,AlonsoSantiago2020} and \citet{Negueruela2021} among those cited in Fig.~\ref{fig:compvmic}, they analyzed optical spectra of RSGs in some young clusters (\object{NGC 6067}, \object{NGC 3105}, \object{NGC 2345}, \object{NGC6649}, \object{NGC 6664}, and \object{Valparaiso 1}). 
The resultant [Fe/H] of all these works except for the work by \citet{AlonsoSantiago2017} are systematically ${\sim }0.2\text{--}0.3\ur{dex}$ lower than the metallicity gradient of Cepheids~(triangles in Fig.~\ref{fig:MetalGrad}), although they used $v_{\mathrm{micro}}$ whose ranges are similar to ours~(Fig.~\ref{fig:compvmic}). 
Since most of their observed targets (spectral types between G--K) are warmer than our target RSGs (spectral types between K--early M) and also have larger $\log g$, it is not trivial to identify the cause of the differences. 

In the work by \citet{Gazak2014}, which is the last one among those cited in Fig.~\ref{fig:compvmic}, they obtained [Fe/H] of RSGs consistent with the metallicity gradient of Cepheids~(diamonds in Fig.~\ref{fig:MetalGrad}). 
Their spectra have relatively low resolution compared to those used in all the other works mentioned here. 
Furthermore, they determined global metallicity, using most of the lines appearing in the \textit{J} band including atomic lines from elements other than iron, molecular lines, and/or strong lines. 
This is in contrast to our approach, which focuses solely on relatively weak \ion{Fe}{i} lines to measure [Fe/H]. 
Given these methodological differences, we do not discuss the cause of the consistency here.

\subsubsection{Radial abundance gradients compared with Cepheids}\label{sssec:AbnGradCep}

\begin{figure}
\centering 
\includegraphics[width=9cm]{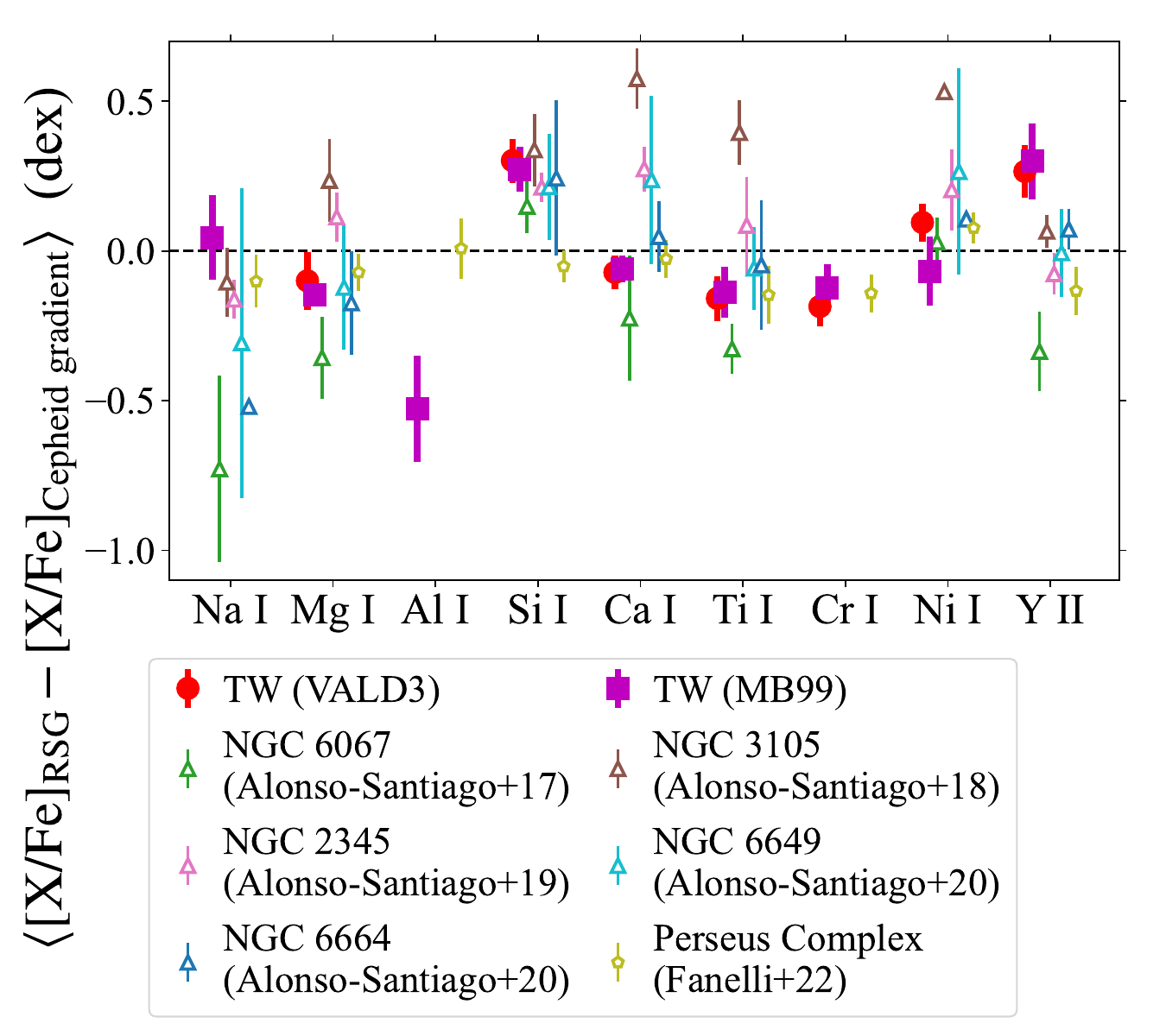}
\caption{Chemical abundances of RSGs after subtracting the radial abundance gradients of Cepheids. Filled red circles and filled magenta squares show the weighted mean and standard error of the derived [X/Fe] of our targets RSGs for VALD3 and MB99, respectively, after subtracting the radial abundance gradients of Cepheids, which are tabulated in Table~\ref{table:AbnFeRes} as Mean. Open symbols show those for RSGs by the works measuring [X/Fe] as well as [Fe/H] among those cited in Fig.~\ref{fig:MetalGrad}: \citet{AlonsoSantiago2017,AlonsoSantiago2018,AlonsoSantiago2019,AlonsoSantiago2020} with green, brown, pink, and cyan/blue triangles, respectively, and \citet{Fanelli2022} with yellow pentagons. We note that we show the results for all the elements for which we determined the abundances of RSGs, except for [K/Fe], as the abundance for Cepheids were not measured by \citet{Luck2018}. }
\label{fig:AbnGradDiff}
\end{figure}

Regarding chemical abundances of elements other than iron, we plot in Fig.~\ref{fig:AbnGradDiff} the weighted means of the derived [X/Fe] of our target RSGs after the radial abundance gradients of Cepheids are subtracted. 
The abundance gradients are calculated as is done for the metallicity gradient. 
As with the case for [Fe/H] discussed in the previous section, the abundance ratios [X/Fe] of both RSGs and Cepheids are expected to follow common gradients. 
Hence, the differences between them, which are plotted in the figure, would be zero when the abundance measurements for both RSGs in this work and Cepheids in the work by \citet{Luck2018} are accurate. 
We note that sodium synthesized inside a star via the NeNa cycle can potentially appear on the surface of evolved stars through mechanism(s) such as dredge-up, rotation, and mass loss~\citep{ElEid1994,Ekstrom2012,Smiljanic2016}. 
Consequently, the current surface abundances of sodium, as well as carbon, nitrogen, and oxygen, of RSGs do not necessarily reflect their initial surface abundances, and by extension, the current surface abundances of Cepheids. 
In other words, the values plotted in Fig.~\ref{fig:AbnGradDiff} for \ion{Na}{i} need not be zero. 

Consequently, we find a good agreement (i.e., within ${\sim }0.1\ur{dex}$) in the abundance ratio of the most representative $\alpha $~element, [Mg/Fe], along with some other elements (e.g., [Ca/Fe] and [Ni/Fe]). 
On the contrary, we find systematic offsets in the obtained abundances for some other species, most notably for [Si/Fe] and [Y/Fe] with offsets of ${\sim }0.3\ur{dex}$. 
Discrepancies of this type were often seen in RSGs' abundances reported by previous papers~(open symbols in Fig.~\ref{fig:AbnGradDiff}). 
The reason for the discrepancies is, however, unknown as of yet and is a remaining problem in the abundance analysis of RSGs. 

In order to assess the possible impact of one of the shortcomings of our analysis, namely the LTE assumption, we derived non-LTE corrections for a part of the lines of \ion{Mg}{i}, \ion{Si}{i}, \ion{Ca}{i}, \ion{Ti}{i}, \ion{Cr}{i}, and \ion{Fe}{i} using the online tool developed by M.~Bergemann's group~\citep{Kovalev2019}\footnote{\url{https://nlte.mpia.de/}}. 
The RSG3 model parameters~(Table~3 of \citetalias{Taniguchi2021}) and the RSG-MARCS grid of model atmospheres were employed in the test. 
For [Fe/H], we find that the non-LTE corrections for $34$ out of $57$ \ion{Fe}{i} lines that were used for either VALD3 or MB99 can be calculated with the tool~\citep{Bergemann2012b,Bergemann2012a}, and all the corrections are negligible ($\lesssim \pm 0.01\ur{dex}$), indicating that the non-LTE effect does not affect our metallicity determination. 
Similarly, the non-LTE corrections for [Ca/H] and [Cr/H] are also negligible. 
For [Ca/H], $5$ out of $6$ \ion{Ca}{i} lines have corrections~\citep{Mashonkina2007}, and for [Cr/H], $10$ out of $15$ \ion{Cr}{i} lines have corrections~\citep{Bergemann2010}, all of which were zero.
In contrast, the non-LTE effect may affect [Mg/H], [Si/H], and [Ti/H]. 
For [Mg/H], the corrections can be calculated for $4$ out of $5$ \ion{Mg}{i} lines~\citep{Bergemann2015,Bergemann2017}: the corrections are zero for two lines ($12417.937\ur{\AA }$ and $12433.45\ur{\AA }$), $-0.040\ur{dex}$ for $12039.822\ur{\AA }$, and $-0.312\ur{dex}$ for $12083.65\ur{\AA }$. 
The rather large correction for the last line, which was only used with the MB99 list, is consistent with the large $\Delta \text{[X/H]}_{i}$ value for the line, $+0.463\ur{dex}$. 
We reiterate that a positive $\Delta \text{[X/H]}_{i}$ value corresponds to the observed line strength being higher than the synthesized one. 
For [Si/H] and [Ti/H], the corrections can be calculated for $14$ out of $18$ \ion{Si}{i} lines~\citep{Bergemann2013} and $18$ out of $25$ \ion{Ti}{i} lines~\citep{Bergemann2011}, with typical corrections of $-0.17$ and $+0.11\ur{dex}$, respectively. 
These may at least partly explain the abundance discrepancies of ${\sim }+0.3\ur{dex}$ for [Si/Fe] and ${\sim }-0.15\ur{dex}$ for [Ti/Fe]. 
In summary, while the non-LTE effect may not affect our abundance results of some elements (\ion{Fe}{i}, \ion{Ca}{i}, and \ion{Cr}{i}), they may have a noticeable impact on some others (\ion{Mg}{i}, \ion{Si}{i}, and \ion{Ti}{i}). 
Nevertheless, we dare not apply the non-LTE corrections to our measurements given the incomplete line list in the tool used. 
Further 3D non-LTE modeling of RSG spectra with a more complete line list is required to better understand the abundance discrepancies observed between RSGs and Cepheids. 

Nevertheless, since most of the Galactic RSGs have stellar parameters within a certain small range~\citep[$3500\lesssim T_{\mathrm{eff}}\lesssim 3900\ur{K}$ and $-9\lesssim M_{\mathrm{bol}}\lesssim -6$;][]{Levesque2005}, we expect that the amount of the systematic error for a given element is nearly constant for any RSGs at least with solar metallicity, as far as the same abundance analysis method, the same line list, the same model atmosphere grid, and the same wavelength coverage is employed. 
To examine if the expectation is genuinely the case with our results, we calculated the weighted standard deviation~(SD) of [Fe/H] and [X/Fe] among the ten RSGs after the radial abundance gradients of Cepheids are subtracted. 
These values basically represent the summations of the statistical and systematic errors in our abundance measurements (without systematic offsets included in the summations), assuming that the chemical abundance of RSGs for each element follows a tight abundance gradient. 
Tables~\ref{table:AbnRes} and \ref{table:AbnFeRes} tabulate the calculated SDs. 
In consequence, we find that the dispersions are within $0.04\text{--}0.12\ur{dex}$ for the elements with the number of measured lines $N_{\mathrm{line}}$ larger than two ([Fe/H], [Mg/Fe], [Si/Fe], [Ca/Fe], [Ti/Fe], [Cr/Fe], and [Ni/Fe]). 
The other elements with a smaller $N_{\mathrm{line}}$, [Na/Fe], [Al/Fe], and [Y/Fe], have the dispersion within $0.09\text{--}0.18\ur{dex}$. 
These dispersions are consistent with the quoted errors, at least for most elements. 
This fact implies good reliability of our procedure of abundance measurement within the quoted error for the relative abundance difference between two objects, although the absolute abundance values for some elements still suffer a significant amount of systematic bias in general.


\section{Summary and future prospects}

In this paper, we establish a procedure for determining the chemical abundances of RSGs using NIR high-resolution spectra in the \textit{YJ} bands. 
We tested the procedure through the analysis of NIR high-resolution spectra of ten nearby RSGs located within $8\lesssim R_{\mathrm{GC}}\lesssim 10\ur{kpc}$, which were obtained with the WINERED spectrograph~($0.97\text{--}1.32\,\micron $; $R=28\,000$). 
In our procedure, we first determined the effective temperature $T_{\mathrm{eff}}$, using LDRs of 11 \ion{Fe}{i}--\ion{Fe}{i} line pairs as in \citetalias{Taniguchi2021}, and calculated the surface gravity $\log g$ using the Stefan-Boltzmann law combined with Geneva's stellar evolution model. 
We then determined the microturbulent velocity $v_{\mathrm{micro}}$ and the metallicity [Fe/H] simultaneously by fitting relatively isolated individual \ion{Fe}{i} absorption lines. 
Finally, we fitted individual lines and determined the relative abundance of elements X to hydrogen [X/H] of \ion{Mg}{i}, \ion{Si}{i}, \ion{Ca}{i}, \ion{Ti}{i}, \ion{Cr}{i}, \ion{Ni}{i}, and \ion{Y}{ii} for the VALD3 and MB99 line lists, and in addition \ion{Na}{i}, \ion{Al}{i}, and \ion{K}{i} for the MB99 list. 
We also estimated the relative precisions of the abundances using the standard deviations for the sample RSGs and found them to be $0.04\text{--}0.12\ur{dex}$ for the elements with a sufficient number of lines analyzed (e.g., \ion{Fe}{i} and \ion{Mg}{i}) and up to $0.18\ur{dex}$ for the elements with fewer than three lines analyzed (e.g., \ion{Na}{i} and \ion{Y}{ii}). 

Our procedure has advantages over previous works with regard to three main points: (1)~the procedure is based on the fitting of observed spectra with synthesized ones on a line-by-line basis, as opposed to simple measurements of EWs as employed in some works, which allows us to circumvent the overestimation of abundances due to contamination by surrounding lines; (2)~the procedure does not use molecular lines for determining stellar parameters, which allows us to circumvent effects related to the complicated outer atmospheres of RSGs; and (3)~the procedure carefully adjusts [C/H], [N/H], and [O/H] to minimize potential systematic bias in the fitting of the lines of interest that originate in contaminating \ce{CN} molecular lines. 

We evaluated the reliability of our results in two ways. 
First, we compared the relation between $\log g$ and $v_{\mathrm{micro}}$ with those derived from previous observational and theoretical results and found no apparent systematic bias in our derived $v_{\mathrm{micro}}$ values~(Fig.~\ref{fig:loggvmicroRSG}). 
Second, we compared the radial abundance gradients of our sample RSGs with those of Cepheids in the literature~(Figs.~\ref{fig:MetalGrad} and \ref{fig:AbnGradDiff}). 
We found good agreement~(${\lesssim }0.1\ur{dex}$) for some abundances, notably [Fe/H] and [Mg/Fe], which are particularly useful abundances in the study of galactic chemical evolution. 
This result markedly differs from those of most previous works, reporting values of ${\sim }0.2\text{--}0.3\ur{dex}$ lower [Fe/H] than those of Cepheids. 
In contrast, we found discrepancies of as large as ${\sim }0.5\ur{dex}$ for some others, such as [Si/Fe], the cause of which may be related to a 3D non-LTE effect and/or uncertainty in the line list used, although were not able to come to any firm conclusions. 
Nevertheless, our procedure should be reliable with regard to its capacity to differentiate the abundances of two RSGs (or two groups of RSGs), considering that the standard deviation of the derived chemical abundances among our sample RSGs is comparable with the quoted precision in the measured abundances. 

RSGs have extremely high luminosities~(${\gtrsim }10^{4}L_{\sun }$), and hence they can be used as good tracers of the chemical abundances of young stars at large distances. 
Indeed, it is expected that we will be able to spectroscopically observe the brightest individual RSGs at a distance of ${\sim }1\ur{Mpc}$, which is the distance to M31, with recently developed and/or near-future NIR high-resolution spectrographs with a very high throughput attached to large-aperture telescopes, such as WINERED/Magellan~\citep{Ikeda2022}. 
Also, RSGs over a large area of the Galactic plane can be observed even with less-sensitive facilities, as long as the dust extinction to the target is not excessively severe. 
Mapping observations of these RSGs would be highly useful for studying the 2D distribution of the chemical abundances on the disks of the Milky Way and nearby galaxies, providing a means to constrain galactic chemical evolution theory.

\section*{Data availability}

The abundance measurements and line list are available as Appendix~\ref{app:tables} at Zenodo (\url{https://doi.org/10.5281/zenodo.14286491}). 
All the tables and reduced spectra are available at the CDS via anonymous ftp to \url{cdsarc.cds.unistra.fr} (\url{130.79.128.5}) or via \url{https://cdsarc.cds.unistra.fr/viz-bin/cat/J/A+A/693/A163}

\begin{acknowledgements}
We acknowledges useful comments from the referee, Alexander Ebenbichler. 
We are grateful to Yuki Moritani, Kyoko Sakamoto, Keiichi Takenaka, and Ayaka Watase for observing a part of our targets. 
We also thank the staff of Koyama Astronomical Observatory for their support during our observations. 
WINERED was developed by the University of Tokyo and the Laboratory of Infrared High-resolution spectroscopy~(LiH), Kyoto Sangyo University under the financial supports of Grants-in-Aid, KAKENHI, from Japan Society for the Promotion of Science~(JSPS; Nos.~16684001, 20340042, and 21840052) and the MEXT Supported Program for the Strategic Research Foundation at Private Universities~(Nos.~S0801061 and S1411028). 

This work has been supported by Masason Foundation. 
DT acknowledges the financial support from Toyota/Dwango AI scholarship, Iwadare Scholarship Foundation, and JSPS Research Fellowship for Young Scientists and accompanying Grants-in-Aid for JSPS Fellows (21J11555 and 23KJ2149). 
BT acknowledges the financial support from the Japan Society for the Promotion of Science as a JSPS International Research Fellow. 
BT acknowledges the financial support from the Wenner-Gren Foundation (WGF2022-0041). 
HS acknowledges the financial support by JSPS KAKENHI Grant Number 19K03917. 

This work has made use of the VALD database, operated at Uppsala University, the Institute of Astronomy RAS in Moscow, and the University of Vienna. 
This research has made use of the SIMBAD database, operated at CDS, Strasbourg, France. 
This work presents results from the European Space Agency~(ESA) space mission \textit{Gaia}. \textit{Gaia} data are being processed by the \textit{Gaia} Data Processing and Analysis Consortium~(DPAC). Funding for the DPAC is provided by national institutions, in particular the institutions participating in the Gaia MultiLateral Agreement (MLA). The Gaia mission website is \url{https://www.cosmos.esa.int/gaia}. The Gaia archive website is \url{https://archives.esac.esa.int/gaia}. 
This publication makes use of data products from the Two Micron All Sky Survey, which is a joint project of the University of Massachusetts and the Infrared Processing and Analysis Center/California Institute of Technology, funded by the National Aeronautics and Space Administration and the National Science Foundation. 
\end{acknowledgements}

\bibliographystyle{aa_url}
\bibliography{RSGabn_taniguchi}

\begin{appendix}

\section{Spectral synthesis in \textsc{Octoman}}\label{app:MOOG}

For the spectral synthesis function in \textsc{Octoman}, we wrapped the spectral synthesis code MOOG~\citep{Sneden1973,Sneden2012}\footnote{We used the February-2017 version of MOOG further modified by M. Jian~(\url{https://github.com/MingjieJian/moog_nosm}). }. 
MOOG synthesizes spectra of late-type stars, assuming the 1D LTE with the plane-parallel geometry. 

\textsc{Octoman} provides users with some choices of the model atmosphere grid, including the \textsc{ATLAS9} grids~\citep{Kurucz1993,Castelli2003,Meszaros2012} and the MARCS grids ~\citep{Gustafsson2008}. 
\textsc{Octoman} obtains the exact model atmosphere for a given set of stellar parameters with linear interpolation (or extrapolation for $\log g<0$) from the grid. 
In this paper, we used the MARCS spherical grid with $M=5M_{\sun }$ and $v_{\mathrm{micro}}=2\ur{\si{km.s^{-1}}}$. 
When no model atmosphere was provided for a grid point with $v_{\mathrm{micro}}=2\ur{\si{km.s^{-1}}}$, we used $v_{\mathrm{micro}}=5\ur{\si{km.s^{-1}}}$ model instead. 
We note that the difference in the geometries between the radiative transfer code (plane-parallel) and the model atmosphere (spherical) only slightly affects the synthesized spectra in general~\citep{Heiter2006}. 

The code provides three choices for the atomic line list: the third release of the Vienna Atomic Line Database~\citep[VALD3;][]{Ryabchikova2015}\footnote{Last downloaded on 2021 May 10 at the time of writing. }, the list of lines in $10,000\text{--}18,000\,\text{\AA }$ with astrophysical $\log gf$ values constructed by \citet[][MB99]{Melendez1999}, and the line list complied by R. Kurucz\footnote{\url{http://kurucz.harvard.edu/linelists/gfnew/}. }. 

The code considers lines of all the molecules included in the VALD3 database except for \ce{TiO}. 
VALD3 contains lines of \ce{^{12}C^{1}H}, \ce{^{13}C^{1}H}, \ce{^{14}N^{1}H}, \ce{^{12}C2}, \ce{^{12}C^{14}N}, \ce{^{12}C^{16}O}, \ce{^{12}C^{17}O}, \ce{^{12}C^{18}O}, \ce{^{13}C^{16}O}, and \ce{TiO} in the \textit{YJ} bands, among which \ce{^{12}C^{14}N} gives the largest contribution to the spectra of our target RSGs. 
A user can select whether to replace the line list of the \ce{^{12}C^{14}N} molecule from the VALD database with the list of \ce{^{12}C^{14}N}, \ce{^{12}C^{15}N}, and \ce{^{13}C^{14}N} calculated by \citet{Sneden2014}. 
The difference in \ce{^{12}C^{14}N} between the two is small, but we found that some lines in the \textit{YJ} bands only appear in the latter list\footnote{We found that some unidentified lines listed in Appendix~B of \citet{Matsunaga2020} are well reproduced by synthesized spectra of either \ce{^{12}C^{14}N} or \ce{^{13}C^{14}N} lines at least for RSGs. \ce{^{12}C^{14}N}: $10163.6$, $10273.1$, $10305.3$, $10338.5$, $10476.5$, $10542.5$, $10549.5$, $10587.1$, $10625.4$, and $10657.4\,\mathrm{\AA }$. \ce{^{13}C^{14}N}: $11050.3$, $11083.7$, $11742.0$, and $11784.9\,\mathrm{\AA }$. }. 
Thus, we used the latter list in this paper. 

The code provides three options for the line lists for metal oxides (e.g., \ce{TiO}), where weaker lines are filtered out of the complete set of the known lines according to a set of threshold conditions, in which the ``$X$ index'' at $3500\ur{K}$ defined in Eq.~(\ref{eq:Xindex}) is utilized. 
This constraint is set because the complete line lists of these molecules contain too many lines to synthesize. 
The three options are (i)~the combination of \ce{^{48}Ti^{16}O} ($X>-4.5$), \ce{^{51}V^{16}O} ($X>-3.0$), and \ce{^{90}Zr^{16}O} ($X>-3.5$) line lists calculated by B. Plez\footnote{\url{https://www.lupm.in2p3.fr/users/plez/}}, (ii)~the ExoMol line lists of \ce{^{48}Ti^{16}O} ($X>-4.5$), \ce{^{46}Ti^{16}O}, \ce{^{47}Ti^{16}O}, \ce{^{49}Ti^{16}O}, and \ce{^{50}Ti^{16}O} ($X>-3.5$)~\citep{McKemmish2019} and \ce{^{51}V^{16}O} ($X>-4.0$)~\citep{McKemmish2016}, and (iii)~basically identical to the second option but with a slight modification, adjusting $\log gf$ values of \ce{TiO} to better reproduce the observed spectra of RSGs by adding $0.3\ur{dex}$ to $\log gf$ of the $\phi $ system ($b{}^{1}\Pi $--$d{}^{1}\Sigma ^{+}$) and subtracting $0.3\ur{dex}$ from $\log gf$ of the $\delta $ system ($b^{1}\Pi $--$a{}^{1}\Delta $). 
In this paper, we adopted the option (iii) to best reproduce the observed spectra of the RSGs. 

In addition, the code adopts the line list of \ce{^{56}Fe^{1}H} calculated by B. Plez\footnotemark[\value{footnote}]. 
With extensive examination, we found that some lines of \ce{FeH} appear in the \textit{YJ}-band spectra of the RSGs with a depth of up to ${\sim }0.05$ and that those lines are well reproduced by synthesized spectra with the dissociation energy of $1.59\ur{eV}$~\citep{Schultz1991}.

\section{Fitting procedure for \ion{Fe}{i} absorption lines in \textsc{Octoman}}\label{app:Octoman}

This section describes the detailed procedure to fit a \ion{Fe}{i} absorption line implemented in the \textsc{Octoman} code. 
In the analysis of fitting of lines of other species presented in this paper, we use mostly the same procedure. 
The procedure mainly follows the algorithm presented by \citet{Takeda1995a} but with some modifications. 

We consider the following four variables during the fitting for a line: (1)~iron abundance, [Fe/H] (or the abundance of another element) --- the parameter of interest, (2)~FWHM (i.e., $v_{\mathrm{broad}}$) of the line broadening including three components of $v_{\mathrm{macro}}$, rotation, and instrumental broadening, (3)~Velocity offset, $\Delta \mathrm{RV}$, and (4)~Continuum normalization factor, $C$. 

Related to variable (1), in the current work, we fix the abundances of carbon, nitrogen, and oxygen to the respective values determined in Sect.~\ref{ssec:CNfitting}. 
We also fix the abundance values of the other elements to the iron abundance in determining [Fe/H]. 
We assume that the global metallicity [M/H] of the model atmosphere is equal to [Fe/H]. 

As for variable (2), the instrumental broadening in our observations ($R=28\,000$ for the WIDE mode of WINERED) is comparable with or smaller than $v_{\mathrm{macro}}$ of RSGs~\citep[e.g., ${\sim }15\ur{\si{km.s^{-1}}}$ by][]{Josselin2007}, and thus both the instrumental broadening and $v_{\mathrm{macro}}$ contribute to, but do not dominate, the \textit{net} broadening. 
The projected rotational velocities of RSGs, e.g., $v\sin i\simeq 5\ur{\si{km.s^{-1}}}$ for \object{Betelgeuse}~\citep{Wheeler2023}, could also slightly contribute to the \textit{net} line broadening, though such a large $v\sin i$ value for RSGs is not expected from single-star evolutionary models~\citep{Wheeler2017,Ma2024}. 
In our analysis, we fit a line with the Gaussian broadening profile, allowing the \textit{net} broadening velocity $v_{\mathrm{broad}}$ in \si{km.s^{-1}} to vary. 
Ideally, we should consider a non-Gaussian broadening profile for the following reasons. 
A macroturbulence profile deviates from the Gaussian~\citep{Gray2008,Magic2014}, especially in cases of stars like RSGs that have a small number of large granules in the photosphere~\citep{Chiavassa2010,Ohnaka2017}. 
A rotational broadening profile, though its contribution is expected to be usually small, does not follow the Gaussian, either, and depends on the limb darkening~\citep{Gray2008}. 
However, it is technically difficult to apply an exact and complicated model broadening profile to fit a line profile in an observed spectrum because most absorption lines in the spectra of RSGs are contaminated with other lines and also because line profiles vary, depending on the atmospheric layers of the origin for the line~\citep{Takeda1995b,Kravchenko2021}. 
This is why we choose a simple Gaussian function for the model fitting of the broadening. 

As for variable (3), though our spectra have been corrected for radial velocities, the observed wavelength of each line has an offset from the theoretical counterpart by up to ${\sim }1\ur{\si{km.s^{-1}}}$, possibly due to imperfect wavelength calibration and/or differences in the radial velocities between different lines~\citep[e.g.,][]{Kravchenko2021}. 
In order to correct the offset, we introduce a small velocity offset as a free parameter in the fitting. 

As for variable (4), though the continua of our spectra have been normalized in advance, the normalized continuum may have an offset from unity by ${\lesssim }1\%$ in some cases. 
In particular, the fitted continuum might be underestimated due to weak (molecular) lines, which may falsely build pseudo continuum. 
For this reason, we introduce a scaling factor $C$ as a free parameter for the observed spectrum around the line of interest. 

In the actual fitting, starting with a given initial guess of the four free parameters, we minimize the residual between the observed and synthesized spectra around an absorption line, using the Newton-Raphson method explained below. 
We define the almost-pre-normalized observed flux and perfectly normalized synthesized flux\footnote{We resample the original synthesized spectra in a way such that the flux is preserved. } of a pixel $i$ as $f_{i}$ and $F_{i}(\{x_{k}\})$, respectively, where $x_{k}$ is the $k$-th free parameter; i.e., $x_{1}=\text{[Fe/H]}$ in dex, $x_{2}=v_{\mathrm{broad}}$ in \si{km.s^{-1}}, and $x_{3}=\Delta \mathrm{RV}$ in \si{km.s^{-1}}. 
In the following description, we omit the variables $\{x_{k}\}$ part in the notation unless ambiguity arises. 
Our goal is to determine the set of parameters $(x_{1},x_{2},x_{3},C)$ that minimizes the difference between the observed and synthesized spectra, given by 
\begin{equation}
\min _{\{x_{k}\},C}\chi ^{2}\equiv \frac{1}{N}\sum _{i}(f_{i}-CF_{i})^{2}=\frac{1}{N}\abs{\vec{f}-C\vec{F}}^{2}\label{eq:chi2}\text{,}
\end{equation}
where $\vec{f}$ and $\vec{F}$ denote the column vectors of sets of $\{f_{i}\}$ and $\{F_{i}\}$, respectively. 
Calculating the partial derivatives of $\chi ^{2}$ with respect to $\{x_{k}\}$ and $C$, we obtain the conditions 
\begin{align}
&(\vec{f}-C\vec{F})^{\mathrm{T}}\vec{F}=0\label{eq:chi2C} \\
&\forall l,\ g_{l}\equiv (\vec{f}-C\vec{F})^{\mathrm{T}}\PD{(C\vec{F})}{x_{l}}=0\label{eq:chi2xl}\text{,}
\end{align}
where the superscript $\mathrm{T}$ indicates the transpose of the vector. 
From Eq.~(\ref{eq:chi2C}), the optimized $C$ value is analytically calculated as 
\begin{equation}
C=\frac{\vec{f}^{\mathrm{T}}\vec{F}}{\abs{\vec{F}}^{2}}\text{,}
\end{equation}
and thus $C$ can be treated as a function of $\{x_{k}\}$, rather than a free parameter of the fitting. 
The problem is thereby reduced to the three equations in Eq.~(\ref{eq:chi2xl}) with three independent variables, $x_{1}$, $x_{2}$, and $x_{3}$. 

In order to solve Eq.~(\ref{eq:chi2xl}) with the Newton-Raphson method, we numerically calculate the Jacobian matrix of $(g_{1}\ g_{2}\ g_{3})^{\mathrm{T}}$, whose elements are 
\begin{equation}
J_{lk}\equiv \PD{g_{l}}{x_{k}}=-\left(\PD{(C\vec{F})}{x_{k}}\right)^{\mathrm{T}}\PD{(C\vec{F})}{x_{l}}+(\vec{f}-C\vec{F})^{\mathrm{T}}\PDPD{(C\vec{F})}{x_{k}}{x_{l}}\label{eq:Jacobian}\text{.}
\end{equation}
The second term on the right-hand side of Eq.~(\ref{eq:Jacobian}), i.e., the second derivative of $C\vec{F}$, is ignored, following the argument by \citet{Takeda1995a}. 
The parameter $J_{lk}$ is numerically approximated according to 
\begin{equation}
\PD{(C\vec{F})}{x_{k}}\simeq \frac{(C\vec{F})(x_{k}+\Delta x_{k})-(C\vec{F})(x_{k}-\Delta x_{k})}{2\Delta x_{k}}\label{eq:differential}\text{,}
\end{equation}
where $\Delta x_{k}$ is a small variation in $x_{k}$, $\Delta x_{1}=0.1\ur{dex}$, and $\Delta x_{2}=\Delta x_{3}=0.001\ur{\si{km.s^{-1}}}$. 
Then, $\{x_{k}\}$ is updated as 
\begin{equation}
x_{k}\mapsto x_{k}+dx_{k},\quad \begin{pmatrix}dx_{1} \\ dx_{2} \\ dx_{3}\end{pmatrix}=-J^{-1}\begin{pmatrix}g_{1} \\ g_{2} \\ g_{3}\end{pmatrix}\text{.}
\end{equation}

The procedure is repeated from the beginning with updated $\{x_{k}\}$ until the end condition, 
\begin{equation}
(dx_{1})^{2}+(0.1dx_{2})^{2}+(0.1dx_{3})^{2}<2\times 10^{-4}\text{,}
\end{equation}
is satisfied. 
Here, the ratios of the weights in $dx_{k}$ ($1$, $0.1$, and $0.1$ for $k=1,2,3$, respectively) roughly correspond to the ratios of $\iPD{(C\vec{F})}{x_{k}}$. 
The threshold of the end condition, $2\times 10^{-4}$, is adopted in order to achieve the numerical error in the [Fe/H] value smaller than $0.01\ur{dex}$ after several tests. 

In the usual cases where no numerical problem arises, the iteration converges within ${\sim }10$ steps. 
In reality, however, $x_{k}$ sometimes oscillates with an amplitude larger than the end condition. 
Such an oscillation most frequently occurs when contaminating line(s) hampers a good reproduction of the observed spectrum with a synthesized counterpart. 
To avoid the oscillation, we introduce a damping parameter~\citep[e.g.,][]{Mansfield1991,Xu2016} in the eighth iteration and later. 
Specifically, we use $0.3$ times smaller steps in the updates than in the standard steps; in other words, we update the variables as $x_{k}\mapsto x_{k}+0.3dx_{k}$ when the number of iterations is eight or larger. 

During an iteration, if one of the following five conditions is met, we regard the iteration as a failure and immediately abort it: (i)~$\vec{F}=\vec{0}$ or $\det J=0$ when $x_{1}$ becomes unrealistically small or large, (ii)~$\abs{x_{1}}>10\ur{dex}$, (iii)~$x_{2}>100\ur{\si{km.s^{-1}}}$, (iv)~$x_{2}\leq 0\ur{\si{km.s^{-1}}}$, or (v)~$\abs{x_{3}}>10\ur{\si{km.s^{-1}}}$. 
In addition, when the number of iterations reaches $40$ (due to oscillations of $x_{k}$ despite the introduction of the damping parameter), we stop the iteration and calculate the root mean square $\sigma _{k}$ of $\{x_{k}\}_{31\leq k\leq 40}$, which indicates the amplitude of the oscillation of $x_{k}$. 
Then, when $\sigma _{k}$ satisfies the condition 
\begin{equation}
(\sigma _{1})^{2}+(0.1\sigma _{2})^{2}+(0.1\sigma _{3})^{2}<2\times 10^{-2}\text{,}
\end{equation}
we judge, albeit with caution, that the iteration converges, and adopt the mean of $\{x_{k}\}_{31\leq k\leq 40}$ as the optimized parameter. 
Otherwise, we judge that the iteration fails. 

We note that in using the MARCS grids of model atmospheres, the model with $\text{[M/H]}=-1.55$ and $+0.95\ur{dex}$ are used when $\text{[Fe/H]}<-1.55$ and $\text{[Fe/H]}>+0.95\ur{dex}$, respectively. 
Therefore, the measurements with $\text{[Fe/H]}\ll -1.55\ur{dex}$ or $\text{[Fe/H]}\gg +0.95\ur{dex}$ would be unreliable, but we do not expect that our sample contains such metal-rich or meta-poor objects.

\section{Dependence of the strengths of \ce{CN} molecular lines on the CNO abundances}\label{app:CN}

We discuss the strengths of the \ce{CN} molecular lines appearing in the \textit{YJ}-band spectra of RSGs and the atmospheric layers from which the lines originate. 

\begin{figure}
\centering 
\includegraphics[width=9cm]{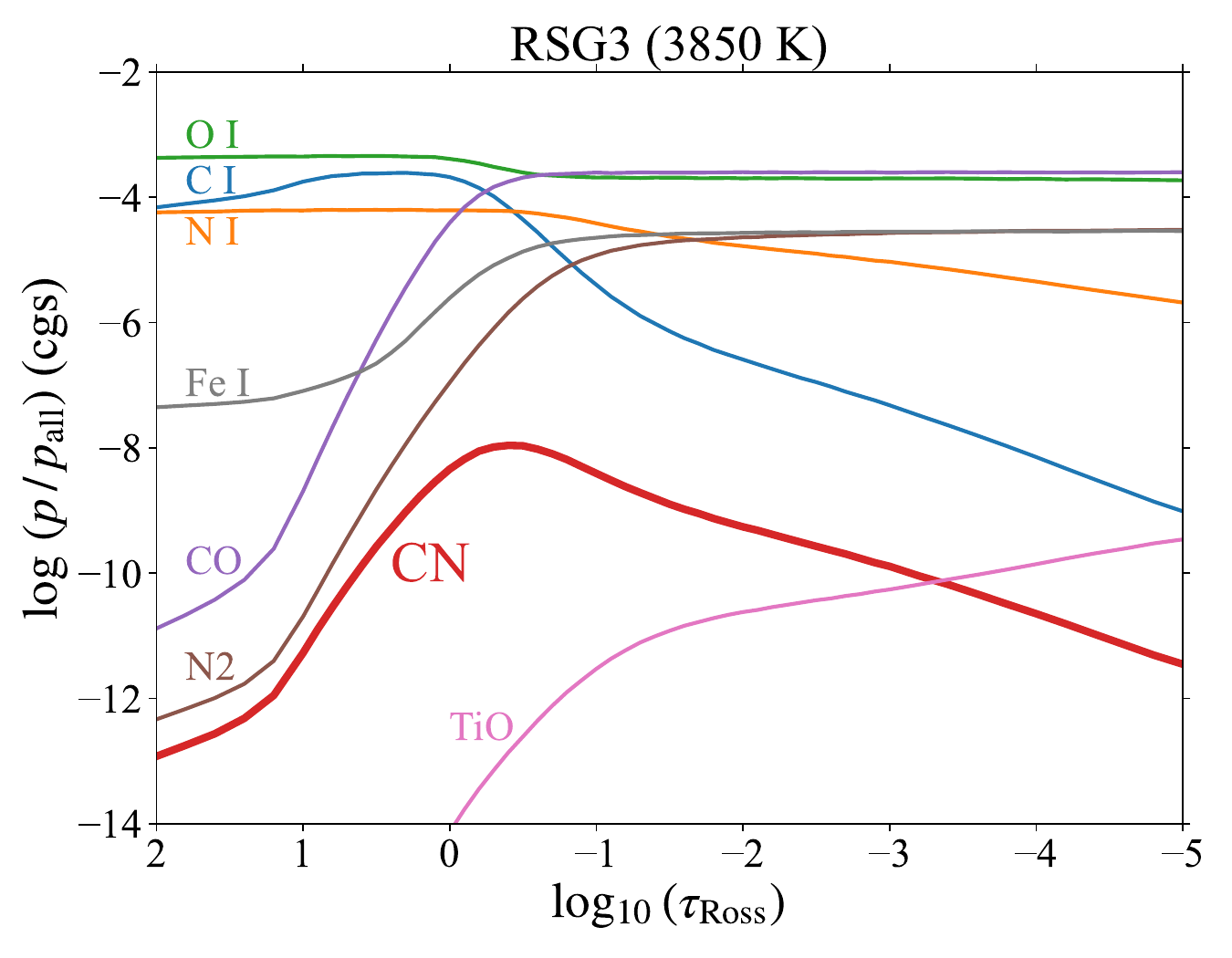}
\caption{Ratio of partial pressures $p$ of some of major molecules and atoms to the total gas pressure $p_{\mathrm{all}}$: \ce{CN} in red, \ce{CO} in purple, \ce{N2} in brown, \ce{TiO} in pink, \ion{C}{i} in blue, \ion{N}{i} in orange, \ion{O}{i} in green, and \ion{Fe}{i} in gray. The results here were calculated using MOOG for the set of stellar parameters of RSG3~(see Sect.~\ref{ssec:LineSelection}). }
\label{fig:molpressure}
\end{figure}

In contrast to some of major molecules whose lines appear in RSGs' spectra like \ce{TiO} and \ce{CO}, the \ce{CN} molecules of origin for lines in the \textit{YJ}-band exist in relatively inner atmospheric layers of RSGs. 
In fact, Fig.~\ref{fig:molpressure} demonstrates that the ratio of the partial pressure $p$ of the \ce{CN} molecules to the total gas pressure $p_{\mathrm{all}}$ is smaller in outer layers except for innermost layers (i.e.,$\log \tau _{\mathrm{Ross}}^{}\gtrsim 0$). 
This is because carbon atoms are mostly contained in \ce{CO} molecules in the outermost layers of oxygen-rich cool stars like RSGs, and thus only a small number of carbon atoms are left to form \ce{CN} molecules. 

The dependence of the \ce{CN} molecule abundance on the CNO atomic abundances varies with the atmospheric layers to which they belong. 
In the innermost layers with $\log \tau _{\mathrm{Ross}}^{}\gtrsim 0$ (corresponding to lines too weak to be detected), where neutral and/or ionized atoms are the dominant form of CNO elements, the number density of the \ce{CN} molecule, $N_{\ce{CN}}$, is approximated as 
\begin{equation}
N_{\ce{CN}}\propto {\varepsilon _{\mathrm{C}}}^{1}{\varepsilon _{\mathrm{N}}}^{1}{\varepsilon _{\mathrm{O}}}^{0}\text{,}
\end{equation}
where $\varepsilon _{\mathrm{C}}$, $\varepsilon _{\mathrm{N}}$, and $\varepsilon _{\mathrm{O}}$ indicate the total abundances of C, N, and O elements, respectively. 
In contrast, in the relatively outer layers with $\log \tau _{\mathrm{Ross}}^{}\lesssim -1.5$ (corresponding to the strongest lines with a depth deeper than ${\sim }0.2$), where \ce{CO} and \ce{N2} molecules and \ion{N}{i} and \ion{O}{i} atoms are the dominant forms of the CNO elements, $N_{\ce{CN}}$ is approximated as 
\begin{equation}
N_{\ce{CN}}\propto {\varepsilon _{\mathrm{C}}}^{a/(a-1)}{\varepsilon _{\mathrm{N}}}^{1/2}{\varepsilon _{\mathrm{O}}}^{-a/(a-1)},\quad a\equiv \varepsilon _{\mathrm{O}}/\varepsilon _{\mathrm{C}}\text{.}
\end{equation}
We note that $a\sim 2$ for RSGs and some other stars having solar C/O ratio~\citep{Asplund2009,Ekstrom2012}. 
Considering these two equations, the strengths of \ce{CN} lines principally depend on [C/O] and [N/H] among the CNO abundances. 

\section{Additional tables}\label{app:tables}

\onecolumn

\begin{sidewaystable*}
\centering 
\caption{Derived chemical abundances [X/H] in dex. }
\label{table:AbnRes}
\scalebox{0.93}{
\begin{tabular}{l c r rrrrrrrrrr} \hline \hline 
Name & List & \multicolumn{1}{c}{[Fe/H]\tablefootmark{e}} & \multicolumn{1}{c}{[Na/H]} & \multicolumn{1}{c}{[Mg/H]} & \multicolumn{1}{c}{[Al/H]} & \multicolumn{1}{c}{[Si/H]} & \multicolumn{1}{c}{[K/H]} & \multicolumn{1}{c}{[Ca/H]} & \multicolumn{1}{c}{[Ti/H]} & \multicolumn{1}{c}{[Cr/H]} & \multicolumn{1}{c}{[Ni/H]} & \multicolumn{1}{c}{[Y/H]} \\ \hline 
$Z$\tablefootmark{a} &  & \multicolumn{1}{c}{$26$} & \multicolumn{1}{c}{$11$} & \multicolumn{1}{c}{$12$} & \multicolumn{1}{c}{$13$} & \multicolumn{1}{c}{$14$} & \multicolumn{1}{c}{$19$} & \multicolumn{1}{c}{$20$} & \multicolumn{1}{c}{$22$} & \multicolumn{1}{c}{$24$} & \multicolumn{1}{c}{$28$} & \multicolumn{1}{c}{$39$} \\ \hline 
\multirow{2}{*}{$N_{\mathrm{line}}$\tablefootmark{b}} & VALD3 & \multicolumn{1}{c}{$39$} &  & \multicolumn{1}{c}{$3$} &  & \multicolumn{1}{c}{$16$} &  & \multicolumn{1}{c}{$6$} & \multicolumn{1}{c}{$24$} & \multicolumn{1}{c}{$12$} & \multicolumn{1}{c}{$3$} & \multicolumn{1}{c}{$1$} \\
 & MB99 & \multicolumn{1}{c}{$36$} & \multicolumn{1}{c}{$2$} & \multicolumn{1}{c}{$4$} & \multicolumn{1}{c}{$1$} & \multicolumn{1}{c}{$17$} & \multicolumn{1}{c}{$1$} & \multicolumn{1}{c}{$6$} & \multicolumn{1}{c}{$16$} & \multicolumn{1}{c}{$14$} & \multicolumn{1}{c}{$3$} & \multicolumn{1}{c}{$1$} \\ \hline 
\multirow{2}{*}{$\zeta $ Cep} & VALD3 & $-0.099^{+0.041}_{-0.038}$ &  & $-0.161^{+0.050}_{-0.050}$ &  & $0.482^{+0.113}_{-0.113}$ &  & $-0.081^{+0.035}_{-0.035}$ & $-0.204^{+0.058}_{-0.060}$ & $-0.292^{+0.081}_{-0.078}$ & $0.029^{+0.079}_{-0.076}$ & $0.318^{+0.090}_{-0.091}$ \\
 & MB99 & $0.087^{+0.042}_{-0.038}$ & $0.470^{+0.081}_{-0.082}$ & $0.017^{+0.059}_{-0.059}$ & $0.378^{+0.114}_{-0.115}$ & $0.671^{+0.099}_{-0.102}$ & $0.081^{+0.116}_{-0.117}$ & $0.052^{+0.054}_{-0.054}$ & $-0.121^{+0.061}_{-0.061}$ & $-0.074^{+0.076}_{-0.075}$ & $0.181^{+0.078}_{-0.078}$ & $0.447^{+0.116}_{-0.116}$ \\
\multirow{2}{*}{41 Gem} & VALD3 & $-0.076^{+0.042}_{-0.037}$ &  & $-0.072^{+0.048}_{-0.048}$ &  & $0.333^{+0.088}_{-0.089}$ &  & $-0.097^{+0.066}_{-0.066}$ & $-0.057^{+0.037}_{-0.036}$ & $-0.104^{+0.069}_{-0.066}$ & $0.045^{+0.099}_{-0.095}$ & $0.286^{+0.117}_{-0.117}$ \\
 & MB99 & $0.065^{+0.050}_{-0.045}$ & $0.646^{+0.051}_{-0.053}$ & $-0.020^{+0.038}_{-0.037}$ & $-0.339^{+0.052}_{-0.052}$ & $0.409^{+0.100}_{-0.102}$ & $0.336^{+0.069}_{-0.068}$ & $0.106^{+0.043}_{-0.044}$ & $0.108^{+0.055}_{-0.053}$ & $0.074^{+0.065}_{-0.060}$ & $0.085^{+0.082}_{-0.082}$ & $0.410^{+0.097}_{-0.099}$ \\
\multirow{2}{*}{$\xi $ Cyg} & VALD3 & $-0.096^{+0.030}_{-0.027}$ &  & $-0.004^{+0.036}_{-0.036}$ &  & $0.280^{+0.081}_{-0.081}$ &  & $-0.113^{+0.026}_{-0.026}$ & $-0.076^{+0.041}_{-0.040}$ & $-0.097^{+0.050}_{-0.051}$ & $0.016^{+0.067}_{-0.065}$ & $0.289^{+0.072}_{-0.072}$ \\
 & MB99 & $0.109^{+0.040}_{-0.035}$ & $0.485^{+0.056}_{-0.056}$ & $-0.002^{+0.041}_{-0.040}$ & $-0.371^{+0.071}_{-0.071}$ & $0.483^{+0.092}_{-0.091}$ & $0.399^{+0.085}_{-0.083}$ & $0.089^{+0.038}_{-0.037}$ & $0.148^{+0.046}_{-0.046}$ & $0.172^{+0.065}_{-0.061}$ & $0.033^{+0.064}_{-0.063}$ & $0.440^{+0.082}_{-0.082}$ \\
\multirow{2}{*}{V809 Cas} & VALD3 & $-0.065^{+0.028}_{-0.024}$ &  & $-0.175^{+0.023}_{-0.022}$ &  & $0.430^{+0.085}_{-0.082}$ &  & $-0.144^{+0.057}_{-0.055}$ & $-0.125^{+0.048}_{-0.037}$ & $-0.119^{+0.061}_{-0.054}$ & $-0.002^{+0.068}_{-0.067}$ & $0.413^{+0.057}_{-0.056}$ \\
 & MB99 & $0.037^{+0.028}_{-0.024}$ & $0.403^{+0.043}_{-0.042}$ & $-0.017^{+0.033}_{-0.031}$ & $-0.458^{+0.046}_{-0.045}$ & $0.483^{+0.084}_{-0.085}$ & $0.227^{+0.068}_{-0.064}$ & $-0.020^{+0.032}_{-0.030}$ & $0.019^{+0.049}_{-0.044}$ & $0.035^{+0.039}_{-0.043}$ & $-0.184^{+0.053}_{-0.053}$ & $0.563^{+0.064}_{-0.063}$ \\
\multirow{2}{*}{V424 Lac} & VALD3 & $-0.039^{+0.039}_{-0.035}$ &  & $-0.061^{+0.081}_{-0.080}$ &  & $0.353^{+0.125}_{-0.129}$ &  & $-0.073^{+0.070}_{-0.071}$ & $-0.109^{+0.055}_{-0.056}$ & $-0.085^{+0.061}_{-0.061}$ & $-0.002^{+0.126}_{-0.121}$ & $0.446^{+0.151}_{-0.151}$ \\
 & MB99 & $0.078^{+0.045}_{-0.039}$ & $0.481^{+0.064}_{-0.065}$ & $0.033^{+0.050}_{-0.048}$ & $-0.276^{+0.081}_{-0.080}$ & $0.465^{+0.113}_{-0.114}$ & $0.306^{+0.098}_{-0.099}$ & $0.043^{+0.067}_{-0.067}$ & $0.082^{+0.054}_{-0.057}$ & $0.107^{+0.052}_{-0.053}$ & $-0.008^{+0.088}_{-0.087}$ & $0.584^{+0.101}_{-0.100}$ \\
\multirow{2}{*}{$\psi ^{1}$ Aur} & VALD3 & $-0.259^{+0.047}_{-0.054}$ &  & $-0.133^{+0.067}_{-0.053}$ &  & $0.293^{+0.155}_{-0.172}$ &  & $-0.278^{+0.050}_{-0.047}$ & $-0.352^{+0.081}_{-0.072}$ & $-0.324^{+0.079}_{-0.081}$ & $-0.261^{+0.125}_{-0.141}$ & $0.112^{+0.168}_{-0.194}$ \\
 & MB99 & $-0.081^{+0.067}_{-0.052}$ & $0.091^{+0.072}_{-0.070}$ & $-0.088^{+0.057}_{-0.052}$ & $-0.538^{+0.110}_{-0.118}$ & $0.420^{+0.143}_{-0.146}$ & $-0.007^{+0.113}_{-0.104}$ & $-0.106^{+0.048}_{-0.045}$ & $-0.204^{+0.079}_{-0.071}$ & $-0.111^{+0.066}_{-0.058}$ & $-0.069^{+0.138}_{-0.151}$ & $0.261^{+0.191}_{-0.217}$ \\
\multirow{2}{*}{TV Gem} & VALD3 & $-0.148^{+0.095}_{-0.107}$ &  & $-0.157^{+0.058}_{-0.050}$ &  & $0.409^{+0.244}_{-0.252}$ &  & $-0.173^{+0.074}_{-0.078}$ & $-0.237^{+0.129}_{-0.119}$ & $-0.168^{+0.133}_{-0.123}$ & $-0.201^{+0.175}_{-0.168}$ & $0.545^{+0.210}_{-0.237}$ \\
 & MB99 & $-0.025^{+0.089}_{-0.065}$ & $0.543^{+0.086}_{-0.082}$ & $-0.045^{+0.075}_{-0.066}$ & $-0.557^{+0.088}_{-0.098}$ & $0.390^{+0.196}_{-0.179}$ & $0.164^{+0.129}_{-0.120}$ & $-0.022^{+0.061}_{-0.056}$ & $-0.089^{+0.115}_{-0.096}$ & $0.069^{+0.112}_{-0.097}$ & $-0.026^{+0.173}_{-0.174}$ & $0.661^{+0.204}_{-0.232}$ \\
\multirow{2}{*}{BU Gem} & VALD3 & $-0.289^{+0.075}_{-0.091}$ &  & $-0.146^{+0.112}_{-0.118}$ &  & $0.032^{+0.189}_{-0.208}$ &  & $-0.217^{+0.064}_{-0.063}$ & $-0.066^{+0.112}_{-0.119}$ & $-0.088^{+0.114}_{-0.133}$ & $-0.444^{+0.134}_{-0.138}$ & $0.440^{+0.188}_{-0.195}$ \\
 & MB99 & $-0.129^{+0.046}_{-0.045}$ & $0.751^{+0.124}_{-0.125}$ & $-0.279^{+0.080}_{-0.079}$ & $-0.565^{+0.170}_{-0.180}$ & $0.168^{+0.160}_{-0.166}$ & $0.328^{+0.187}_{-0.186}$ & $-0.059^{+0.049}_{-0.049}$ & $0.042^{+0.101}_{-0.102}$ & $0.114^{+0.087}_{-0.086}$ & $-0.208^{+0.151}_{-0.154}$ & $0.590^{+0.222}_{-0.232}$ \\
\multirow{2}{*}{Betelgeuse} & VALD3 & $-0.111^{+0.076}_{-0.061}$ &  & $-0.285^{+0.112}_{-0.111}$ &  & $0.425^{+0.117}_{-0.112}$ &  & $-0.214^{+0.061}_{-0.060}$ & $-0.251^{+0.053}_{-0.052}$ & $-0.135^{+0.091}_{-0.071}$ & $-0.081^{+0.152}_{-0.149}$ & $0.499^{+0.210}_{-0.209}$ \\
 & MB99 & $-0.064^{+0.050}_{-0.042}$ & $0.297^{+0.067}_{-0.063}$ & $-0.185^{+0.052}_{-0.046}$ & $-0.455^{+0.073}_{-0.072}$ & $0.410^{+0.105}_{-0.104}$ & $0.216^{+0.104}_{-0.096}$ & $-0.093^{+0.047}_{-0.046}$ & $-0.151^{+0.057}_{-0.049}$ & $-0.014^{+0.064}_{-0.055}$ & $-0.022^{+0.094}_{-0.092}$ & $0.562^{+0.108}_{-0.108}$ \\
\multirow{2}{*}{NO Aur} & VALD3 & $-0.078^{+0.050}_{-0.046}$ &  & $-0.180^{+0.081}_{-0.081}$ &  & $0.370^{+0.095}_{-0.088}$ &  & $-0.302^{+0.115}_{-0.114}$ & $-0.193^{+0.049}_{-0.043}$ & $-0.105^{+0.089}_{-0.080}$ & $-0.050^{+0.121}_{-0.117}$ & $0.530^{+0.156}_{-0.155}$ \\
 & MB99 & $-0.056^{+0.055}_{-0.050}$ & $0.210^{+0.051}_{-0.048}$ & $-0.092^{+0.045}_{-0.042}$ & $-0.570^{+0.056}_{-0.055}$ & $0.341^{+0.082}_{-0.078}$ & $0.131^{+0.081}_{-0.075}$ & $-0.105^{+0.067}_{-0.066}$ & $-0.069^{+0.064}_{-0.061}$ & $-0.005^{+0.050}_{-0.052}$ & $-0.244^{+0.070}_{-0.070}$ & $0.580^{+0.085}_{-0.084}$ \\
\hline 
\multirow{2}{*}{Mean\tablefootmark{c}} & VALD3 & $-0.125$ &  & $-0.242$ &  & $+0.181$ &  & $-0.216$ & $-0.285$ & $-0.318$ & $-0.047$ & $+0.141$ \\
 & MB99 & $-0.004$ & $+0.041$ & $-0.155$ & $-0.538$ & $+0.266$ & \multicolumn{1}{c}{\tablefootmark{d}} & $-0.066$ & $-0.147$ & $-0.139$ & $-0.064$ & $+0.297$ \\
\multirow{2}{*}{SD\tablefootmark{c}} & VALD3 & $0.048$ &  & $0.075$ &  & $0.086$ &  & $0.048$ & $0.085$ & $0.076$ & $0.105$ & $0.097$ \\
 & MB99 & $0.060$ & $0.174$ & $0.074$ & $0.181$ & $0.096$ & \multicolumn{1}{c}{\tablefootmark{d}} & $0.073$ & $0.118$ & $0.076$ & $0.139$ & $0.086$ \\
\hline 
\end{tabular}
}
\tablefoot{
\tablefoottext{a}{Atomic number. }
\tablefoottext{b}{Number of the absorption lines used to determine abundances. The same as in Table~\ref{table:AbnFeRes}. }
\tablefoottext{c}{The weighted mean and standard deviation of [X/H] of the target RSGs after subtracting the radial abundance gradient traced with Cepheids using the Cepheids' abundances presented by \citet{Luck2018}. }
\tablefoottext{d}{[K/H] of Cepheids were not given by \citet{Luck2018}. }
\tablefoottext{e}{Iron abundance or global metallicity, which are also listed in Table~\ref{table:AtmosRes}. }
}
\end{sidewaystable*}

\begin{sidewaystable*}
\centering 
\caption{Derived chemical abundances [X/Fe] in dex. }
\label{table:AbnFeRes}
\scalebox{0.93}{
\begin{tabular}{l c rrrrrrrrrr} \hline \hline 
Name & List & \multicolumn{1}{c}{[Na/Fe]} & \multicolumn{1}{c}{[Mg/Fe]} & \multicolumn{1}{c}{[Al/Fe]} & \multicolumn{1}{c}{[Si/Fe]} & \multicolumn{1}{c}{[K/Fe]} & \multicolumn{1}{c}{[Ca/Fe]} & \multicolumn{1}{c}{[Ti/Fe]} & \multicolumn{1}{c}{[Cr/Fe]} & \multicolumn{1}{c}{[Ni/Fe]} & \multicolumn{1}{c}{[Y/Fe]} \\ \hline 
$Z$\tablefootmark{a} &  & \multicolumn{1}{c}{$11$} & \multicolumn{1}{c}{$12$} & \multicolumn{1}{c}{$13$} & \multicolumn{1}{c}{$14$} & \multicolumn{1}{c}{$19$} & \multicolumn{1}{c}{$20$} & \multicolumn{1}{c}{$22$} & \multicolumn{1}{c}{$24$} & \multicolumn{1}{c}{$28$} & \multicolumn{1}{c}{$39$} \\ \hline 
\multirow{2}{*}{$N_{\mathrm{line}}$\tablefootmark{b}} & VALD3 &  & \multicolumn{1}{c}{$3$} &  & \multicolumn{1}{c}{$16$} &  & \multicolumn{1}{c}{$6$} & \multicolumn{1}{c}{$24$} & \multicolumn{1}{c}{$12$} & \multicolumn{1}{c}{$3$} & \multicolumn{1}{c}{$1$} \\
 & MB99 & \multicolumn{1}{c}{$2$} & \multicolumn{1}{c}{$4$} & \multicolumn{1}{c}{$1$} & \multicolumn{1}{c}{$17$} & \multicolumn{1}{c}{$1$} & \multicolumn{1}{c}{$6$} & \multicolumn{1}{c}{$16$} & \multicolumn{1}{c}{$14$} & \multicolumn{1}{c}{$3$} & \multicolumn{1}{c}{$1$} \\ \hline 
\multirow{2}{*}{$\zeta $ Cep} & VALD3 &  & $-0.062^{+0.068}_{-0.070}$ &  & $0.581^{+0.116}_{-0.117}$ &  & $0.018^{+0.055}_{-0.057}$ & $-0.105^{+0.072}_{-0.076}$ & $-0.193^{+0.093}_{-0.091}$ & $0.128^{+0.078}_{-0.076}$ & $0.417^{+0.093}_{-0.095}$ \\
 & MB99 & $0.383^{+0.090}_{-0.092}$ & $-0.070^{+0.065}_{-0.067}$ & $0.291^{+0.113}_{-0.113}$ & $0.584^{+0.098}_{-0.102}$ & $-0.006^{+0.123}_{-0.125}$ & $-0.035^{+0.068}_{-0.071}$ & $-0.208^{+0.076}_{-0.078}$ & $-0.161^{+0.086}_{-0.087}$ & $0.094^{+0.074}_{-0.076}$ & $0.360^{+0.115}_{-0.116}$ \\
\multirow{2}{*}{41 Gem} & VALD3 &  & $0.003^{+0.062}_{-0.064}$ &  & $0.408^{+0.076}_{-0.080}$ &  & $-0.022^{+0.079}_{-0.082}$ & $0.019^{+0.049}_{-0.050}$ & $-0.028^{+0.072}_{-0.071}$ & $0.120^{+0.079}_{-0.079}$ & $0.362^{+0.102}_{-0.105}$ \\
\hline 
\end{tabular}
}
\tablefoot{
\tablefoottext{a}{Atomic number. }
\tablefoottext{b}{Number of the absorption lines used to determine abundances. The same as in Table~\ref{table:AbnRes}. }
\tablefoottext{c}{The weighted mean and standard deviation of [X/Fe] of the target RSGs after subtracting the radial abundance gradient traced with Cepheids using the Cepheids' abundances presented by \citet{Luck2018}. }
\tablefoottext{d}{[K/Fe] of Cepheids were not given by \citet{Luck2018}. }
}
\end{sidewaystable*}

\begin{landscape}
\begin{small}
\begin{longtable}{cccccccccc cccccccccc}
\caption{Line list from VALD3~\citep{Ryabchikova2015} and MB99~\citep{Melendez1999}. See main text for the definitions of the listed quantities. }
\label{table:LineListVALDMelendez} \\ \hline \hline 
\multicolumn{10}{c}{VALD3} & \multicolumn{10}{c}{MB99} \\ \cmidrule(rl){1-10} \cmidrule(rl){11-20}
$\lambda _{\mathrm{air}}$ & EP & $\log gf$ & $\log \tau _{\mathrm{Ross}}^{}$ & $d_{\texttt{OnlyOne}}$ & $\beta _{1}$ & $\beta _{2}$ & $N_{i}$ & $\Delta \text{[X/H]}_{i}$ & Used & $\lambda _{\mathrm{air}}$ & EP & $\log gf$ & $\log \tau _{\mathrm{Ross}}^{}$ & $d_{\texttt{OnlyOne}}$ & $\beta _{1}$ & $\beta _{2}$ & $N_{i}$ & $\Delta \text{[X/H]}_{i}$ & Used \\
{(\AA )} & {(eV)} & {(dex)} & & & & & & {(dex)} & & {(\AA )} & {(eV)} & {(dex)} & & & & & & {(dex)} & \\ \hline 
\endfirsthead 
\caption{continued.} \\ \hline \hline 
\multicolumn{10}{c}{VALD3} & \multicolumn{10}{c}{MB99} \\ \cmidrule(rl){1-10} \cmidrule(rl){11-20}
$\lambda _{\mathrm{air}}$ & EP & $\log gf$ & $\log \tau _{\mathrm{Ross}}^{}$ & $d_{\texttt{OnlyOne}}$ & $\beta _{1}$ & $\beta _{2}$ & $N_{i}$ & $\Delta \text{[X/H]}_{i}$ & Used & $\lambda _{\mathrm{air}}$ & EP & $\log gf$ & $\log \tau _{\mathrm{Ross}}^{}$ & $d_{\texttt{OnlyOne}}$ & $\beta _{1}$ & $\beta _{2}$ & $N_{i}$ & $\Delta \text{[X/H]}_{i}$ & Used \\
{(\AA )} & {(eV)} & {(dex)} & & & & & & {(dex)} & & {(\AA )} & {(eV)} & {(dex)} & & & & & & {(dex)} & \\ \hline 
\endhead 
\hline 
\endfoot 
\hline 
\endlastfoot 
\multicolumn{10}{l}{\ion{Fe}{i} ($39$ out of $83$ lines used)} & \multicolumn{10}{l}{\ion{Fe}{i} ($36$ out of $72$ lines used)} \\ \cmidrule(rl){1-10} \cmidrule(rl){11-20}
$9800.30752$ & $5.086$ & $-0.453$ & $-1.83$ & $0.199$ & $0.404$ & $0.493$ & $10$ & $+0.078$ & \checkmark & & & & & & & & & & \\
$9820.24079$ & $2.424$ & $-5.073$ & $-1.54$ & $0.123$ & $0.310$ & $0.727$ & $3$ & $+0.066$ & & & & & & & & & & & \\
$9861.73375$ & $5.064$ & $-0.142$ & $-2.16$ & $0.235$ & $0.495$ & $0.350$ & $10$ & $-0.064$ & \checkmark & & & & & & & & & & \\
$9868.18574$ & $5.086$ & $-0.979$ & $-1.33$ & $0.141$ & $0.316$ & $0.267$ & $10$ & $+0.317$ & \checkmark & & & & & & & & & & \\
$9889.03509$ & $5.033$ & $-0.446$ & $-1.92$ & $0.208$ & $0.256$ & $0.720$ & $10$ & $+0.304$ & \checkmark & & & & & & & & & & \\
$9937.08979$ & $4.593$ & $-2.442$ & $-0.80$ & $0.060$ & $0.380$ & $0.574$ & $10$ & $-0.187$ & \checkmark & & & & & & & & & & \\
$9980.46288$ & $5.033$ & $-1.379$ & $-1.07$ & $0.105$ & $0.483$ & $0.551$ & $10$ & $-0.075$ & \checkmark & & & & & & & & & & \\
\end{longtable}
\end{small}
\end{landscape}

\end{appendix}

\end{document}